\documentclass[traditabstract]{aa} 
\usepackage{graphicx}
\usepackage{txfonts}
\usepackage{rotating}
\usepackage{multirow}
\usepackage{url}
\usepackage{txfonts}
\usepackage{pstricks}
\usepackage{natbib}
\usepackage{subfigure}
\usepackage{enumerate}

\newcommand{\avgg}[1]{\left< #1 \right>} 

\def\OIII{[\ion{O}{III}]}

\def\OIIIb{[\ion{O}{III}]\,$\lambda$ 1663}
\def\Hbeta{\ion{H}{$\beta$}}
\def\Halpha{\ion{H}{$\alpha$}}
\def\HeII{[\ion{He}{II}]}
\def\AlIII{\ion{Al}{III}}
\def\SiIII{\ion{Si}{III}]}

\def\OVI{[\ion{O}{VI}]}
\def\MgII{\ion{Mg}{II}}

\def\Lyalpha{\ion{Ly}{$\alpha$}}
\def\CIII{\ion{C}{III]}}
\def\CIV{\ion{C}{IV}}
\def\NeV{[\ion{Ne}{V}]}

\def\OII{[\ion{O}{II}]}

\def\FeII{\ion{Fe}{II}}

\begin{document}
   \title{Microlensing of the broad line region in 17 lensed quasars\thanks{Based on observations made 
 with the ESO-VLT Unit Telescope \#~2 Kueyen 
 (Cerro Paranal, Chile; Proposals 074.A-0563, 075.A-0377, 077.A-0155, PI: G. Meylan).}$^{\rm{,}}$ \thanks{The new spectra presented in this paper are available in electronic form at the CDS via anonymous ftp to cdsarc.u-strasbg.fr (130.79.128.5) or via {\url{http://cdsweb.u-strasbg.fr/cgi-bin/qcat?J/A+A/}} and via the German virtual observatory {\url{http://dc.g-vo.org/mlqso/q/web/form}}.}}

\author{ D.\,Sluse\inst{1,2} \and D.\,Hutsem\'ekers\inst{3} \and F.\,Courbin\inst{4} \and G.\,Meylan \inst{4} \and J.\,Wambsganss \inst{1}}

\institute{ Astronomisches Rechen-Institut am Zentrum f\"ur Astronomie
  der Universit\"at Heidelberg M\"onchhofstrasse 12-14, 69120
  Heidelberg, Germany 
\and 
Argelander-Institut f\"ur Astronomie, Auf dem H\"ugel 71, 53121 Bonn, Germany
\email{dsluse@astro.uni-bonn.de}
\and 
  F.R.S.-FNRS, Institut d'Astrophysique et
  de G\'eophysique, Universit\'e de Li\`ege, All\'ee du 6 Ao\^ut 17,
  B5c, 4000 Li\`ege, Belgium
\and
  Laboratoire d'Astrophysique, Ecole
  Polytechnique F\'ed\'erale de Lausanne (EPFL), Observatoire de
  Sauverny, 1290 Versoix, Switzerland  
 }

\date{Received 28 February 2012 ; accepted 25 May 2012}

  \abstract
{When an image of a strongly lensed quasar is microlensed, the different components of its spectrum are expected to be differentially magnified owing to the different sizes of the corresponding emitting region. Chromatic changes are expected to be observed in the continuum while the emission lines should be deformed as a function of the size, geometry and kinematics of the regions from which they originate. Microlensing of the emission lines has been reported only in a handful of systems so far. In this paper we search for microlensing deformations of the optical spectra of pairs of images in 17 lensed quasars with bolometric luminosities between $10^{44.7-47.4}$\,erg/s and black hole masses $10^{7.6-9.8} M_{\sun}$. This sample is composed of 13 pairs of previously unpublished spectra and four pairs of spectra from literature. Our analysis is based on a simple spectral decomposition technique which allows us to isolate the microlensed fraction of the flux independently of a detailed modeling of the quasar emission lines. Using this technique, we detect microlensing of the continuum in 85\% of the systems. Among them, 80\% show microlensing of the broad emission lines. Focusing on the most common emission lines in our spectra (\CIII~and \MgII) we detect microlensing of either the blue or the red wing, or of both wings with the same amplitude. This observation implies that the broad line region is not in general spherically symmetric. In addition, the frequent detection of microlensing of the blue and red wings independently but not simultaneously with a different amplitude, does not support existing microlensing simulations of a biconical outflow. Our analysis also provides the intrinsic flux ratio between the lensed images and the magnitude of the microlensing affecting the continuum. These two quantities are particularly relevant for the determination of the fraction of matter in clumpy form in galaxies and for the detection of dark matter substructures via the identification of flux ratio anomalies. }


   \keywords{Gravitational lensing: micro, strong, quasars: general, quasars: emission lines, line: formation }

   \maketitle

\section{Introduction}


Soon after the discovery of the first gravitationally lensed quasar, it has been realised that microlensing (ML) produced by compact objects in the lensing galaxy towards a multiply imaged quasar could be used as an astrophysical tool to probe the inner parsecs of distant quasars \citep{Chang1979, Kayser1986, Paczynski1986, Grieger1988}. Microlenses typically magnify regions of the source on scales similar to or smaller than a few micro-arcsecs, the size of their angular Einstein radius $R_E$ \citep{Wambsganss1998, Wambsganss2006, Schmidt2010}. Hence, the quasar continuum region (accretion disc) and the broad line region are likely to be microlensed. Because the magnification varies with the source size, the power law continuum emission of quasars is expected to be more magnified as the wavelength, and hence the continuum size, decreases. Microlensing is therefore expected to produce significant color changes in macro-lensed quasar images \citep{Wambsganss1991}. The latter have indeed been observed \citep[e.g.][]{Wisotzki1993, Claeskens2001, Burud2002a, Wucknitz2003}. The effect on the broad line emitting region (BLR) has been first addressed by \cite{Nemiroff1988} who calculated the changes produced by a single microlensing-star on the emission lines. This analysis has been refined soon after by \cite{Schneider1990a} who considered the more realistic case of ML by a population of microlenses. These papers have demonstrated that microlensing of the BLR could be significant and does depend on the BLR geometry and its kinematics. They also showed that ML of a spherically symmetric BLR (in geometry and velocity field) would lead to symmetric variations of the emission lines (i.e. of the blue and red components) while ML of a keplerian disc would lead in general to asymmetric variations of the emission lines and possible shift of the line centroid. Microlensing affecting more peculiar line profiles from e.g. broad absorption lines quasars, or generated in a relativistic disc have been discussed in \cite{Hutsemekers1993, Hutsemekers1994, Lewis1998b, Belle2000, Popovic2001}. Despite these promising results, detection of microlensing in the emission lines remained elusive \citep{Fillipenko1989a, Lewis1998} or invoked to explain differences between the spectra of candidate lensed quasars \citep{Steidel1991, Small1997}. The interest in BLR microlensing got revived with the papers of \cite{Abajas2002} and of \cite{Lewis2004a}, who re-investigated this question after the discovery that the BLR was smaller than previously thought \citep{Wandel1999, Kaspi2000}. Based on the size of the BLR measured in NGC5548 and using the scaling relation $R_{BLR} \propto L^{0.7}$, \cite{Abajas2002} estimated that the BLR should be significantly microlensed in about $\sim$ 30\% of the systems. They also extended the work of \cite{Nemiroff1988}, and calculated the microlensing by a single lens for various BLR geometries and kinematics, considering BLR models described in \cite{Robinson1995}. \cite{Lewis2004a} extended the work of \cite{Abajas2002} by using more realistic microlensing patterns. Finally, microlensing of a biconical BLR, already presented in \cite{Abajas2002} and in \cite{Lewis2004a}, for two peculiar orientations of the axis of the bicone, has been discussed for more general bicone configurations in \cite{Abajas2007}. These papers have confirmed most of the earlier findings and made more detailed predictions on the line-shifts and asymetries induced by ML. They also showed that the line deformation depends only weakly on the value of the surface density $\kappa$ and shear $\gamma$ at the position of the lensed images, but more strongly on the orientation on the BLR w.r.t. to the direction of the shear. 

Many papers dedicated to the detection and interpretation of a microlensing signal have focused on the Einstein Cross $\equiv$ Q2237+0305, which is probably the most favourable object for microlensing studies. Indeed, this system has a negligible time-delay between the lensed images, which enables one to easily disentangle microlensing and intrinsic variability, and a low-redshift lensing galaxy which leads to a small $R_E$ and to relatively fast microlensing variations \citep{Mosquera2011}. After the first detection of microlensing in the continuum and in the broad line \citep{Irwin1989, Fillipenko1989a}, microlensing has started to be used as a tool to constrain the size of the accretion disc and of the broad line emitting region \citep{Lewis1998, Wyithe2000a, Yonehara2000, Wyithe2000b}. In the last decade, important progress in observational techniques allowed photometric and spectrophotometric monitoring of the individual lensed images to be successfully carried out \citep{Wozniak2000, Dai2003a, Anguita2008a, Eigenbrod2008a, Zimmer2011}. On the other hand, the development of more advanced numerical techniques allowed quicker calculation of magnification maps and more sophisticated analysis \citep{Kochanek2004a, Poindexter2010a, Poindexter2010b, Bate2010a,  Mediavilla2011b, Garsden2011b, Bate2012a}. Owing to these two ingredients, tight constraints on the size of the accretion disc and on its temperature profile, on the size of the broad line region and on its kinematics have been obtained for Q2237+0305 \citep{Eigenbrod2008a, Poindexter2010a, Odowd2011, Sluse2011a}. 

Most of the recent papers focused on the study of the quasar accretion disc, which can be done using broad band photometry from X-ray to optical and near-infrared wavelengths \citep[e.g.][]{Pooley2007a, Floyd2009a, Hutsemekers2010, Dai2010a, Blackburne2011a, Munoz2011}. Studies of BLR microlensing are more sparse and detections have been reported only for a handful of systems \citep{Richards2004a, Wayth2005, Keeton2006a, Sluse2007, Hutsemekers2010}. In this paper, we present a careful re-extraction and analysis of archive spectra of a sample of 13 lensed quasars initially observed with the aim of measuring the redshift of the lensing galaxy. Our systematic analysis allow us to characterise the microlensing-induced deformation of the emission lines. To get a more complete overview of the microlensing signal, we also discuss the signal detected in four objects we presented elsewhere. 


From our spectra, we also derive flux ratios corrected for microlensing which are closer to the intrinsic flux ratios between pairs of images. This is important for the study of doubly imaged quasars for which the flux ratios are mandatory to constrain the lens models, because of the few observational constraints in these systems \citep[e.g.][]{Chantry2010, Sluse2011b}. Intrinsic flux ratios are also particularly relevant for the identification of flux ratio anomalies possibly produced by dark matter substructures or ML.  A popular technique to study the amount of massive dark matter substructures in galaxies relies on the identification of flux ratios between lensed image pairs which deviate from lens model prediction, i.e. the so called flux ratio anomaly \citep[][ and references therein]{Mao1998, Metcalf2001, Dalal2002, Keeton2003, Fadely2011a, Zackrisson2010}. One of the current limitation of this technique is the small number of reliable flux ratios which may be used to identify an anomaly. Indeed, most of the lensed quasars are observed at visible and near-infrared wavelengths where microlensing and differential extinction significantly contaminate the flux ratios, while only a handful of systems are detected in the mid-infrared or at radio wavelengths where these two effects are negligible. In this paper we discuss how flux ratios derived using spectra may provide a good proxy to intrinsic flux ratios, allowing one to significantly extend the sample of objects where flux ratio anomalies can be studied. 

The structure of our paper is as follows. In Sect.~\ref{sec:data}, we present the extraction and flux calibration of the archive spectra, and the spectra from literature. In Sect.~\ref{sec:properties}, we derive the physical properties (bolometric luminosity, black hole mass, Eddington ratio) of the lensed quasars. We also discuss there the published redshifts of the quasar and provide an alternative value based on the \MgII~emission line. Sect.~\ref{sec:decomposition} is devoted to the analysis of the microlensing in the spectra. It includes a presentation of the technique used to isolate the microlensed fraction of the flux, a description of the microlensing signal observed in each object and a discussion on the accuracy of our microlensing-corrected flux ratios. In Sect.~\ref{sec:discussion}, we discuss the microlensing signal observed in the continuum, the occurrence and variety of microlensing of the broad lines and the consequences for the structure of the BLR. Finally, we summarize our main results in Sect.~\ref{sec:conclusions}.

\section{Observations and data processing}
\label{sec:data}

\subsection{Observations}
\label{subsec:obs}

We have gathered spectroscopic observations of 13 gravitationally lensed quasars previously presented in \citet{Eigenbrod2006b, Eigenbrod2007}. The spectra were obtained in multi-object mode with the FORS1 instrument mounted on the ESO-VLT Unit Telescope 2 Kueyen (Cerro Paranal, Chile). The object names and main observational characteristics of the data are presented in Table~\ref{tab:data}. The following information is provided: name of the object, observing date, total exposure time, average seeing, average airmass, and reference where more details regarding the observations and data reduction can be found. All the spectra were obtained through 1$''$ slit width with the G300V grism and the GG435 order sorting filter. This setup provides us with spectra from 4400 to 8650\,$\AA$. All the objects, except SDSS~J1226-0006, were observed with the High Resolution collimator of FORS, leading to a spectral resolution $R = \lambda / \Delta \lambda \sim $ 210 at 5900\,$\AA$, and to a pixel scale of 2.9\,$\AA$ in the spectral direction and of 0.1$''$ in the spatial direction. For SDSS J1226-006, the Standard Resolution mode of the FORS instrument was used, leading to 0.2$''$ per pixel in the spatial direction and to $R \sim $ 400 at 5900\,$\AA$. 

\begin{table*}[t!]
\begin{center}
\begin{tabular}{l|cc|ccccccl}
\hline
Object & $z_s$ & $z_l$ & Images & Date & Date (MJD) & Exp.(s) & Seeing [$''$] &  Airmass & Ref. \\
\hline
(a) HE~0047-1756      & 1.678$^\dagger$ & 0.407 & B-A & 18-07-2005 & 53569 & 2$\times$1400 & 0.51 & 1.24 & 1 \\
(b) Q0142-100         & 2.719 & 0.491 & B-A & 11-08-2006 & 53958 & 2$\times$1400 & 0.80 & 1.10 & 2 \\
                  &       &       &     & 19-08-2006 & 53966 & 2$\times$1400 & 0.79 & 1.61 & 2 \\
(c) SDSS~J0246-0825   & 1.689$^\dagger$ & 0.723 & B-A & 22-08-2006 & 53969 & 6$\times$1400 & 0.64 & 1.23 & 2 \\
(d) HE~0435-1223       & 1.693$^\dagger$ & 0.454 & B-D & 11-10-2004 & 53289 & 4$\times$1400 & 0.48 & 1.03 & 1 \\
                  &       &       &     & 11-11-2004 & 53320 & 2$\times$1400 & 0.57 & 1.11 & 1 \\
(e) SDSS~J0806+2006   & 1.540 & 0.573 & B-A & 22-04-2006 & 53847 & 2$\times$1400 & 0.91 & 1.56 & 2 \\
(f) FBQ~0951+2635     & 1.247$^\dagger$ & 0.260 & B-A & 31-03-2006 & 53825 & 4$\times$1400 & 0.65 & 1.60 & 2 \\
(g) BRI~0952-0115     & 4.426 & 0.632 & B-A & 23-04-2006 & 53848 & 6$\times$1400 & 0.53 & 1.13 & 2 \\ 
(h) SDSS J1138+0314   & 2.438 & 0.445 & C-B & 10-05-2005 & 53500 & 5$\times$1400 & 0.70 & 1.15 & 1 \\
(i) J1226-0006        & 1.123$^\dagger$ & 0.517 & B-A & 16-05-2005 & 53506 & 8$\times$1400 & 0.88 & 1.25 & 1 \\
(j) SDSS~J1335+0118   & 1.570 & 0.440 & B-A & 03-02-2005 & 53404 & 2$\times$1400 & 0.72 & 1.15 & 1 \\
                  &       &       &     & 03-03-2005 & 53432 & 4$\times$1400 & 0.69 & 1.15 & 1 \\
(k) Q1355-2257        & 1.370$^\dagger$ & 0.701 & B-A & 05-03-2005 & 53434 & 2$\times$1400 & 0.71 & 1.03 & 1 \\
                  &       &       &     & 20-03-2005 & 53449 & 4$\times$1400 & 0.58 & 1.08 & 1 \\
(l) WFI~2033-4723     & 1.662$^\dagger$ & 0.661 & C-B & 13-05-2005 & 53503 & 5$\times$1400 & 0.54 & 1.16 & 1 \\
(m) HE~2149-2745      & 2.033 & 0.603 & B-A & 04-08-2006 & 53951 & 6$\times$1400 & 0.62 & 1.48 & 2 \\
\hline
\end{tabular}
\end{center}
\vspace{0.2cm}
{\tiny{Notes: $\dagger$ New measurements, see Sect.~\ref{subsec:redshfits}}}
\tablebib{ 
$(1)$ \citet{Eigenbrod2006b}; $(2)$ \citet{Eigenbrod2007}
}
\caption{Data summary for our sample of 13 lenses. We give the object name, the redshift of the QSO lensed images ($z_s$), of the lensing galaxy ($z_l$), the names of the lensed images in the slit, the date of observation and exposure time, average seeing, airmass and reference to the paper where the data where first presented.}
\label{tab:data}
\end{table*}

\subsection{Reduction and deconvolution}
\label{sec:reduction}

\begin{figure*}
\centering
\includegraphics[scale=0.45]{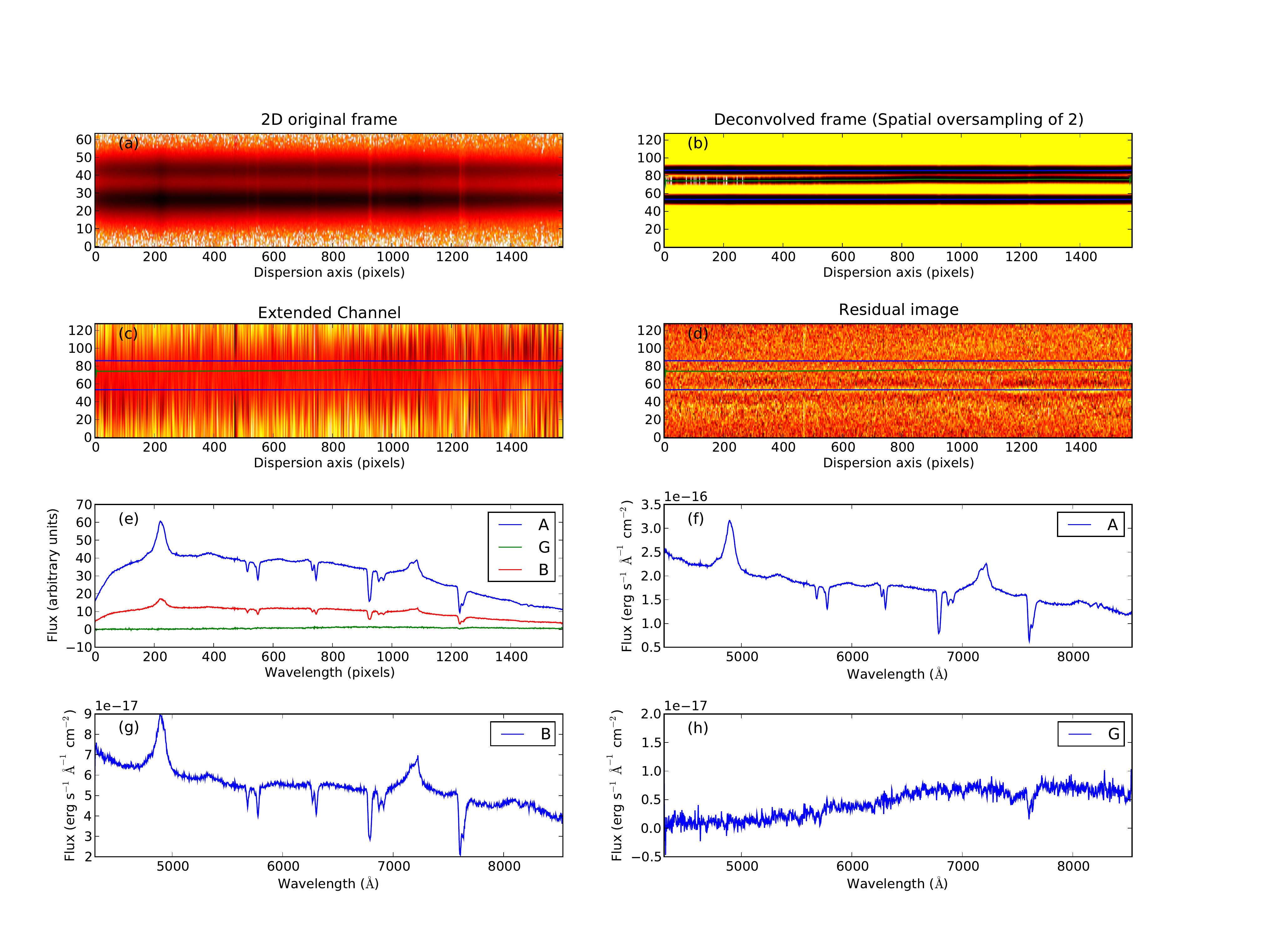}
\caption{Example of deconvolution for SDSS~J1335+0118. The first two lines show the original frame (a) and its deconvolved version spatially oversampled by a factor 2 (b), the deconvolved flux which is not PSF-like (c) and the residual frame (d) with upper/lower cuts at $\pm 3 \sigma$. The location of the centroid of the deconvolved PSFs is indicated with solid lines in panels (b, c, d). Panel (e) shows the extracted spectra of the lensed QSO images and of the lensing galaxy derived from the deconvolution. The last three panels (f, g, h) show the calibrated spectra of these components following the procedure explained in Sect.~\ref{sec:reduction}.}
\label{fig:decexample}
\end{figure*}

We used the combined 2D spectra of each object as obtained after the standard reduction procedure presented in \citet{Eigenbrod2006a, Eigenbrod2007}. An example of such a 2-D spectrum is displayed in Fig.~\ref{fig:decexample}a. The extraction of the spectra and the flux calibration, as described hereafter, have been updated compared to the original publication. Note that the differences arise because the scientific goal of the original papers was the determination of the redshift of the lensing galaxy, while we are interested here in a detailed study of the unpublished quasar spectra. Like \citet{Eigenbrod2006a, Eigenbrod2007}, we use MCS deconvolution algorithm adapted to spectra \citep{COU00} in order to deblend the components of the lens system. This algorithm deconvolves the observed frame into a frame of finite resolution with a Gaussian PSF chosen to have $FWHM =$ 2 pixels. During the deconvolution process, the flux is separated in 2 channels, one channel (i.e. ``the point-source channel'') containing only the deconvolved flux of the point-like sources, and one numerical channel (i.e. the ``extended channel'', Fig.~\ref{fig:decexample}c) containing the remaining flux correlated over several pixels. This extended channel contains mostly signal from the lensing galaxy and sometimes improperly subtracted sky signal. We improved the extraction of the spectra with respect to the former deconvolution by systematically assuming 3 point sources in the 2D deconvolved spectra (Fig.~\ref{fig:decexample}b): two for the QSO lensed images and one for the lens galaxy which has a relatively peaked profile in the center. Flux from the galaxy which deviates from a PSF is included in the ``extended channel''. The residuals are systematically inspected (Fig.~\ref{fig:decexample}d) to identify possible deconvolution problems. For objects where the deconvolution was not satisfactory, we also tested alternative PSFs and chose the one leading to the lower residuals. Finally, the 1-D spectra of the QSO and of the lensing galaxy (Fig.~\ref{fig:decexample}e) are corrected from the response curve based on standard stars, and from differential extinction using the updated Paranal extinction published by \citet{Patat2011}. An example of 1-D flux calibrated spectra for SDSS~J1335+0118 is displayed in Fig.~\ref{fig:decexample}f, g, h. The extracted spectra are available on electronic form via CDS, while 1-D and 2-D spectra are available via the German Virtual Observatory\footnote{\url{http://dc.g-vo.org/mlqso/q/web/form}}.

When possible, we checked that the flux ratios derived from our spectra are compatible with nearly simultaneous R-band ratios from literature or from the COSMOGRAIL monitoring project. We could not obtain simultaneous R-band flux ratios for the following systems: SDSS~J0246-0825, SDSS~J0806+2006, FBQ~0951+2635, BRI~0952-0115, SDSS~J1138+0314 and Q1355-2257. Photometric and spectro-photometric data always agree within the error bars except for WFI~2033-4723 for which we find $C/B\sim$ 0.85 while \cite{Vuissoz2008}, retrieved $C/B\,=\,$0.7 in the r-band on about the same date (MJD$=$53500). Because of that disagreement, we have multiplied the spectrum of image B in WFI~2033-4723 by 1.2 in order to match the photometric flux ratios. Our comparison with photometric flux ratios implies that the systematic error on the spectral ratio introduced by slit losses is $<$ 5 \%. 

\subsection{The extended sample}
\label{subsec:addsys}

In order to provide a more global overview of the variety of signals which may be observed, we completed our main sample with five systems we studied elsewhere:  SDSS~J0924+0219 \citep{Eigenbrod2006a}, H1413+117 \citep{Hutsemekers2010}, J1131-1231 \citep{Sluse2007},  HE~2149-2745 \citep{Burud2002a}, and Q2237+0305 \citep{Eigenbrod2008a, Sluse2011a}.  This is not a complete list of known microlensed quasars \cite[see e.g.][]{Richards2004a, Anguita2008b}, but these systems were among the first quasars lensed by a single lensing galaxy where significant microlensing of the BLR has been unveiled and discussed, with the advantage that the fully processed spectra are at our disposal. In the following, we refer to this sample as the ``extended sample''. 
In order to derive the luminosity of J1131-1231 and H1413+117 as for the other systems, we performed an approximate flux calibration of their spectra, by matching the broad-band magnitudes derived from the spectra with published absolute photometry. We used the $I-$band magnitude of \cite{Sluse2006} for J1131-1231 and the $H$-band 2MASS photometry $H$=14.531$\pm$0.054 for H1413+117. 

\section{Physical properties of the sample}
\label{sec:properties}

In this section, we calculate the black hole mass, the absolute luminosity, and the BLR size of each of the lensed quasars. We explain in Sect.~\ref{subsec:BH}, \ref{subsec:models}, \& \ref{subsec:BLRsize} how we derive these quantities. In Sect.~\ref{subsec:redshfits}, we present updated redshifts of the quasars based on the \MgII~line. Finally,  we present the results of our calculation and the error estimates in Sect.~\ref{subsec:phys}. We emphasize that the determination of the luminosity makes use of macro-magnification ratios presented in Sect.~\ref{sec:decomposition}. 

\subsection{Black Hole mass measurement}
\label{subsec:BH}

We estimate the virial black-hole mass using the width of the \MgII~emission line, following the original prescription of \cite{McLure2004}. In order to convert line width and continuum luminosity into black hole masses, we use a relation of the form
\begin{equation}
\log \left(\frac{M_{BH}}{10^6 M_{\sun}}\right) = a + \beta \log\left(\frac{L_{3000}} {10^{44} {\rm{erg/s}}}\right) + \gamma \log\left(\frac{FWHM}{1000 {\rm {km/s}}}\right), 
\label{equ:MBH}
\end{equation}

\noindent where $M_{BH}$ is the black hole mass, $L_{3000} = \lambda L_\lambda (3000 \AA)$ in erg/s and $FWHM$ is the Full Width at Half Maximum of the \MgII~emission line. We use $a=1.13\pm0.27$, $\beta=0.5$ and $\gamma=1.51\pm0.49$ as derived by \citet[][ Eq. 10]{Wang2009} from an analysis of a sample of about 500 SDSS spectra of intermediate redshift quasars. We derive $L_{3000}$ and $FWHM($\MgII$)$ by simultaneously fitting two Gaussians to the \MgII\,line on top of a pseudo-continuum component. The latter is the sum of a power-law continuum and of a broadened \FeII~template obtained by convolving (in the velocity space) the \FeII~template derived from I Zw 1 by \citet{Vestergaard2001}, with a Gaussian of adequate width to reproduce the observed \FeII~emission. The fit is always performed in the range [2200, 2675]\AA~, excluding regions affected by the sky or intervening absorption. In order to use a methodology similar to the one commonly used, and despite the fact that the evidence for a narrow \MgII~ component in AGNs remains debated \citep[e.g. ][]{McLure2004, Sluse2007, Fine2008, Shen2008, Wang2009}, we model the \MgII~emission with a narrow ($FWHM <$ 1200 km/s) and a broad component ($FWHM >$ 1200 km/s). Such a fitting procedure is similar to those used by \citet{McLure2004} and \citet{Shen2008} but differs slightly from the one used by \cite{Wang2009} who included a Balmer continuum component in their fit and explicitly model \MgII~as a doublet. Because of the low resolution of our spectra, it is not meaningful to split the \MgII~$\lambda 2798\,\AA$~line into its individual components $\lambda 2796/\lambda 2803$ which are spectroscopically unresolved. On the other hand, we have not fitted a Balmer continuum because of the degeneracy between this component and the power law index. Not accounting for the Balmer continuum will not affect the measurement of the $FWHM$ but could lead to an overestimate of $L_{3000}$. The Balmer continuum is commonly assumed to contribute to $\sim$ 10\% of  $L_{3000}$, but recent results suggest a contribution of $\sim$25\% in average, with a large scatter \citep[Wang, private communication,][]{Jin2011}. This remains in general small enough to weakly affect $M_{BH}$ since $M_{BH} \propto L_{3000}^{0.5}$. During the fit, we test for three different \FeII~templates{\footnote{Our \FeII\,template from \cite{Vestergaard2001} includes a small fraction of the Balmer continuum from I Zw 1. This is not the case of the template constructed by \cite{Tsuzuki2006} used by \cite{Wang2009} to derive Eq.~\ref{equ:MBH}.}}: (a) a "low level" template which assumes that there is no \FeII~emission under \MgII, (b) a "fiducial" template which assumes that the intensity of \FeII~in the range $\sim$ [2757, 2825]\,\AA~equals the average intensity of \FeII~ measured in the range [2930, 2970]\,\AA~\citep{Fine2008}, (c) a "high level" template which assumes that the intensity of \FeII~in the range $\sim$ [2757, 2825]\,\AA~is 40\% larger than in case (b), such that the intensity of \FeII~under \MgII~equals the amount of \FeII~on each side of the line. We use the fiducial template to estimate $FWHM$(\MgII). Because the fraction of \FeII~ under \MgII~is the main source of error in the measurement of the $FWHM$, we use the spread between (a), (b) and (c) to set $\sigma_{FWHM}$. The luminosity $L_{3000}$ is estimated from the power law continuum and corrected from the macro-magnification associated to lensing (see Sect.~\ref{subsec:models}).

\begin{figure} 
\centering
\includegraphics[scale=0.45]{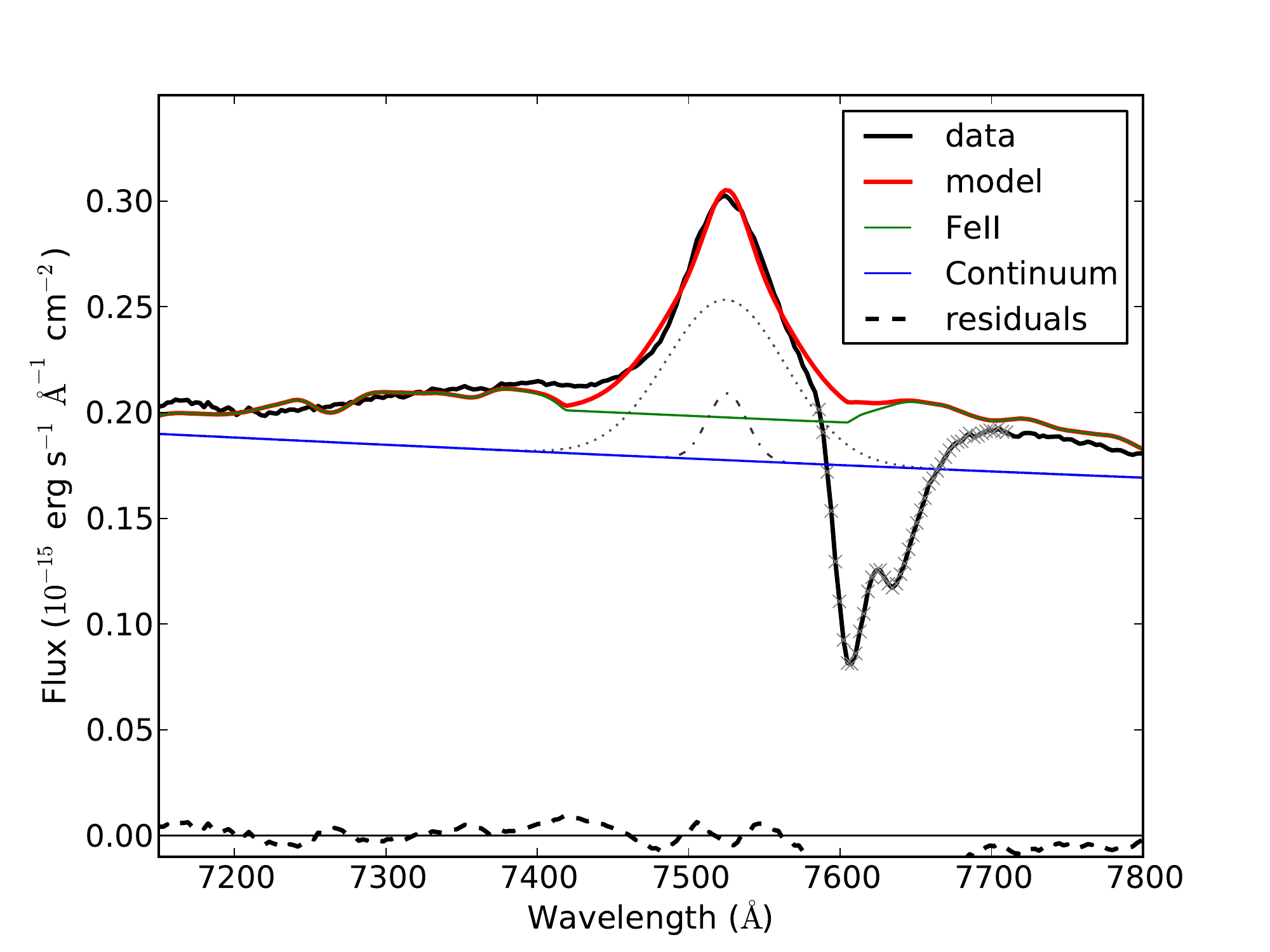} 
\caption{Example of fit of the MgII line for SDSS~J0246-0825 (black solid line) with a model (red) using the procedure described in Sect.~\ref{subsec:BH}. The model is the sum of a power-law continuum (blue), a \FeII~template (green), a broad \MgII~component (dotted gray) and a narrow \MgII~(dashed gray). The gray crosses indicate the region of the spectrum affected by atmospheric absorption and not used for the fit. The dashed black line shows the residual of the fit (data$-$model). }
\label{fig:MgIIfit}
\end{figure}

In four objects, we were unable to measure the black hole mass from \MgII, because this line falls outside of the wavelength range covered by our spectra or is located less than 300\,\AA~from the edge of the spectra (HE~2149-2745). For these systems we use black hole mass estimates from literature obtained using a relation similar to Eq.\ref{equ:MBH} but for the \CIV~line. Black-hole mass estimates derived from that line are more controversial because of uncertainties on the geometry of the \CIV\,emitting region, because of the contamination from a putative narrow \CIV~emission and because virial equilibrium is not expected to hold perfectly for such high ionization lines \citep{Richards2002, Bachev2004, Marziani2006, Fine2011, Assef2011, Wang2011, Richards2011, Marziani2012a}. For two systems (H1413+117 and J1131-1231), we also used  \Hbeta~ to calculate $M_{BH}$. For this purpose, we used the relation (A7) of \cite{McLure2004}. 

\subsection{Lens modeling and intrinsic luminosities}
\label{subsec:models}

In order to derive the macro-magnification of the lensed quasar images of our main sample, we have modeled the lensing galaxy with an isothermal mass distribution using \texttt{lensmodel} v1.99o \citep{Keeton2001}. The doubly-imaged systems have been modeled with a Singular Isothermal Sphere (SIS) and the quads with a Singular Isothermal Ellipsoid (SIE). For both doubles and quads, we accounted for the lens environment with an external shear term $\gamma$. We used the same modeling technique as in \cite{Chantry2010} with slightly different observational constraints for the quads. In particular, we constrained the model with the relative astrometry of the lensed images and lensing galaxy, and with the flux ratio of the lensed images. We used the flux ratios $M$ derived from our spectra (see Sect.~\ref{sec:decomposition} and Table~\ref{tab:macro}), accounting for 10\% error bars on the flux ratio. Contrary to what we did in \citet{Chantry2010}, the flux ratios from our spectra have also been used to model the quads. We provide in Table~\ref{tab:models} the values of the convergence $\kappa$, shear $\gamma$, shear position angle $\theta$, and macro-magnification $\mathcal{M}$ (with the sign reflecting the image parity) predicted by the models at the location of the lensed images. Note that for two objects, published HST-based astrometry was lacking and we used instead CASTLES\footnote{\url{http://www.cfa.harvard.edu/castles/}} measurements. We should mention that the flux ratios derived for the doubly imaged quasar SDSS~0246-0825 cannot be reproduced with a SIS+$\gamma$ model. This result was also found in the discovery paper \citep{Inada2005} and suggests that a companion object may strongly modify the lens potential. The visible companion galaxy G1 located $\sim 1.8 \arcsec$ from the lens cannot produce the anomaly and another invisible companion has to be included in the model. For this system, we searched for models with a second SIS located $\pm 1$\arcsec\,from the position of the putative companion derived by \cite{Inada2005}. 

For the extended sample (Sect.~\ref{subsec:addsys}), we have not calculated new lens models. Instead we used the following macro-magnifications factors $\mathcal{M}_A^{J0924} =$ 26.2 \citep{Keeton2006a}, $\mathcal{M}_B^{J1131} =$ 11.6  \citep{Sluse2011b}, $\mathcal{M}_{AB}^{H1413} =$ 10.3 \citep{McLeod2009},  $\mathcal{M}_{D}^{Q2237} =$ 3.9 \citep{Kochanek2004a}.

We converted intrinsic luminosities into bolometric luminosities using $L_{\rm{bol}} = BC \times \,L_{\rm {ref}}$, where BC is the bolometric correction and $L_{\rm ref}$ is the reference wavelength. We used the bolometric correction BC=(3.81, 5.15, 9.6) corresponding to $L_{\rm ref}=$($L_{1350}, L_{3000}, L_{5100}$) from \cite{Shen2008}. We used $L_{\rm ref}=L_{3000}$ when we detect \MgII~and $L_{\rm ref}=L_{5100}$ when we observe \Hbeta. We used $L_{\rm ref}=L_{1350}$ in the other cases.

\begin{table}
\begin{center}
\begin{tabular}{l|c|cccc}
\hline
Lens system & image & $\kappa$ & $\gamma$ & $\theta$ & $\mathcal{M}$ \\
\hline
(a) HE~0047-1756$^1$		&A	&0.45 	&0.48 	&-14.3 	&13.78 	\\
		&B	&0.63 	&0.68 	&1.0 	&-3.10 	\\
(b) Q0142-100$^2$		&A 	&0.32 	&0.39  	&-77.9  &3.21 	\\
		&B 	&1.56 	&1.64 	&-85.0 	&-0.42 	\\
(c) SDSS~0246-0825$^3$	&A	&0.57 	&0.50  	&-23.3  &-14.55 \\
		&B	&0.43 	& 0.35 	&-74.8 	& 4.96 	\\
(d) HE~0435-1223$^4$		&B   	&0.55 	&0.60  	& 15.2  &-6.46 	\\
		&D   	&0.59 	&0.64 	& 12.7 	&-4.21 	\\
(e) SDSS~J0806+2006$^5$		&A	&0.35 	& 0.35  & 61.9  &3.42 	 \\
		&B	&0.85 	& 0.85 	& 65.4 	&-1.44 	\\
(f) FBQ~0951+2635$^6$		&A	&0.32 	& 0.31  &-49.8  & 2.68 	 \\
		&B	&1.23 	& 1.27 	& -38.2 &-0.64 	\\
(g) BRI~0952-0115$^2$		&A	&0.40 	& 0.40  &41.4  	& 5.00 	 \\
		&B	&0.70 	& 0.74 	& 57.3 	& -2.20 \\
(h) SDSS~J1138+0314$^1$		&B	&0.54 	& 0.66  &38.9  	& -4.53  \\
		&C	&0.46 	& 0.36 	& -72.0 &6.24 	\\
(i) J1226-006$^1$		&A	&0.67 	& 0.61  &87.7  	& -3.83 \\
		&B	&0.36 	& 0.29 	& 87.4 	& 3.07 	\\
(j) SDSS~J1335+0118$^7$		&A	&0.35 	& 0.39  &-40.4  &  3.77  \\
		&B	&1.01 	& 1.07 	&-30.9 	&-0.87 	\\
(k) Q1355-2257$^7$		&A	&0.31 	& 0.28  &75.8  	& 2.50 	\\
		&B	&1.11 	& 1.09 	& 82.1 	& -0.85 \\
(l) WFI~2033-4723$^8$		&B	&0.41 	& 0.18  &-80.1  &  3.13  \\
		&C	&0.72 	& 0.57 	& 46.1 	& -4.11 \\
(m) HE~2149-2745$^6$		&A	&0.31 	& 0.32  &31.0  	& 2.71 	 \\
		&B	&1.25 	& 1.25 	& 29.6 	& -0.66 \\
\hline
\end{tabular}
\end{center}
\vspace{0.2cm}
{\tiny{{\bf {References:}} (1) \citet{Chantry2010}, (2) \citet{Lehar2000}, (3) \citet{Inada2005}, error bars increased to 0.003\arcsec due to saturation in the PSF, (4) \citet{Courbin2010}, (5) \citet{Sluse2008}, (6) \citet{Sluse2011b}, (7) CASTLES, (8) \citet{Vuissoz2008}}}
\caption{Lens models for each system of our main sample: convergence $\kappa$, shear $\gamma$, shear position angle $\theta$ and macro-magnification $\mathcal{M}$ at the position of the lensed images. The references correspond to the lens astrometry used to perform the model.}
\label{tab:models}
\end{table}	

\subsection{Size of the BLR and Einstein radius}
\label{subsec:BLRsize}

We calculate the size $R_{\rm {BLR}}$ of the BLR using the virial relation:
\begin{equation}
R_{\rm {BLR}}= \frac{G \, M_{BH} }{f \, FWHM^2}
\label{equ:virial}
\end{equation}
\noindent where $G$ is the universal constant of gravitation and $f$ is a factor of order of one which encodes assumptions regarding the BLR geometry \citep{Vestergaard2006, Collin2006, Decarli2008, Lamura2009, Graham2011}. As we perform only relative comparison between objects, the exact value of $f$ does not matter. We therefore assume the same value $f=1$ for the whole sample. Because occurrence of ML does not only depend on the source size but also on the Einstein radius of the lens, we calculate $R_{\rm {BLR}}/R_{E}$, where $R_E$ is the angular Einstein radius of a microlens projected onto the source plane as 

\begin{equation}
R_E = \sqrt{\frac{4G\avgg{M}}{c^2}\frac{D_{os}D_{ls}}{D_{ol}}},
\end{equation}

\noindent where the $D$'s are angular diameter distances, and the indices $o$, $l$, $s$ refer to observer, lens and source. ${\avgg{M}}$ is the average mass of microlenses that we assume $\avgg{M} = 1\,M_{\sun}$.

\subsection{Redshifts}
\label{subsec:redshfits}

For nine out of thirteen objects re-analyzed here, we were able to measure the redshift of the source based on the \MgII~ emission line, which is thought to give a good proxy towards the systemic redshift of the object. We found measurement compatible with literature data (i.e. within $\delta z = 0.003$) for six objects (SDSS~J0806+2006 and SDSS~J1335+0118, SDSS~J0246-0825, SDSS~J1226-006, Q1355-2257, WFI2033-4723). The differences for the other objects probably arise from the use of different emission lines used for the redshift calibration. This is clearly the case of HE~0435-1223 whose redshift was measured based on the \CIV~ line which is known to be prone to systematic blueshifts in many quasars. The redshifts we measured based on \MgII~are reported in Table~\ref{tab:data}.

\subsection{Physical properties of the sample: final remarks and uncertainties}
\label{subsec:phys}

\begin{table*}[t!]
\begin{center}
\begin{tabular}{l|cc|ccccccc}
\hline
Object & Pair & Line & $\log(L_{\rm {bol}, 1}^{\rm {erg/s}})$ & $\log(L_{\rm {bol}, 2}^{\rm {erg/s}})$ & FWHM ($\AA$) & $\log(M_{BH}/M_{\sun})$ & $L/L_{Edd}$& $\log(R_{\rm {BLR}}^{\rm {cm}})$ & $R_{\rm {BLR}}/R_E$  \\ 
\hline
(a) HE~0047-1756     & A-B & \MgII	 & 46.3	 & 46.3	 &4145$\pm$365	 & 8.86$\pm$0.23	 & 0.23$\pm$0.13	 & 17.74$\pm$0.54	 & 9.53 \\  
(b) Q0142-100        & A-B & \CIV	 & 47.4	 & 47.4	 &(4750$\pm$220)	 & (9.51$\pm$0.3)3	 & 0.58$\pm$0.45	 & 18.28$\pm$0.76	 & 35.85 \\ 
(c) SDSS~J0246-0825  & A-B & \MgII	 & 46.0	 & 45.8	 &3700$\pm$670	 & 8.59$\pm$0.36	 & 0.17$\pm$0.15	 & 17.57$\pm$0.85	 & 9.00 \\  
(d) HE~0435-1223     & B-D & \MgII	 & 45.7	 & 46.0	 &4930$\pm$195	 & 8.76$\pm$0.44	 & 0.11$\pm$0.11	 & 17.50$\pm$1.00	 & 5.67 \\  
(e) SDSS~J0806+2006  & A-B & \MgII	 & 45.8	 & 46.0	 &3370$\pm$430	 & 8.53$\pm$0.35	 & 0.20$\pm$0.17	 & 17.59$\pm$0.81	 & 8.34 \\  
(f) FBQ~0951+2635    & A-B & \MgII	 & 46.5	 & 46.6	 &5850$\pm$133	 & 9.21$\pm$0.26	 & 0.18$\pm$0.11	 & 17.80$\pm$0.59	 & 9.04 \\  
(g) BRI~0952-0115    & A-B & \CIV	 & 46.2	 & 46.5	 &(5210$\pm$1300)	 & (9.14$\pm$0.40)	 & 0.14$\pm$0.13	 & 17.83$\pm$0.95	 & 16.07 \\ 
(h) SDSS~J1138+0314  & B-C & \CIV	 & 45.5	 & 45.3	 &(1990$\pm$180)$^\dagger$	 & (7.69$\pm$0.33)	 & 0.43$\pm$0.36	 & 17.22$\pm$0.77	 & 2.92 \\  
(i) J1226-0006       & A-B & \MgII	 & 45.8	 & 45.5	 &7840$\pm$550	 & 8.96$\pm$0.43	 & 0.04$\pm$0.04	 & 17.30$\pm$0.99	 & 4.54 \\  
(j) SDSS~J1335+0118  & A-B & \MgII	 & 46.4	 & 46.5	 &6110$\pm$205	 & 9.19$\pm$0.26	 & 0.15$\pm$0.10	 & 17.74$\pm$0.61	 & 9.92 \\  
(k) Q1355-2257       & A-B & \MgII	 & 46.5	 & 46.3	 &5035$\pm$140	 & 9.04$\pm$0.34	 & 0.19$\pm$0.16	 & 17.76$\pm$0.79	 & 14.89 \\ 
(l) WFI~2033-4723    & B-C & \MgII	 & 46.0	 & 45.8	 &3960$\pm$465	 & 8.63$\pm$0.35	 & 0.15$\pm$0.13	 & 17.56$\pm$0.81	 & 8.25 \\  
(m) HE~2149-2745     & A-B & \CIV	 & 46.0	 & 46.0	 &(7470$\pm$1865)	 & (9.82$\pm$0.40)	 & 0.01$\pm$0.01	 & 18.20$\pm$0.95	 & 32.47 \\ 
\hline
(n) J0924+0219       & A   & \MgII       & 44.7	 & -	 &3660$\pm$310	 & 7.93$\pm$0.34	 & 0.04$\pm$0.03	 & 16.93$\pm$0.78	 & 1.45 \\ 
(o) J1131-1231       & B   & \MgII       & 45.0	 & -	 &5630$\pm$165	 & 8.32$\pm$0.62	 & 0.03$\pm$0.04	 & 16.94$\pm$1.42	 & 1.89 \\ 
                     & B   & \Hbeta      & 45.0	 & -	 &4545$\pm$255	 & 7.90$\pm$0.60	 & 0.07$\pm$0.10	 & 16.71$\pm$1.37	 & 1.12 \\ 
(p) H1413+117        & AB  & \Hbeta      & 46.7	 & -	 &5170$\pm$250	 & 9.12$\pm$0.01	 & 0.28$\pm$0.06	 & 17.82$\pm$0.05	 & 17.76$^\ddagger$ \\
(q) Q2237+0305       & D   & \MgII       & 46.5	 & -	 &2900$\pm$565	 & 8.68$\pm$0.36	 & 0.44$\pm$0.39	 & 17.88$\pm$0.86	 & 4.06 \\ 
                     & C-D & \CIV        & 46.1	 & -	 &(3780$\pm$120)	 & (8.63$\pm$0.32)	 & 0.25$\pm$0.19	 & 17.60$\pm$0.74	 & 2.13 \\

\hline
\end{tabular}
\end{center}
\vspace{0.2cm}
{\tiny{Notes: $\dagger$ Lower limit, see Appendix A of \cite{Assef2011}; $\ddagger$ the redshift of this object is unsecure \citep{Kneib1998, Goicoechea2010} and we used $z_l=$1.0}} 

\caption{Physical quantities associated to the source quasar. Values in parentheses are from \cite{Peng2006b} and \cite{Assef2011}. The last four systems under the horizontal line are part of the extended sample which gather data from literature (Sect.~\ref{subsec:addsys}).}
\label{tab:MBH}
\end{table*}

We present in Table~\ref{tab:MBH} the physical properties of all the lensed sources discussed in this paper. Because we only measure differential ML between two macro-images, we report values of $L_{\rm{bol}, 1}$ and  $L_{\rm{bol}, 2}$ derived for each of the lensed image. We use the average between $L_1$ and $L_2$ for the calculation of $M_{BH}$ and $L/L_{Edd}$.

The errors on the quantities reported in Table~\ref{tab:MBH} were calculated in the following way. The error on the luminosity $L$ is caused by a) microlensing, b) uncertainty on the absolute image magnification associated to the lens model, c) flux calibration and, d) intrinsic variability. We assumed that the microlensing budget error can be estimated from the spread between $L_1$ and $L_2$ and that the other sources of errors (b+c+d) correspond to 20\% of the flux of $L$. We added these errors quadratically. The calculation of the error on the $FWHM$ is explained in Sect.~\ref{subsec:BH}. In case of measurements from literature, when no error bars on the FWHM were provided, we conservatively assumed $\sigma_{FWHM}$ = 0.25\,$FWHM$. The errors on the other quantities were derived using error propagation formulae. When we report $M_{BH}$ from literature, we used a typical 0.4 dex error when no error bar was reported.

\section{Microlensing analysis of the spectra}
\label{sec:decomposition}

Several techniques can be used to unveil microlensing (ML) in lensed quasars. The most common method is the measurement of the emission lines equivalent width but this mostly allows one to assess microlensing of the continuum. Alternative methods exists such as ``multi-variable'' scaling of the spectra \citep[e.g.][]{Angonin1990, Burud2002a, Wucknitz2003}, the technique introduced by \cite{Popovic2005} based on multi-epoch spectra, flux ratio measurements after a multi-component decomposition of the spectra \citep{Wucknitz2003, Sluse2007, Eigenbrod2008a} or the so-called Macro-micro decomposition \citep{Sluse2007}. The latter technique, that we recall in Sect.~\ref{subsec:FMFMmu}, is the best suited for our goals because it enables an easy visualisation of the differential ML affecting the QSO spectrum, including partial ML of an emission line. It also provides a reliable estimate of the amount of ML affecting the continuum and of the intrinsic flux ratio $M$ between the lensed images \citep[see ][ for comparison with other methods]{Sluse2007, Sluse2008b, Sluse2011a, Hutsemekers2010}.

\subsection{Macro-micro decomposition (MmD)}
\label{subsec:FMFMmu}

Following \citet{Sluse2007} and \citet{Hutsemekers2010}, we assume that the spectrum of an observed lensed image $F_i$ is the superposition of 2 components, one component $F_M$ which is only macro-lensed and another one, $F_{M\mu}$, both macro- and micro-lensed. According to this procedure, it is possible to extract $F_M$ and $F_{M\mu}$ by using pairs of observed spectra. Defining $M=M_1/M_2 \,(>0)$ as the macro-magnification ratio between image 1 and image 2 and $\mu$ as the relative micro-lensing factor between image 1 and 2, we have:
\begin{equation}
\begin{array}{l}
F_1 = M \times F_M + M \times \mu \times F_{M\mu}\\
F_2 = F_M + F_{M\mu} \,.\\
\end{array}
\label{eq:decomp1}
\end{equation}
\noindent

To extract $F_M$ and $F_{M\mu}$ when $M$ is not known a priori, these equations can be conveniently rewritten
\begin{equation}
\begin{array}{l}
F_M = \frac{-A}{A-M}\left(\frac{F_1}{A}-F_2\right) \\
\vspace{2mm} F_{M \mu} = \frac{M}{A-M}\left(\frac{F_1}{M}-F_2\right), \\
\end{array}
\label{eq:decomp2}
\end{equation}

\noindent where $A = M \times \mu$. As explained in more detail in Hutsem\'ekers et al. (2010, Sect. 4.1 and Appendix A)\nocite{Hutsemekers2010}, the factor $A$ can be accurately determined as the value for which $F_M (A) = 0$ in the continuum adjacent to the emission line. The factor $M$ is chosen so that there is no visible emission above the continuum in $F_{M\mu}(M)$ at the wavelength of the narrow emission lines which originate from regions too large to be microlensed. Often, narrow emission lines are absent of the observed spectrum and then $M$ is chosen to minimise the emission above the continuum in $F_{M\mu}$ at the location of the broad line (see Sect.~\ref{subsec:example} for an illustration). If only a portion of the broad emission line is not microlensed, then $M$ can be determined using this portion of the line \citep[see][]{Hutsemekers2010}. The micro-(de)magnification factor $\mu$ is simply derived using $\mu= A/M$. 

The factor $M$ is generally wavelength dependent because differential extinction between the lensed images produced by the lensing galaxy may take place. In addition, chromatic ML sometimes occurs due to wavelength dependence of the source size, the blue continuum emitting region being smaller and more microlensed than the red one, we may also have a wavelength dependence of $\mu$. For this reason it is necessary to estimate $A$ and $M$ in the vicinity of each emission line. Because differential extinction affects both the emission line and the continuum, it does not modify the decomposition outlined in Eq.~\ref{eq:decomp2} as far as $A$ and $M$ are derived at the location of each line. 

Our method should ideally be applied to pairs of spectra separated by the time delay. In Appendix~\ref{appendixC}, we show that intrinsic variability generally leads to small errors on $M$ and $\mu$ and could rarely mimic microlensing deformation of the emission lines, only when the time delay between the lensed images is large (typically $>$50 days). Consequently, we apply the MmD to spectra obtained at one epoch, like those described in Sect.~\ref{sec:data}, with the caveat that in rare occasions intrinsic variability could be an issue.

\subsection{Example}
\label{subsec:example}

\begin{figure}[tb!]
{\includegraphics[scale=0.45]{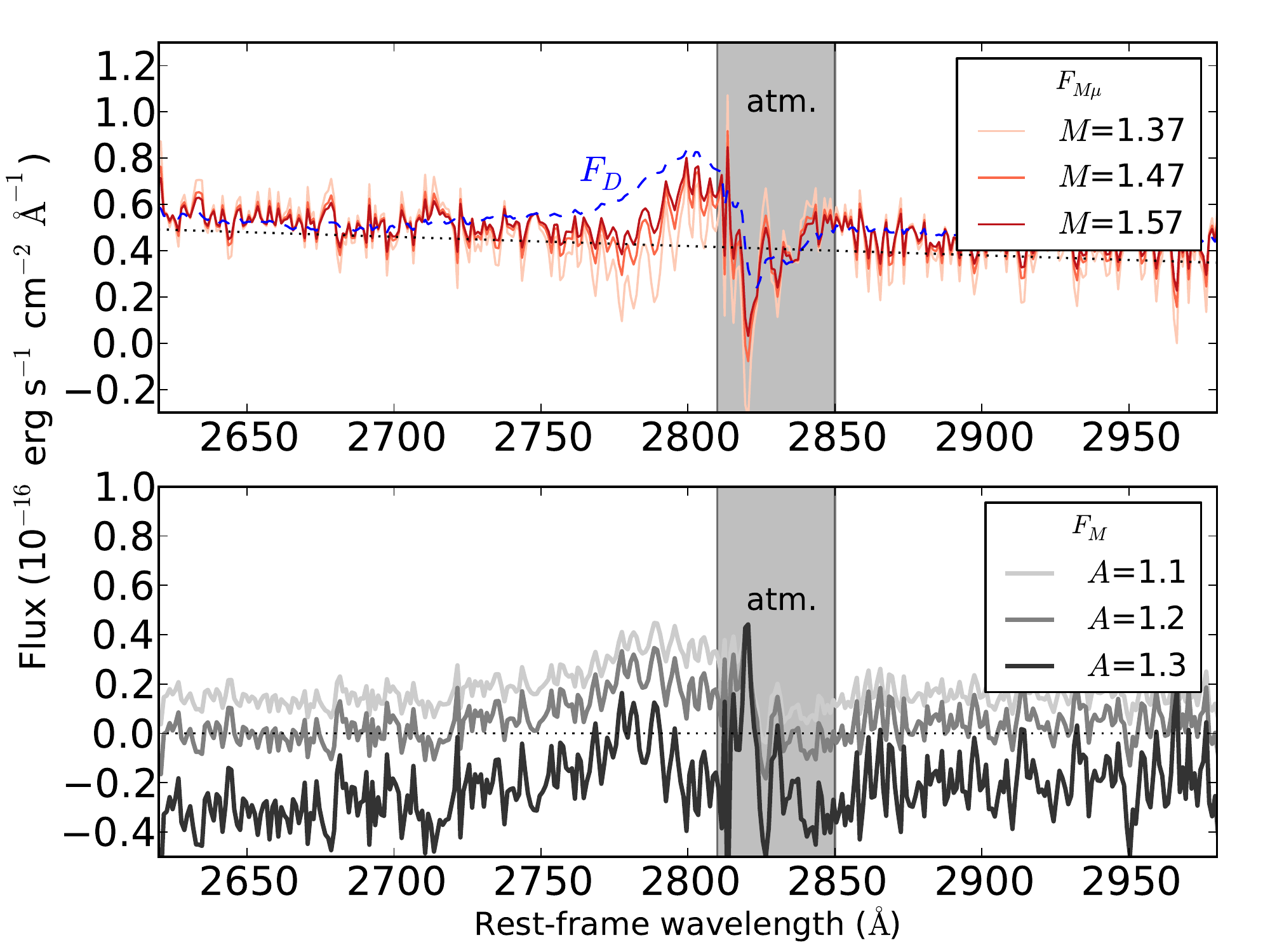}}

\caption{Macro-micro decomposition (MmD) applied to the \MgII~line in HE~0435-1223 using the spectra of images B ($F_1$) and D ($F_2$). The bottom pannel shows $F_M$ for three different values of $A$ and an arbitrary value of $M$. The best value of $A$ is $A\sim$1.2 because it leads to $F_M=0$ in the continuum regions blueward and redward of the emission line. The upper pannel shows the decomposition for 3 different values of $M$. The best value is $M=$1.47 because it minimizes the emission {\it {in the line}}, i.e. keeping the flux at the location of the line just above the apparent local continuum depicted as a dotted black line. }
\label{fig:MmDexample}
\end{figure}

We illustrate in this section how the MmD is used to derive $A, M$ and $\mu$ and to unveil microlensing of the emission lines. In Fig.~\ref{fig:MmDexample}, we show the MmD applied, at the location of the \MgII~ emission, to the spectra of images $B$ and $D$ of HE~0435-1223.
The MmD is performed in the following way. Up to a scale factor, $F_M$ only depends on $A$, and $F_{M\mu}$ on $M$. Therefore we first measure the factor $A$ by making $F_M = 0$ in regions of the continuum blueward and redward of important emission lines, in this case \MgII\,(cf. Eq.~\ref{eq:decomp2}). Which spectrum corresponds to $F_1$ or $F_2$ is arbitrary. We choose $F_1 = F_B$ and $F_2 = F_D$. In the bottom pannel of Fig.~\ref{fig:MmDexample}, we see that $A \sim 1.2$ leads to $F_M \sim 0$ in the continuum regions. Alternatively, it is also possible to measure $A$ in the continuum regions from $F_1/F_2$. If the values blueward and redward of the line are different, we average them. Second, the factor $M$ is chosen so as to minimize the emission above the continuum in $F_{M\mu}$ at the location of the broad lines. The upper pannel of Fig.~\ref{fig:MmDexample} shows the decomposition for three values of $M$. The value $M=1.37$ is not adequate because it leads to a dip in $F_{M\mu}$ in the range 2750-2800$\AA$. Only values $M \geq 1.47$ lead to a valid decomposition. Finally $F_M$ and $F_{M\mu}$ are scaled according to Eq.~\ref{eq:decomp2} so as to have $F_2 = F_M+F_{M\mu}$. The amount of microlensing $\mu$ is derived with $\mu = A/M$. For the example of Fig.~\ref{fig:MmDexample}, we find ($A$, $M$, $\mu$)=(1.2, 1.47, 0.82).  Larger amplitudes of ML are conceivable provided $M$ is modified such that $A = M \times \mu$ still holds. We also emphasize that, with only a pair of spectra, it is not possible to know if image \#1 is magnified (resp. de-magnified) by $\mu$ or if image \#2 is de-magnified (resp. magnified) by $1/\mu$. The errors on $A$ and $M$ are determined from the range of values which provide acceptable solutions to $F_M$ and $F_{M\mu}$. The quality of the spectral decomposition, and errors strongly depend on the S/N of the spectra and on the strength of the microlensing effect (i.e. if $\mu \to 1$, then $M \to A$ in Eq.~\ref{eq:decomp2}). Another illustration of the method is proposed in Appendix~\ref{appendixD} based on simulated spectra.

\subsection{Results}
\label{subsec:results}

\begin{figure*} 
\centering
\setcounter{subfigure}{0}
\renewcommand{\thesubfigure}{(\alph{subfigure}1)}
  \subfigure[HE~0047-1756 (\CIII)]{\includegraphics[scale=0.44]{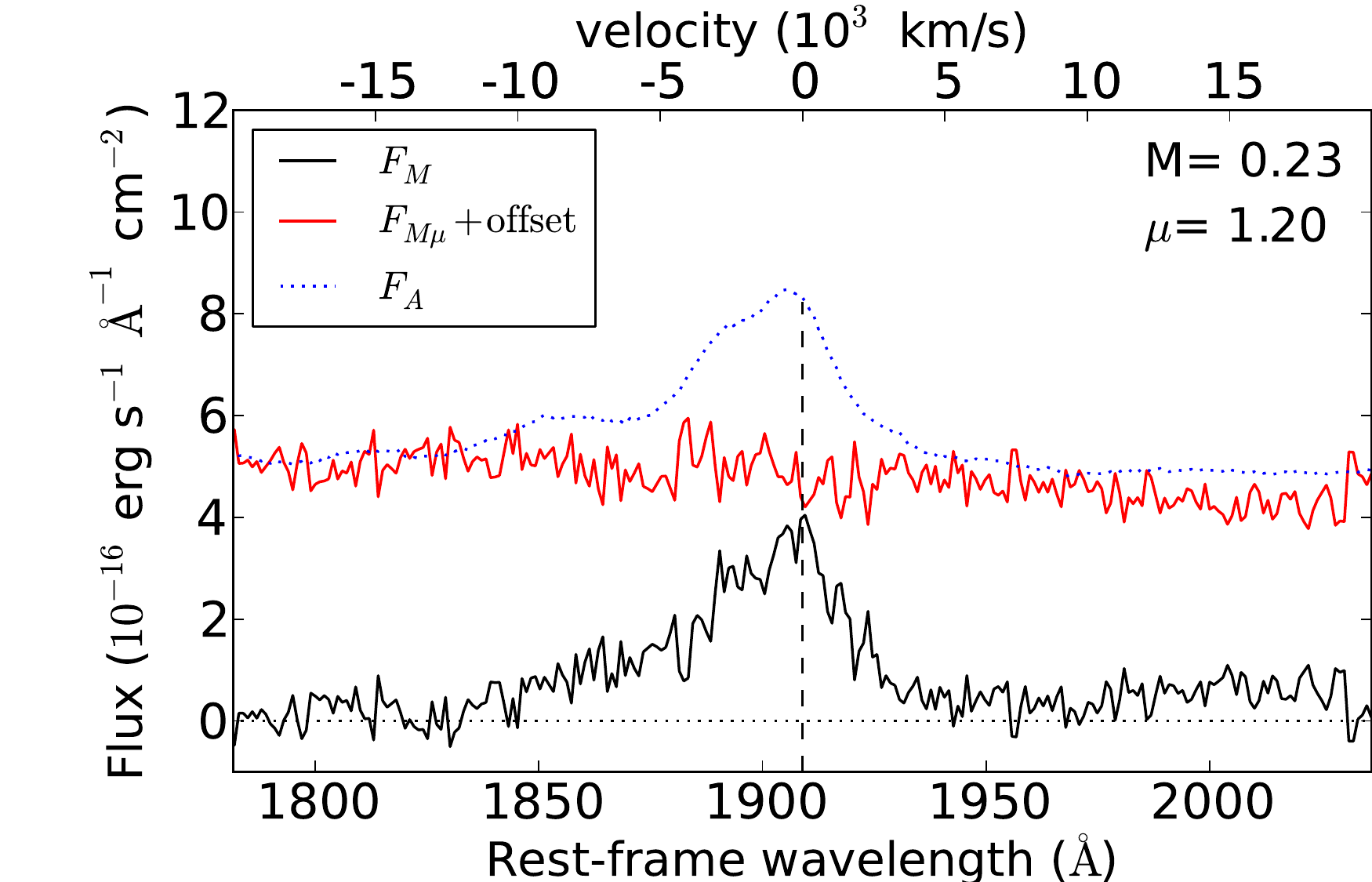}} 
\setcounter{subfigure}{0}
\renewcommand{\thesubfigure}{(\alph{subfigure}2)}
  \subfigure[HE~0047-1756 (\MgII\,+\,atm)]{\includegraphics[scale=0.44]{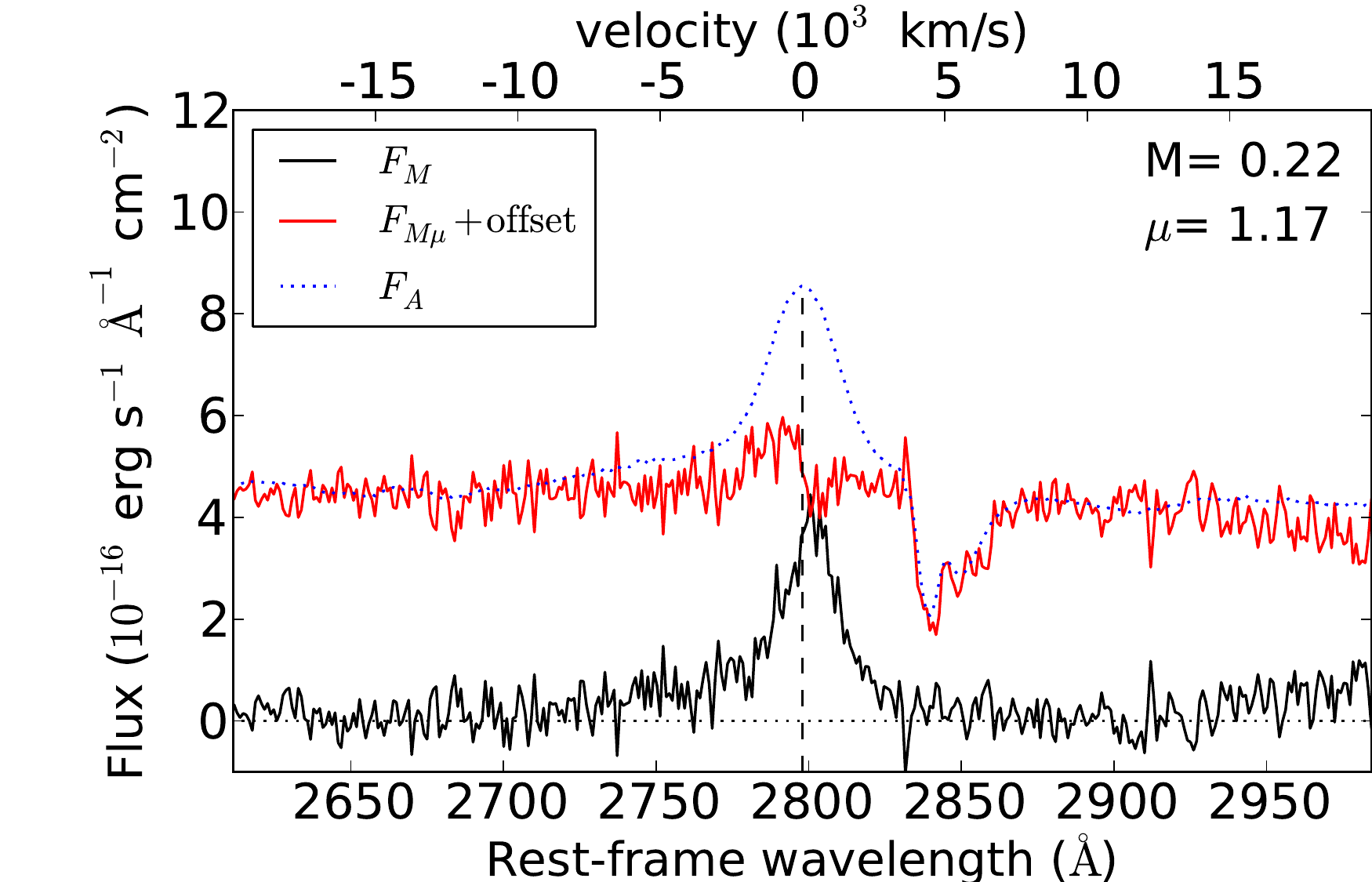}}              

\setcounter{subfigure}{2}
\renewcommand{\thesubfigure}{(\alph{subfigure}1)}
  \subfigure[SDSS J0246-0825 (\CIII)]{\includegraphics[scale=0.44]{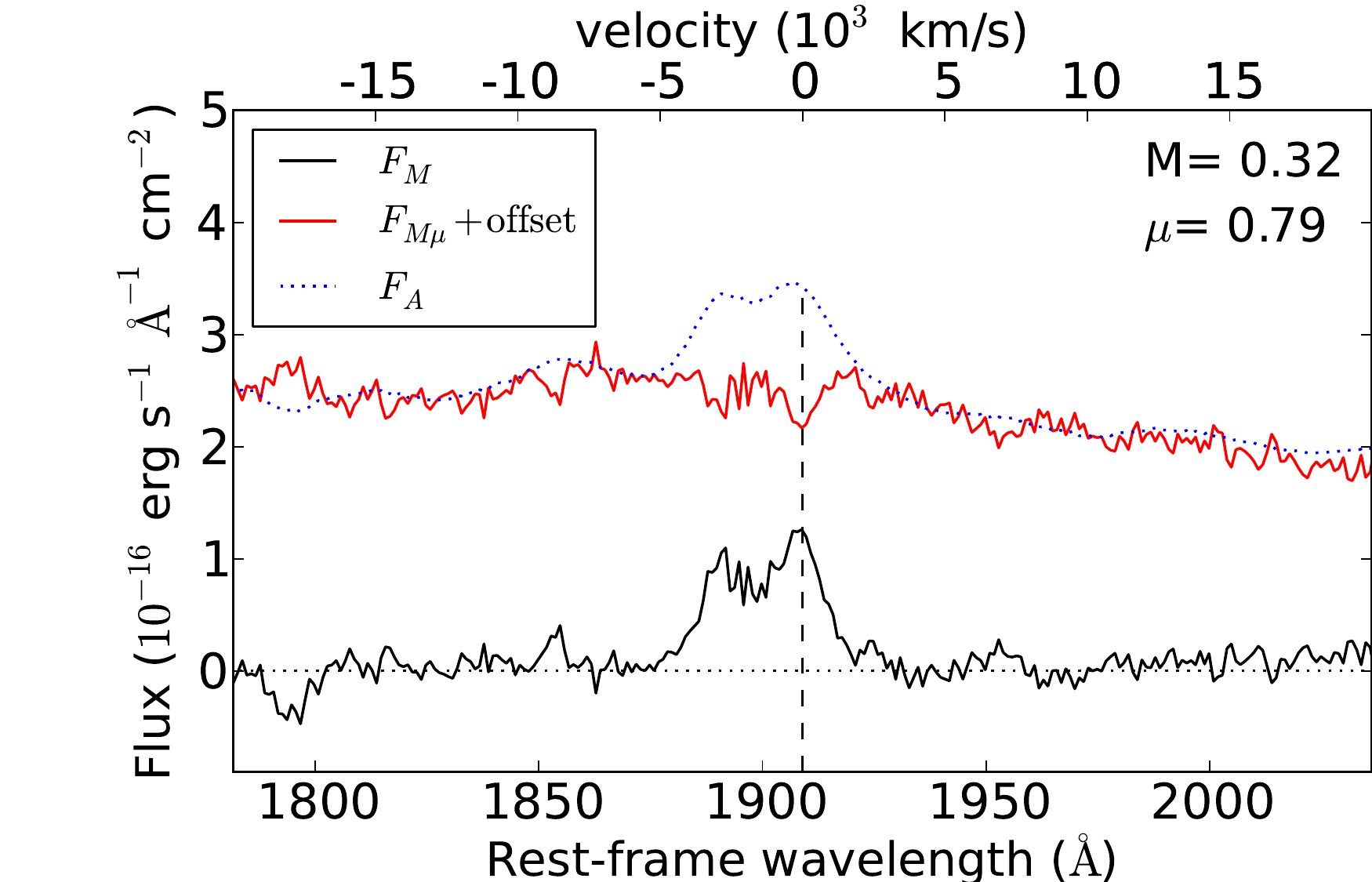}}
\setcounter{subfigure}{2}
\renewcommand{\thesubfigure}{(\alph{subfigure}2)}
  \subfigure[SDSS J0246-0825 (\MgII\,+\,atm)]{\includegraphics[scale=0.44]{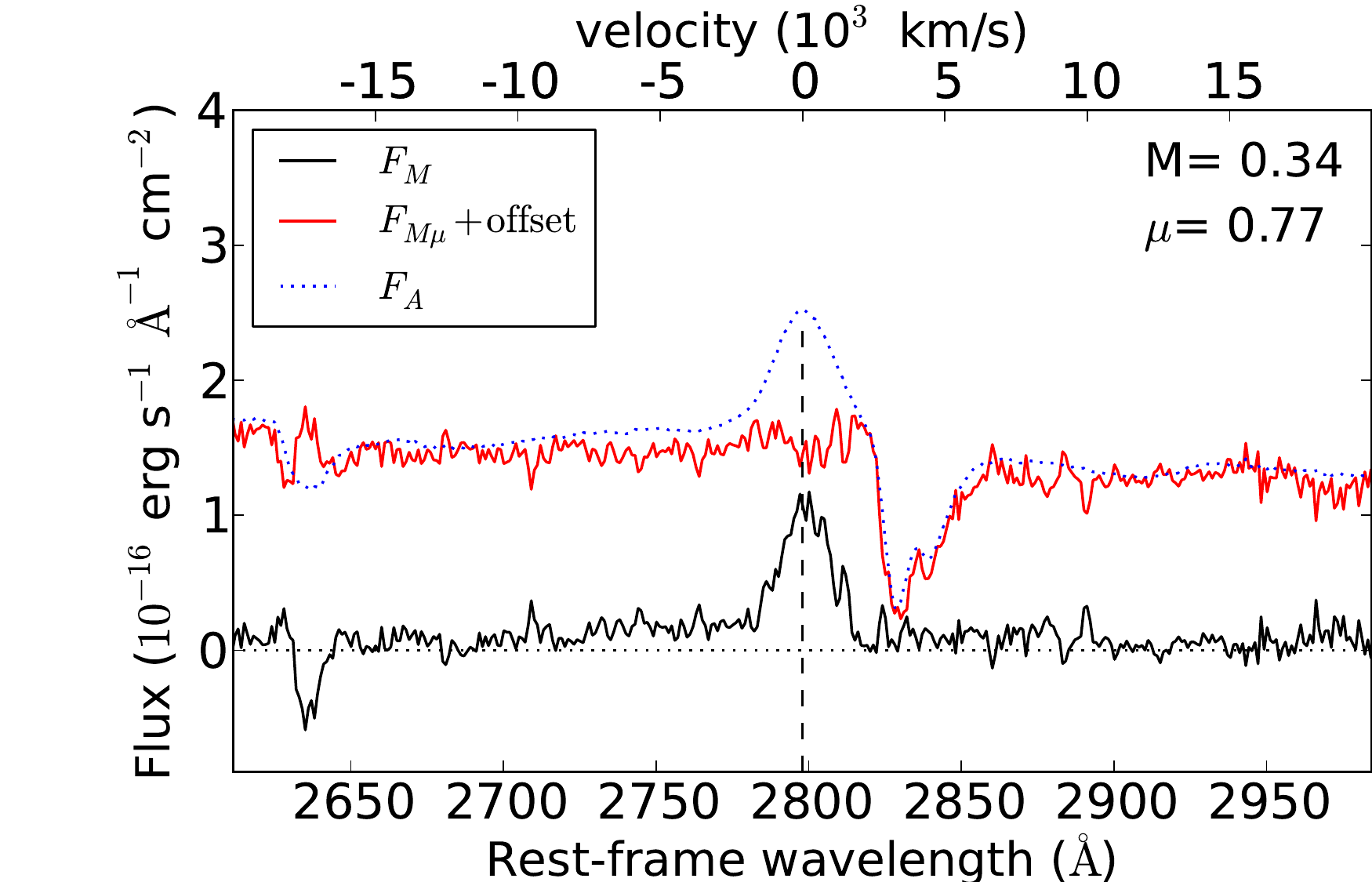}}

\setcounter{subfigure}{3}
\renewcommand{\thesubfigure}{(\alph{subfigure}1)}
  \subfigure[HE 0435-1223 (\CIII)]{\includegraphics[scale=0.44]{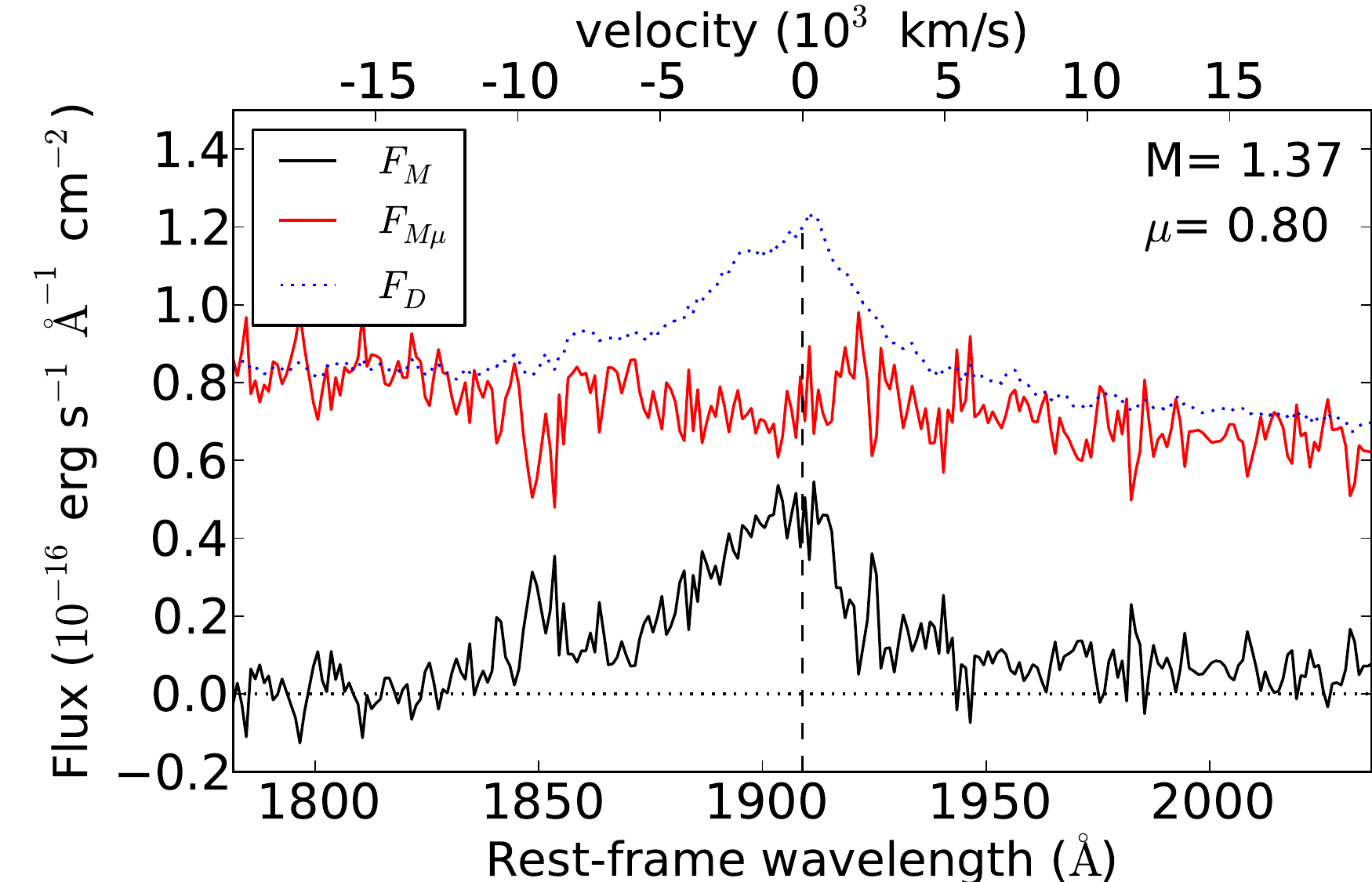}} 
\setcounter{subfigure}{3}
\renewcommand{\thesubfigure}{(\alph{subfigure}2)}
\subfigure[HE 0435-1223 (\MgII\,+\,atm)]{\includegraphics[scale=0.44]{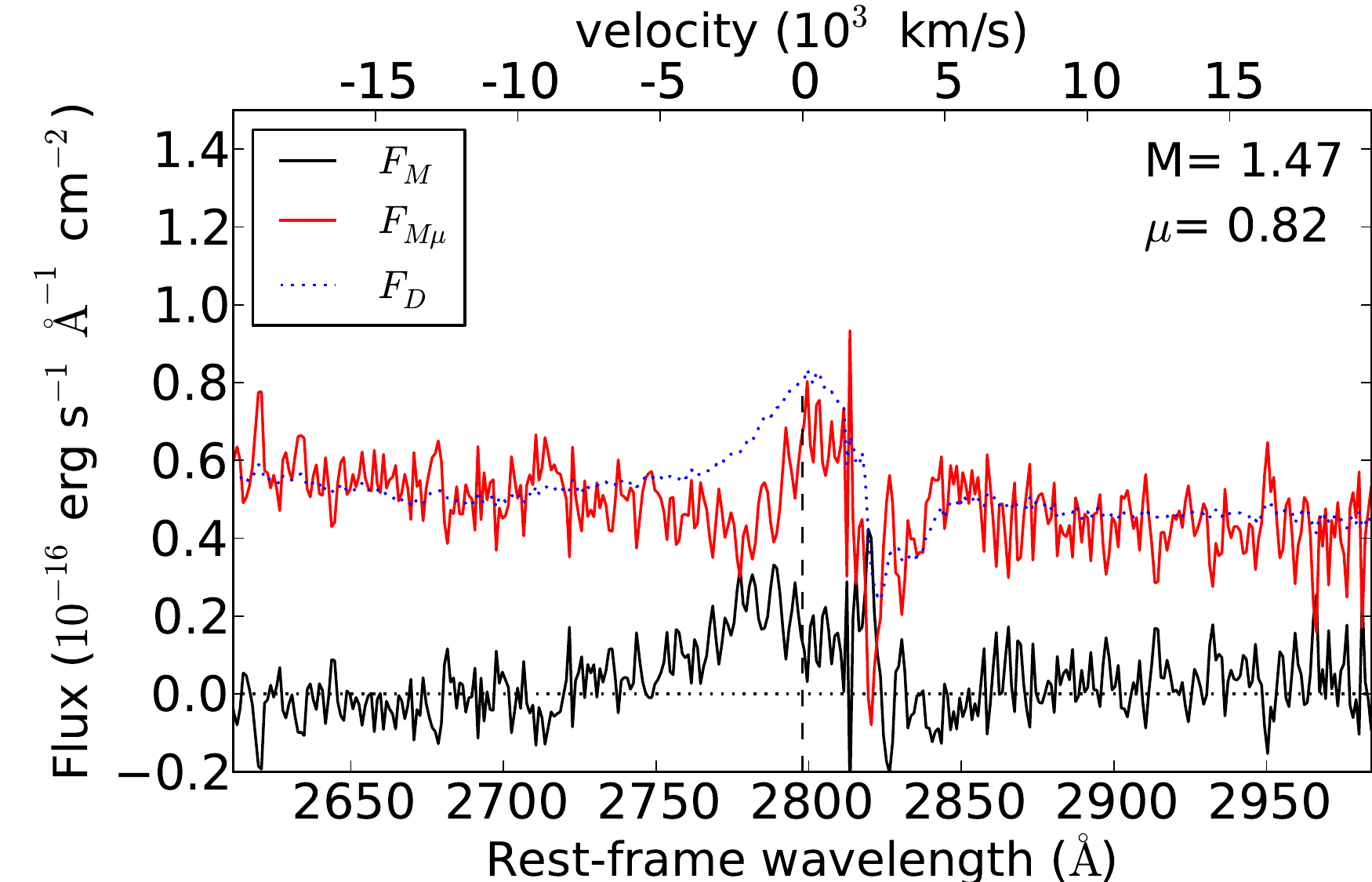}} 

\setcounter{subfigure}{4}
\renewcommand{\thesubfigure}{(\alph{subfigure}1)}
  \subfigure[SDSS 0806+2006 (\CIII)]{\includegraphics[scale=0.44]{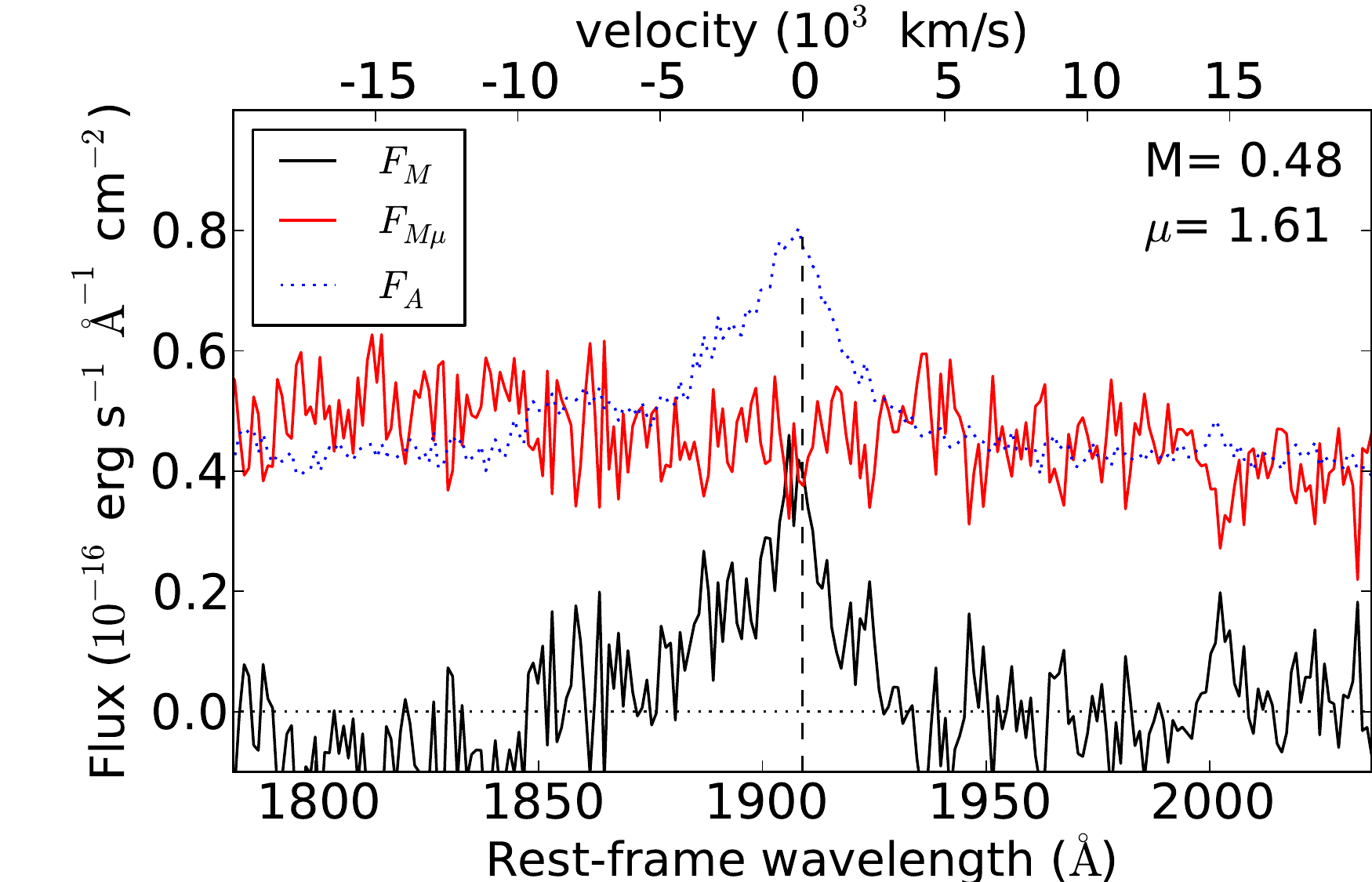}} 
\setcounter{subfigure}{4}
\renewcommand{\thesubfigure}{(\alph{subfigure}2)}
\subfigure[SDSS 0806+2006 (\MgII)]{\includegraphics[scale=0.44]{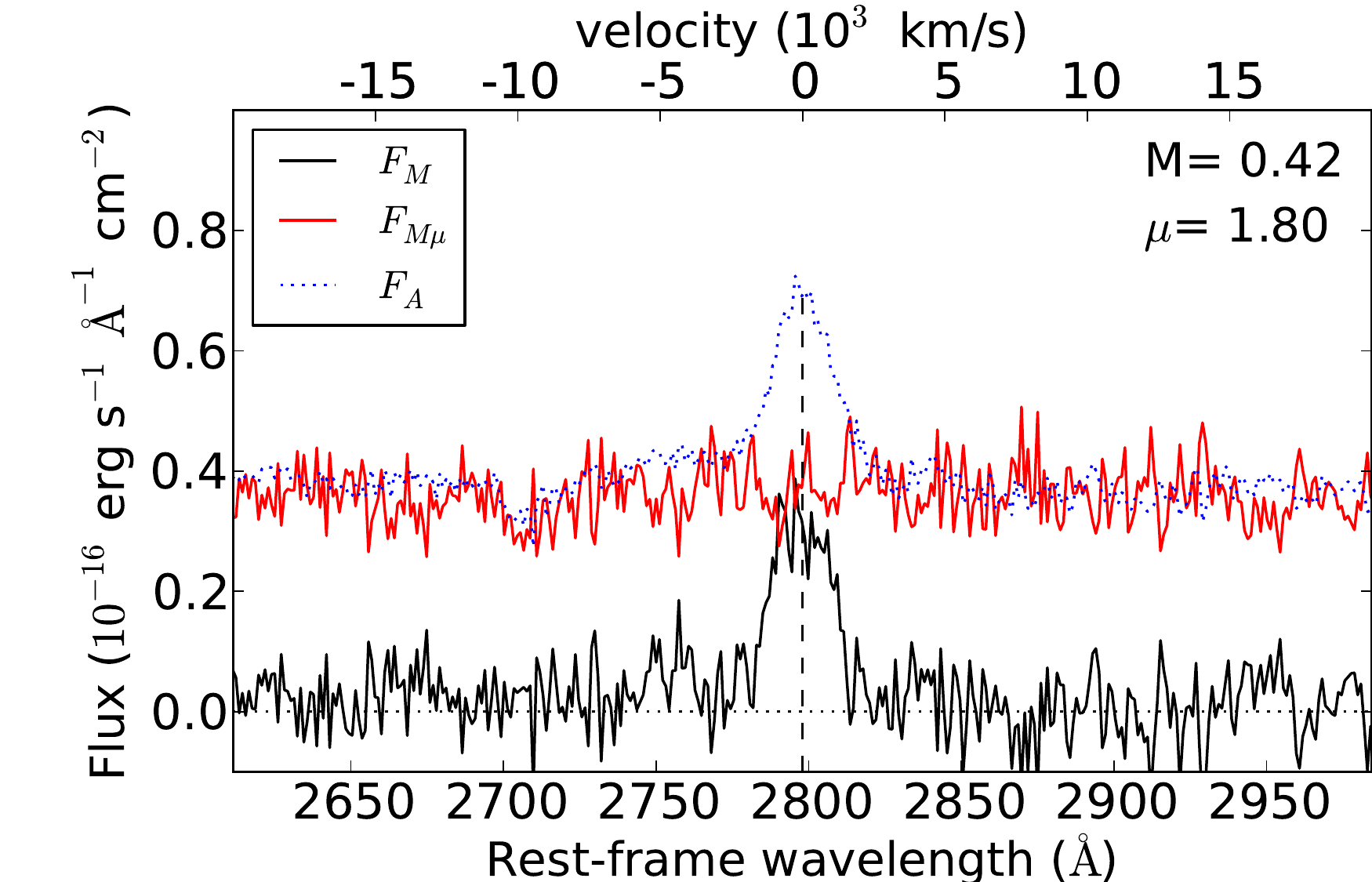}} 
\caption{Macro-micro decomposition (MmD) around the main emission lines. In each pannel, we show the part of the line profile which is only macrolensed, $F_M$, and the part of the line profile which is both macro- and micro-lensed, $F_{M\mu}$. The spectrum of Image \#2 ($F_2 = F_M + F_{M\mu}$) is superimposed (dotted line). The vertical dotted line denotes the rest wavelength of the emission line (using $z_s$ from Table~\ref{tab:data}). The values of $M$ and $\mu$ are given for each decomposition (Sect.~\ref{subsec:FMFMmu}). $F_{M\mu}$ contains the micro-magnified continuum and, quite often, a part of the emission line profiles, while $F_M$ contains the bulk of the emission lines unaffected by microlensing.}
\label{fig:MmD}
\end{figure*}

\begin{figure*} 
\centering
\setcounter{subfigure}{5}
  \subfigure[FBQ 0951+2635 (\MgII)]{\includegraphics[scale=0.44]{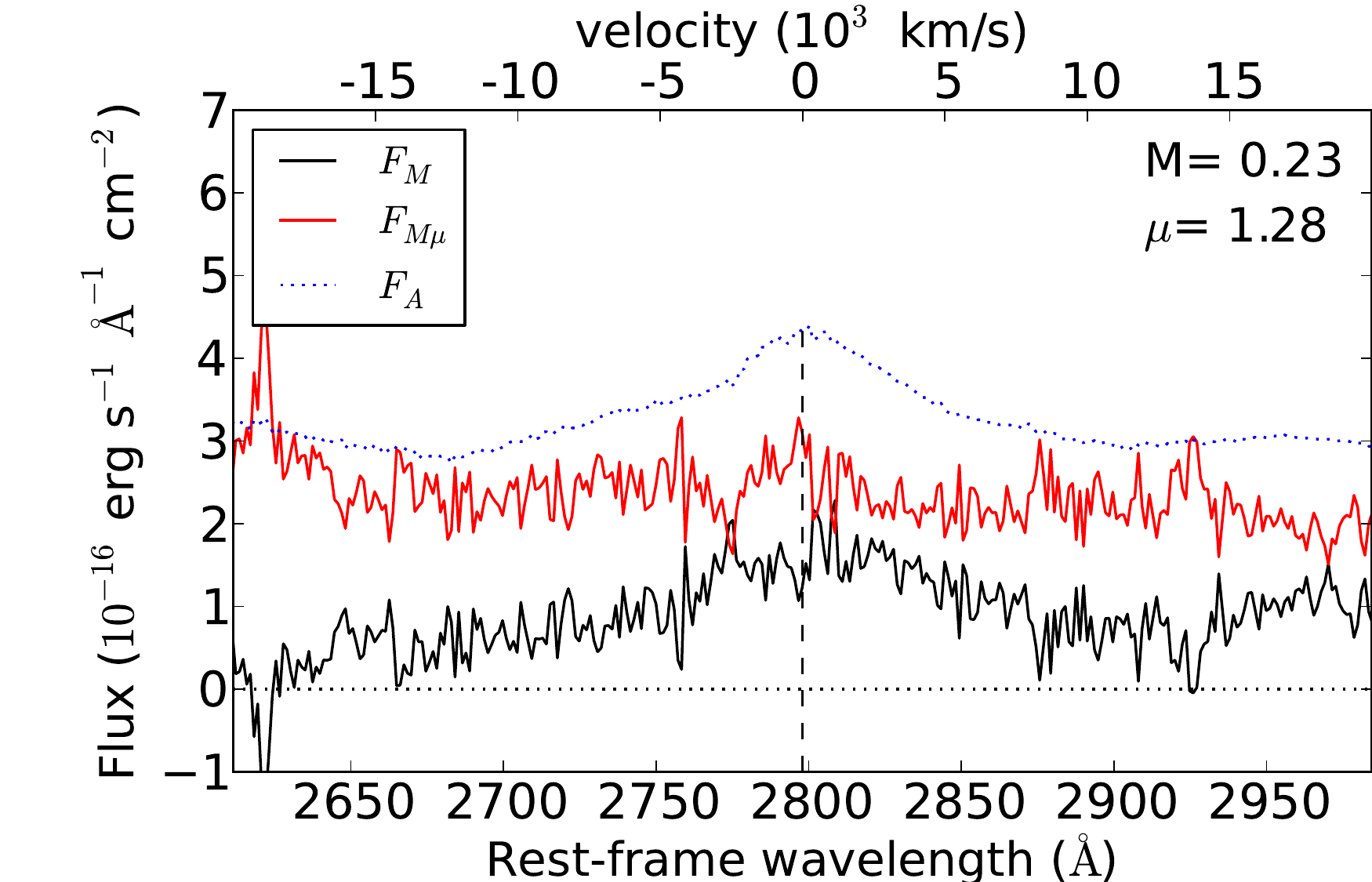}} 
  \subfigure[BRI~0952-115 (\Lyalpha)]{\includegraphics[scale=0.44]{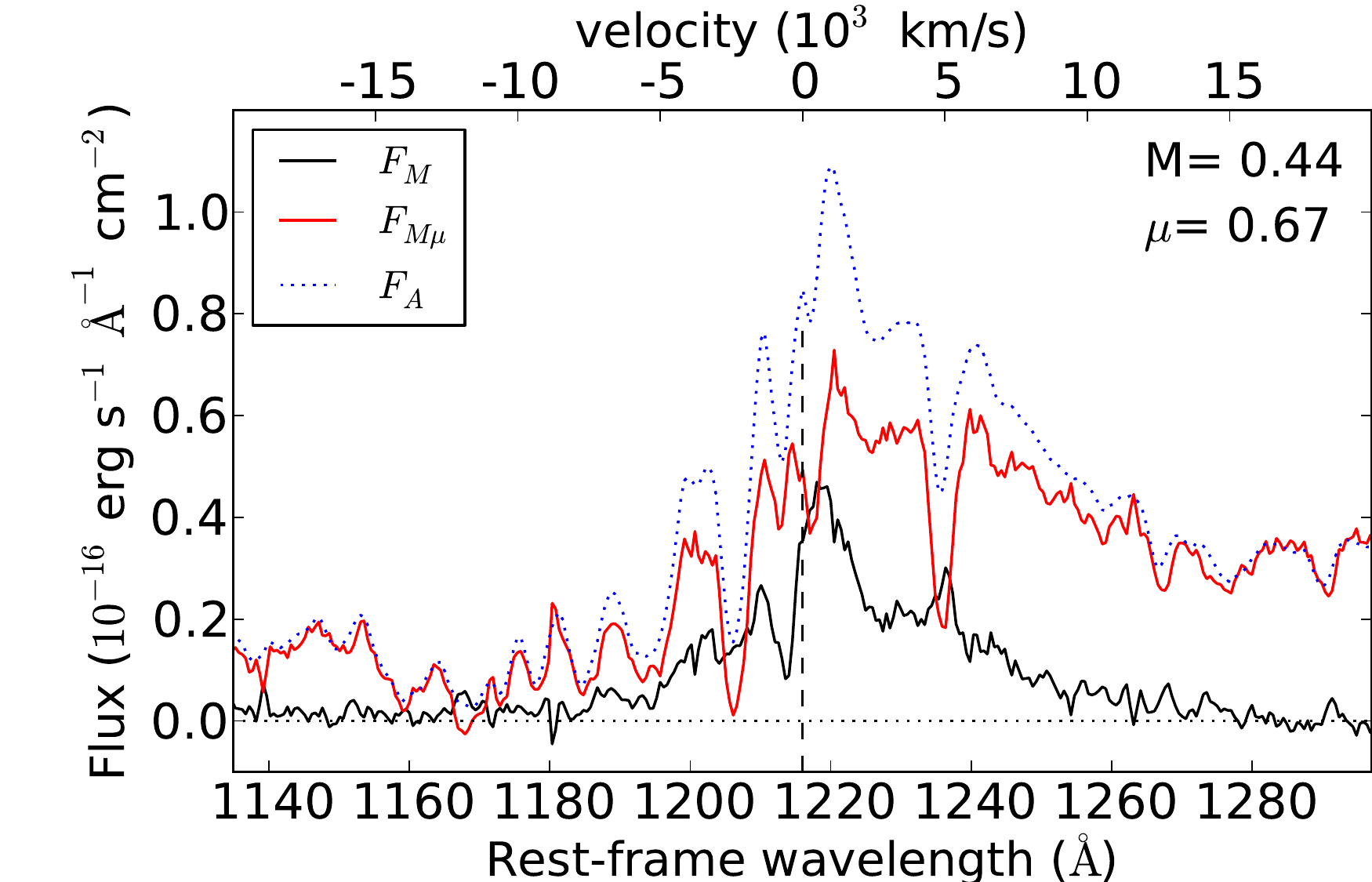}} 

\setcounter{subfigure}{7}
\renewcommand{\thesubfigure}{(\alph{subfigure}1)}
  \subfigure[SDSS 1138+0314 (\CIV)]{\includegraphics[scale=0.44]{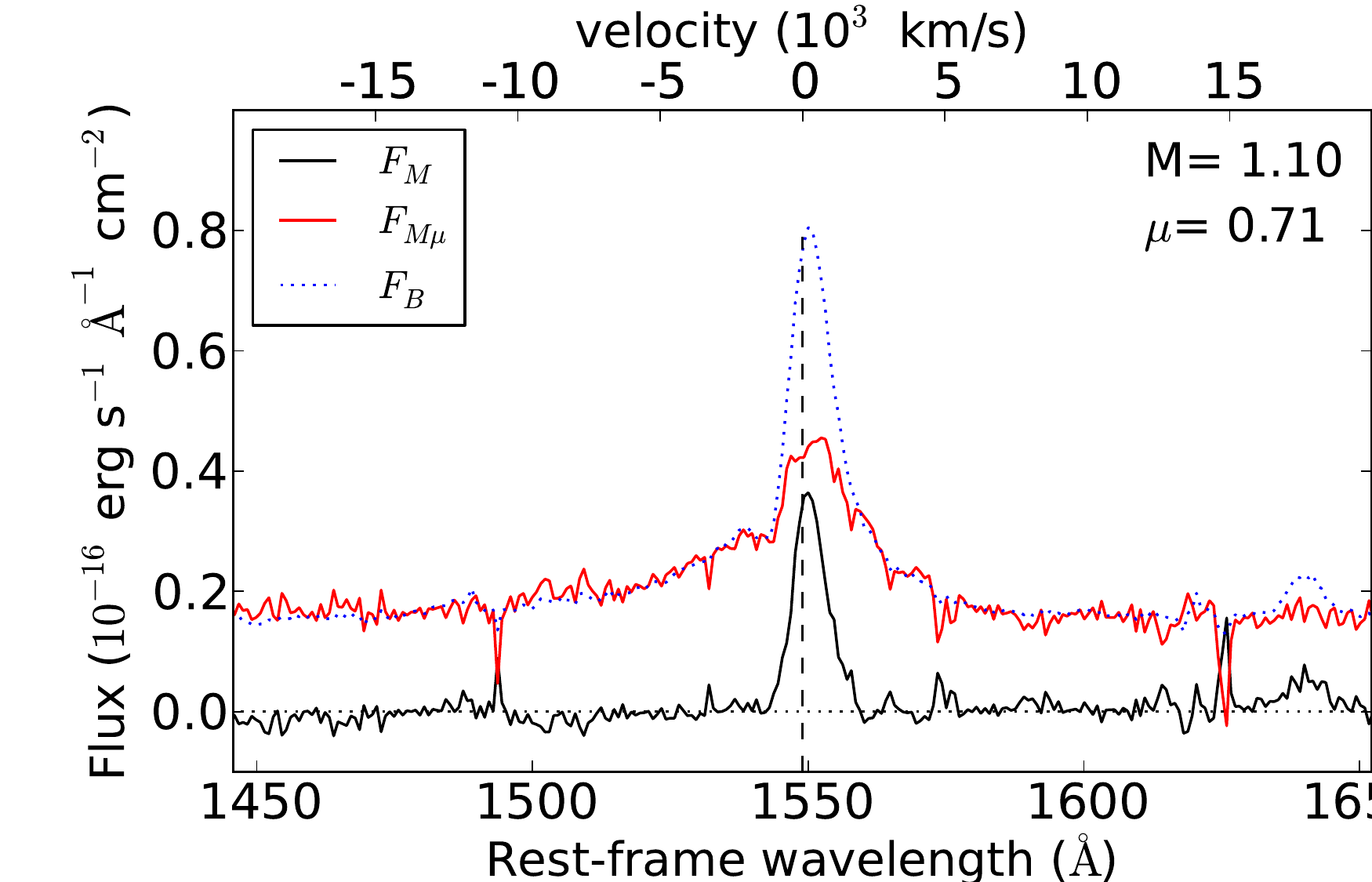}} 
\setcounter{subfigure}{7}
\renewcommand{\thesubfigure}{(\alph{subfigure}2)}
\subfigure[SDSS 1138+0314 (\CIII)]{\includegraphics[scale=0.44]{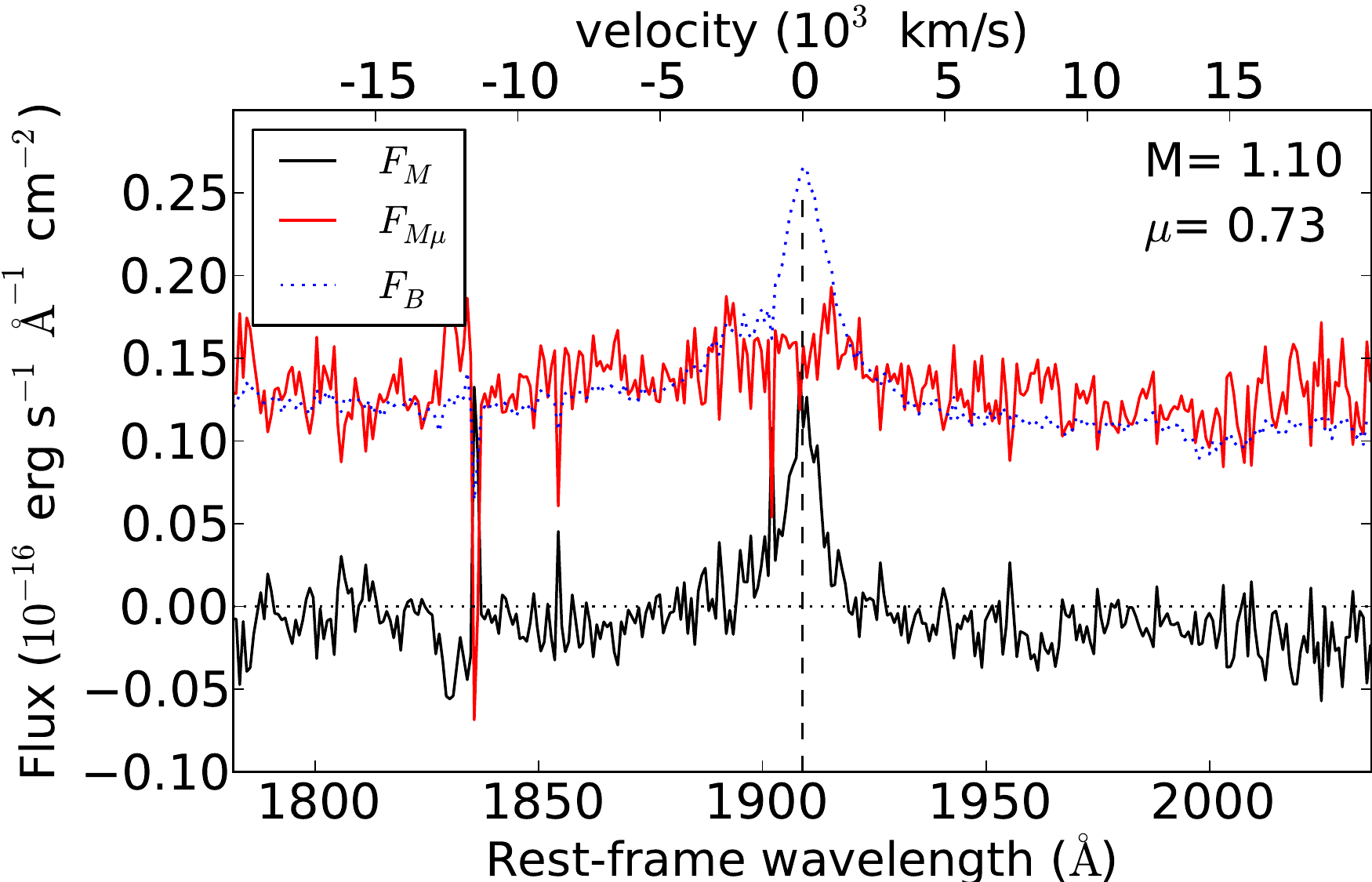}} 

\setcounter{subfigure}{8}
\renewcommand{\thesubfigure}{(\alph{subfigure}1)}
  \subfigure[J1226-006 (\MgII)]{\includegraphics[scale=0.44]{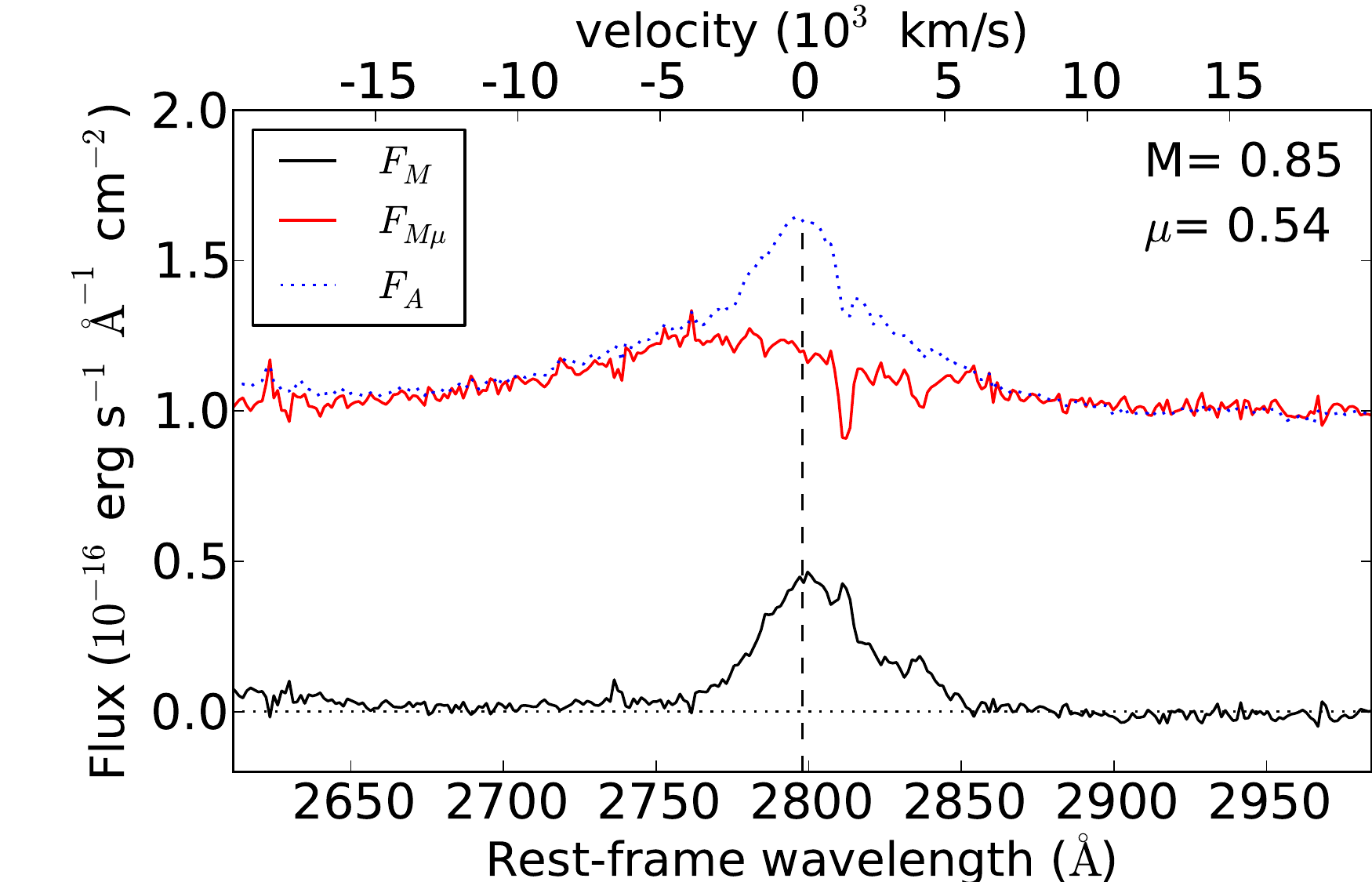}} 
\setcounter{subfigure}{8}
\renewcommand{\thesubfigure}{(\alph{subfigure}2)}
\subfigure[J1226-006 (\OII\,redshifted)]{\includegraphics[scale=0.44]{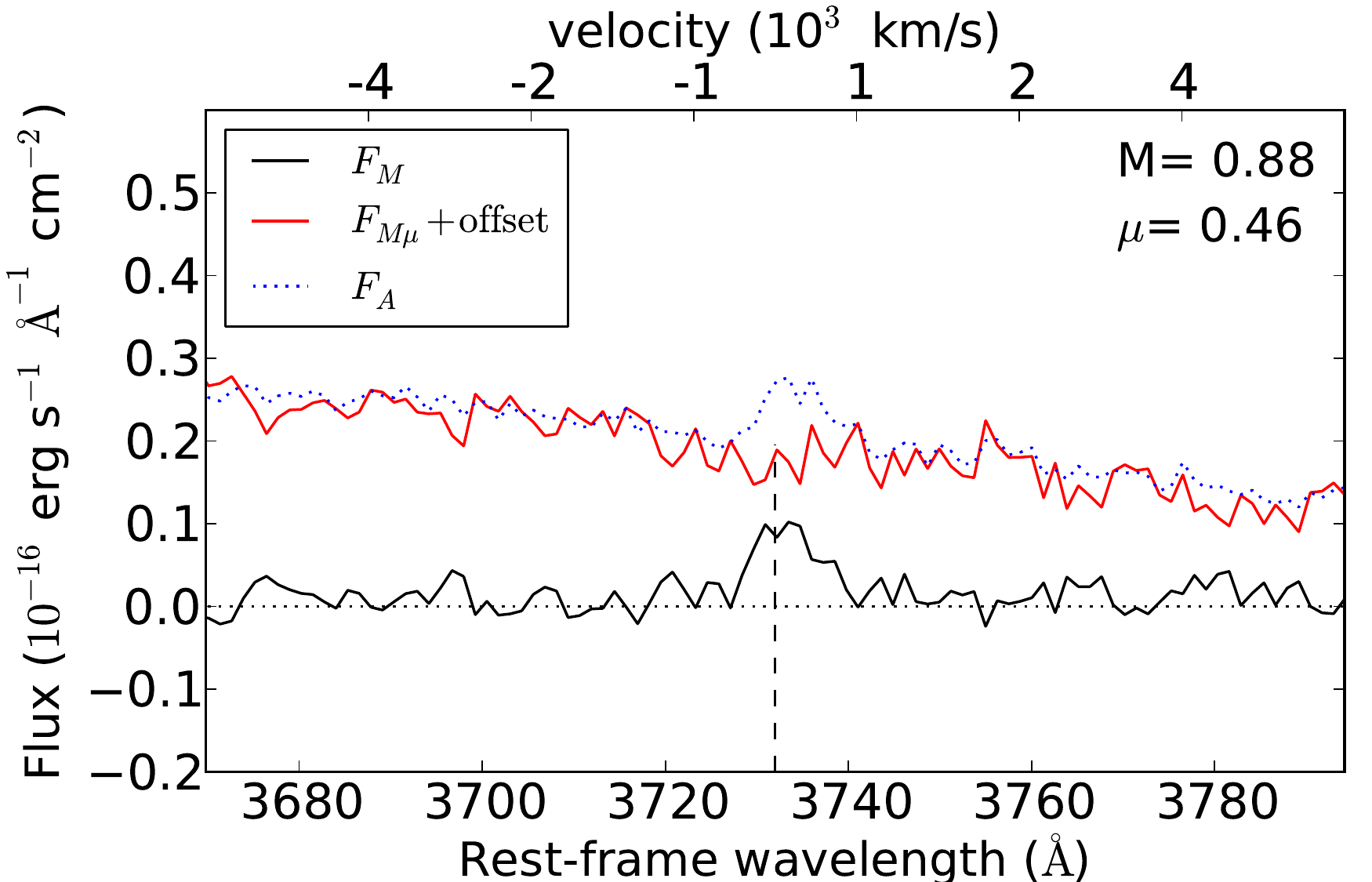}} 

\setcounter{subfigure}{9}
\renewcommand{\thesubfigure}{(\alph{subfigure}1)}
  \subfigure[SDSS J1335+0118 (\CIII)]{\includegraphics[scale=0.44]{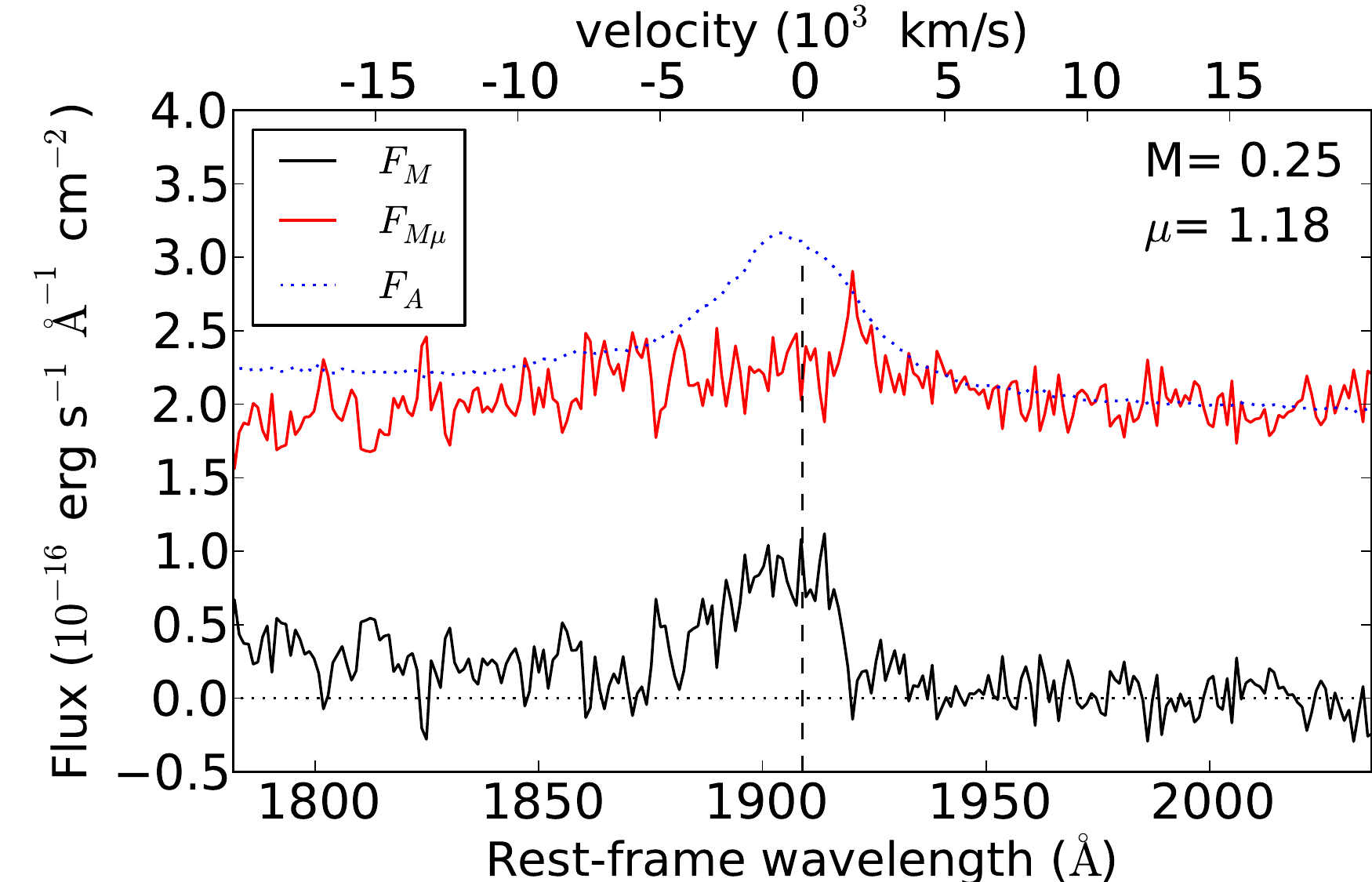}} 
\setcounter{subfigure}{9}
\renewcommand{\thesubfigure}{(\alph{subfigure}2)}
\subfigure[SDSS J1335+0118 (\MgII)]{\includegraphics[scale=0.44]{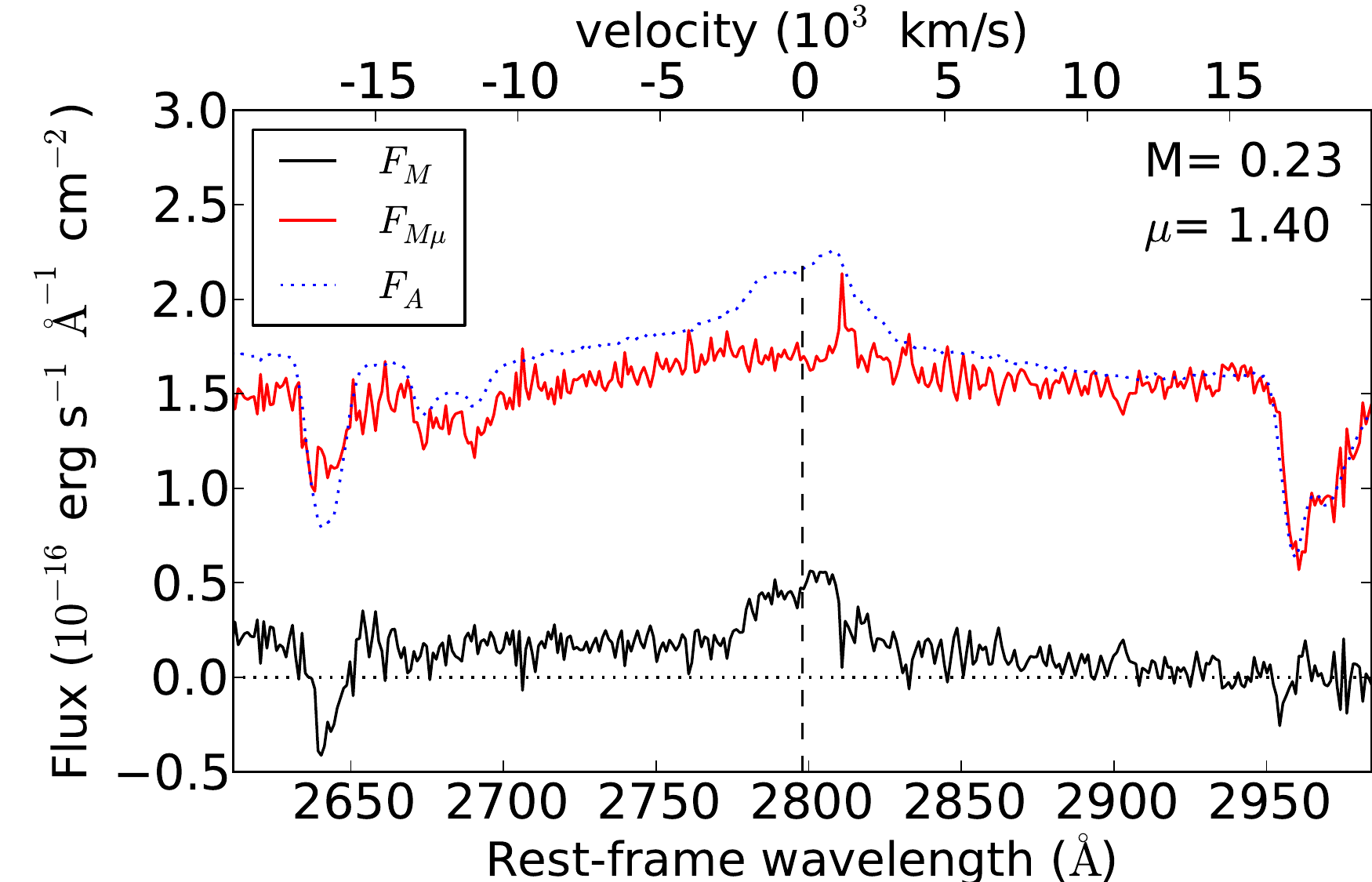}} 

\setcounter{figure}{3}
\caption[]{continued}
\end{figure*}
\begin{figure*} 
\centering

\setcounter{subfigure}{10}
\renewcommand{\thesubfigure}{(\alph{subfigure}1)}
  \subfigure[Q1355-2257 (\CIII; decomposition uncertain)]{\includegraphics[scale=0.44]{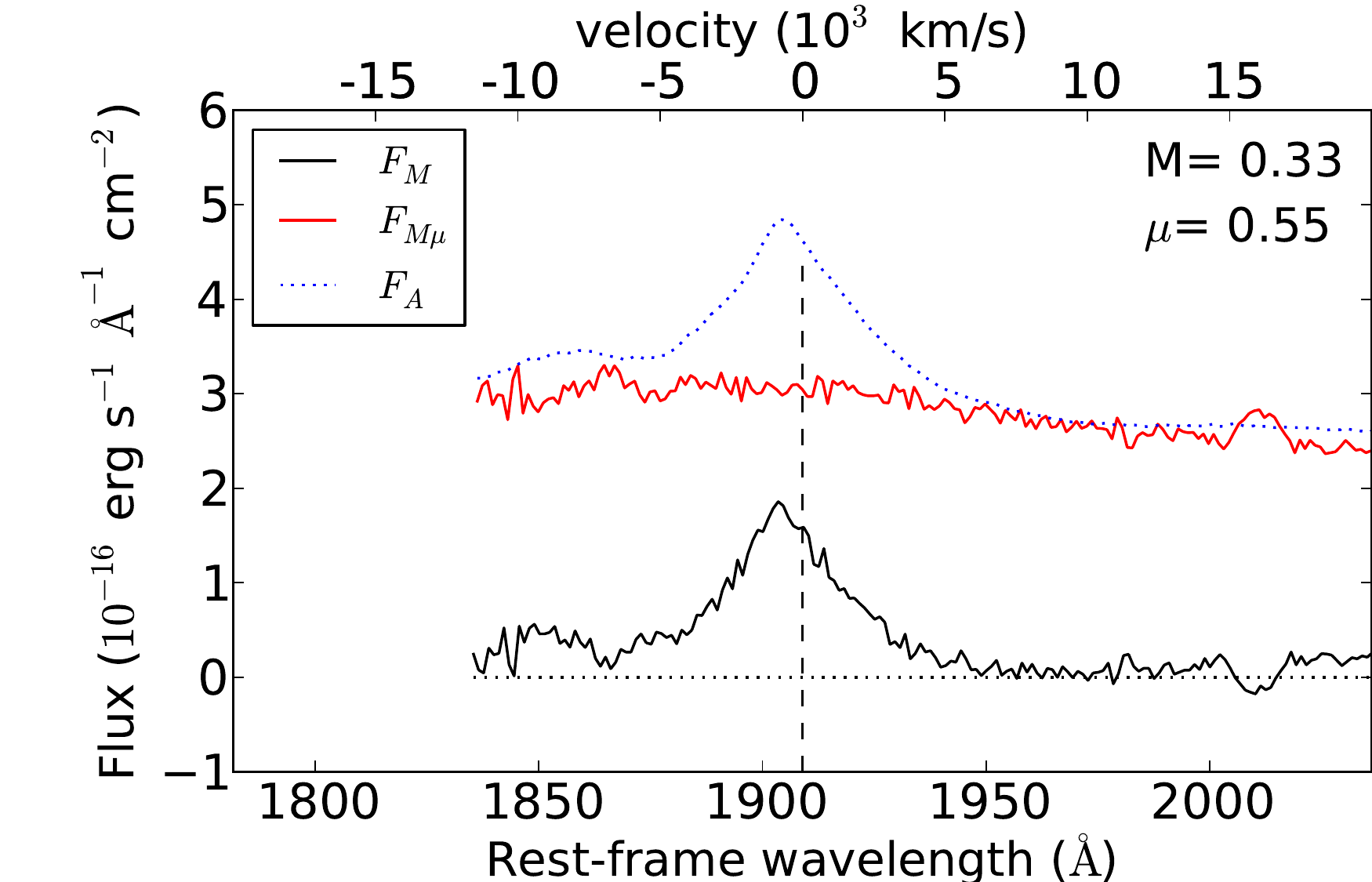}} 
\setcounter{subfigure}{10}
\renewcommand{\thesubfigure}{(\alph{subfigure}2)}
\subfigure[Q1355-2257 (\MgII)]{\includegraphics[scale=0.44]{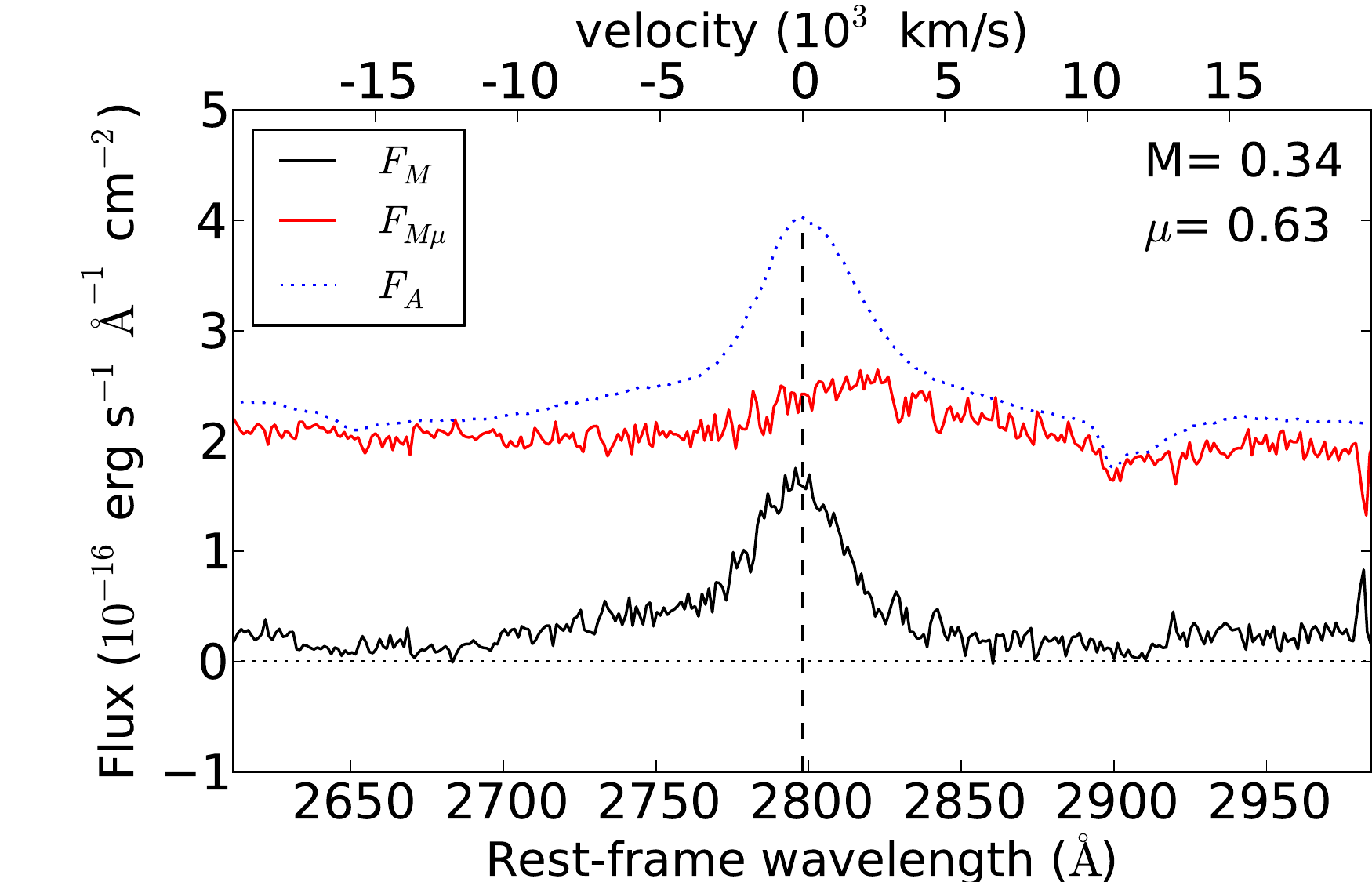}} 

\setcounter{subfigure}{11}
\renewcommand{\thesubfigure}{(\alph{subfigure}1)}
  \subfigure[WFI2033-4723 (\CIII)]{\includegraphics[scale=0.44]{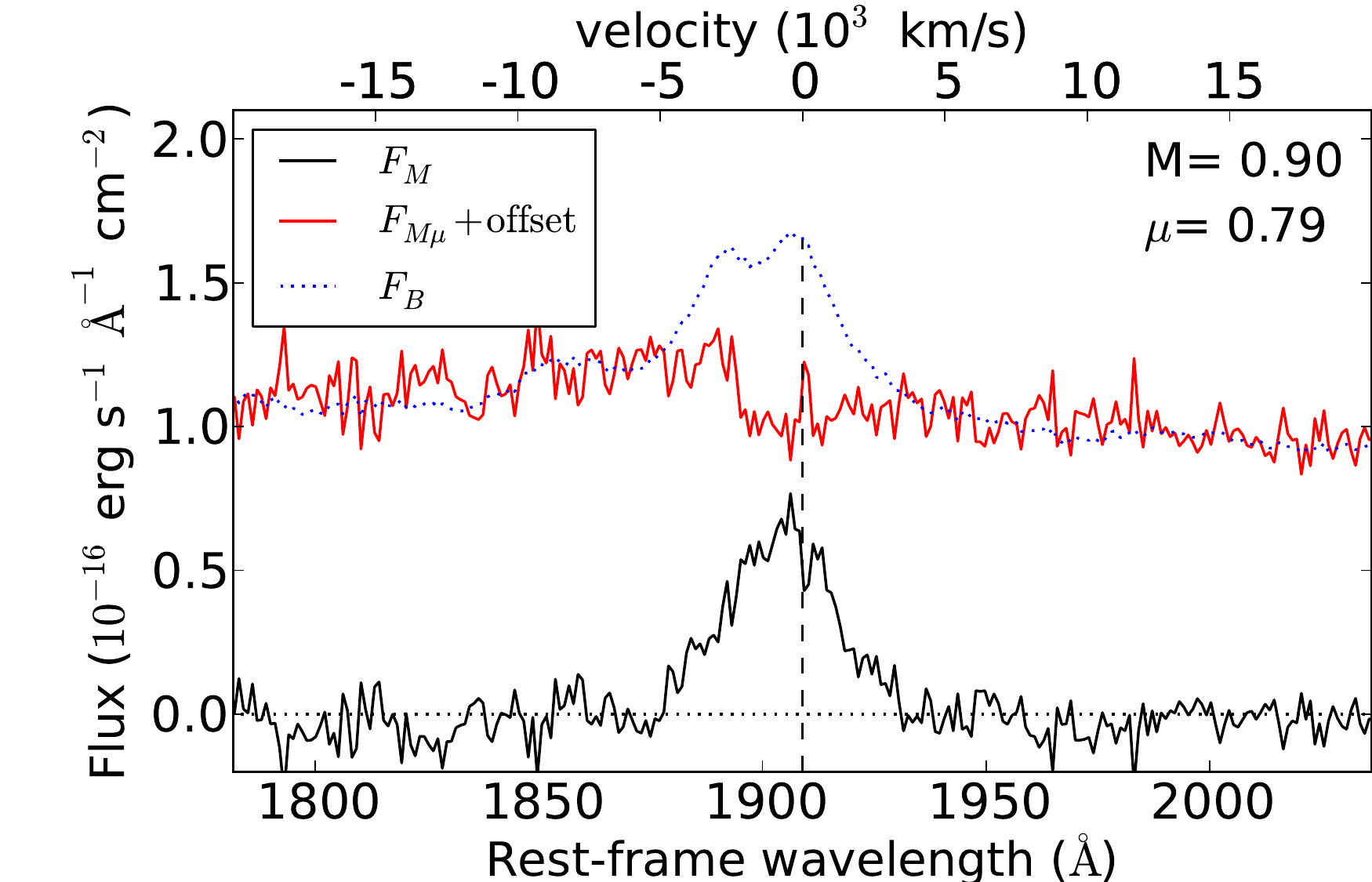}} 
\setcounter{subfigure}{11}
\renewcommand{\thesubfigure}{(\alph{subfigure}2)}
\subfigure[WFI2033-4723 (\MgII\,+\,atm)]{\includegraphics[scale=0.44]{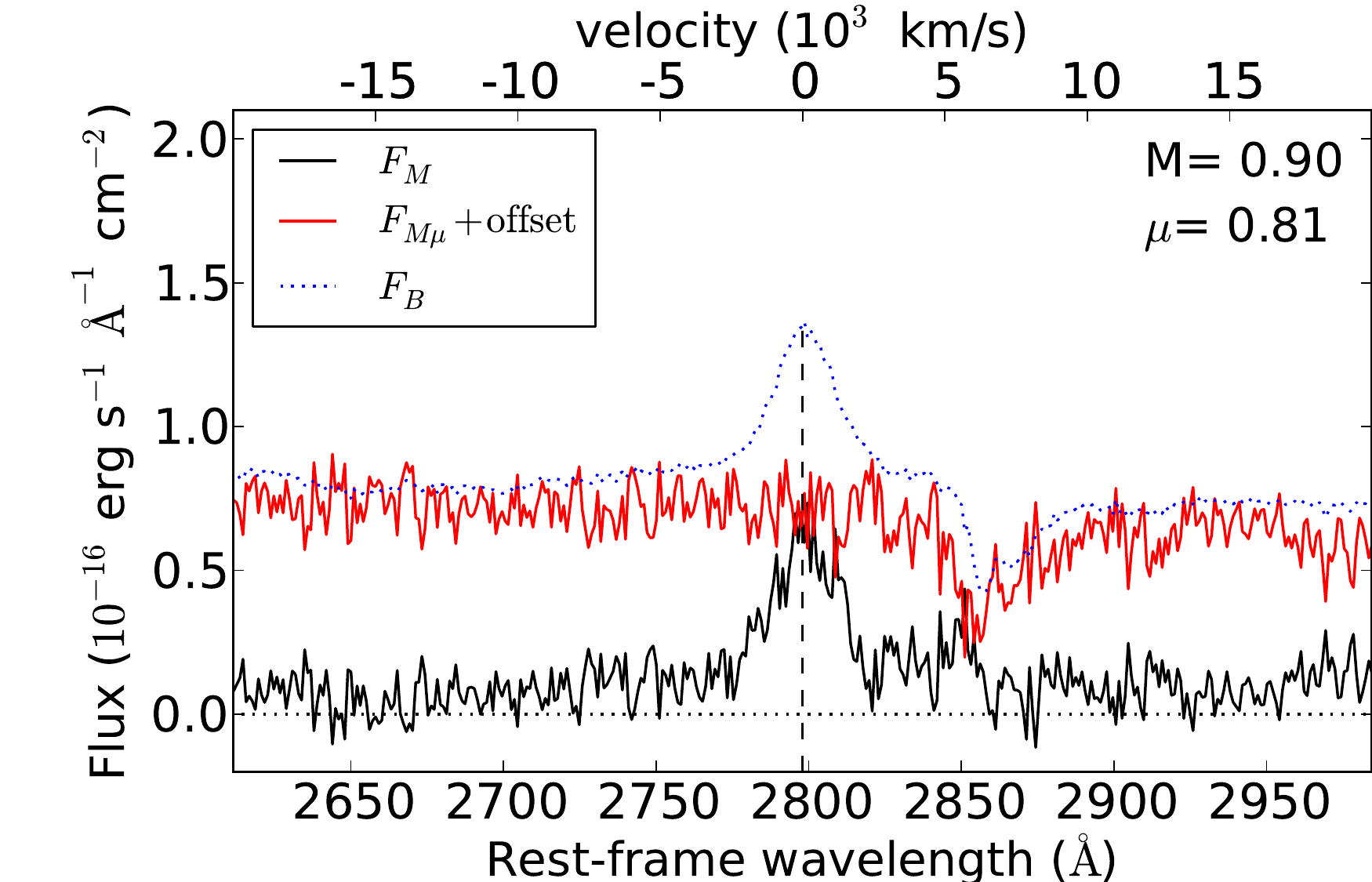}} 

\setcounter{subfigure}{13}
\renewcommand{\thesubfigure}{(\alph{subfigure}1)}
  \subfigure[J0924+0219 (\CIII)]{\includegraphics[scale=0.44]{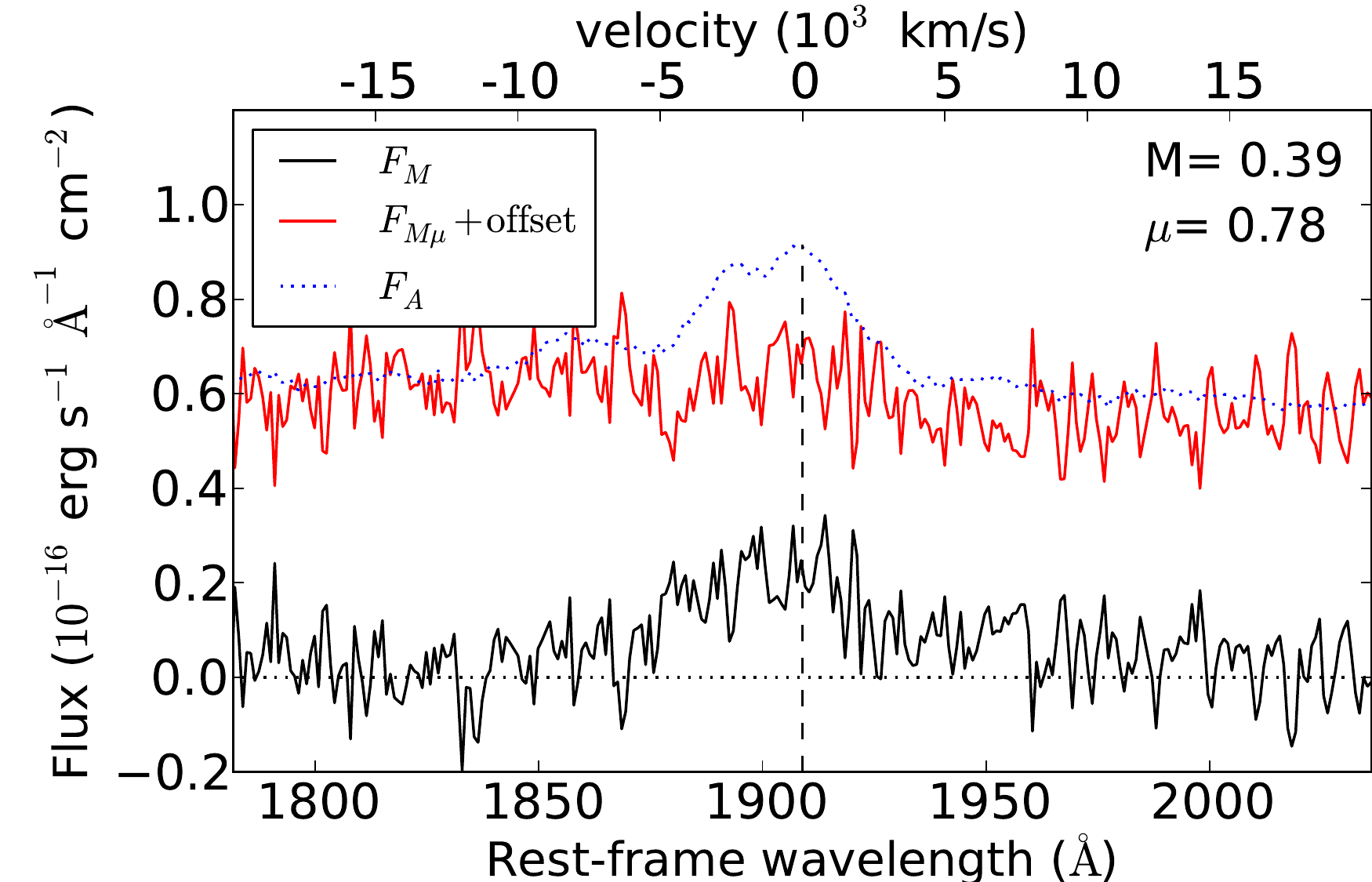}} 
\setcounter{subfigure}{13}
\renewcommand{\thesubfigure}{(\alph{subfigure}2)}
\subfigure[J0924+0219 (\MgII)]{\includegraphics[scale=0.44]{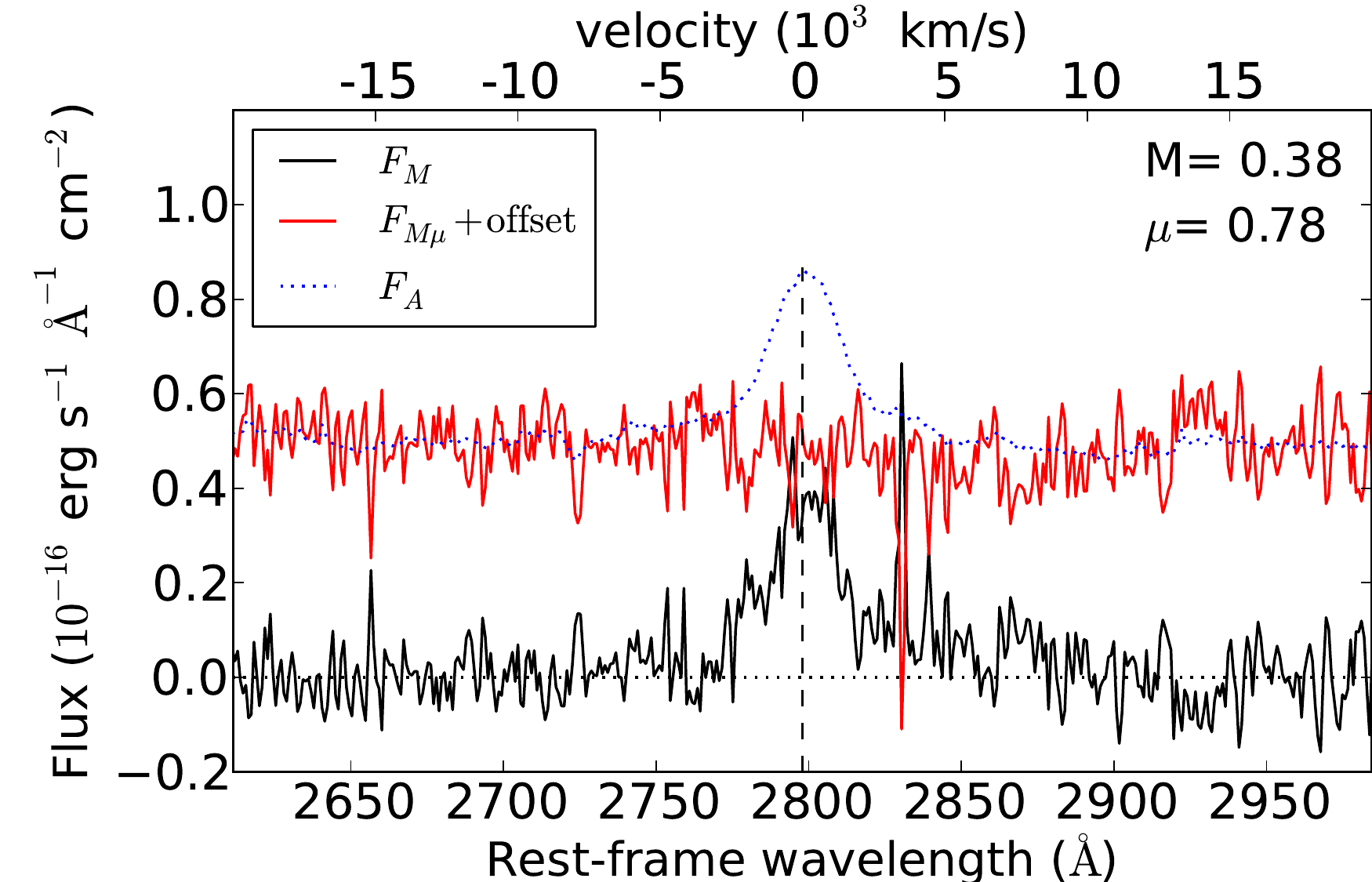}} 

\setcounter{subfigure}{14}
\renewcommand{\thesubfigure}{(\alph{subfigure}1)}
  \subfigure[J1131-1231, image pair B-C (\MgII)]{\includegraphics[scale=0.44]{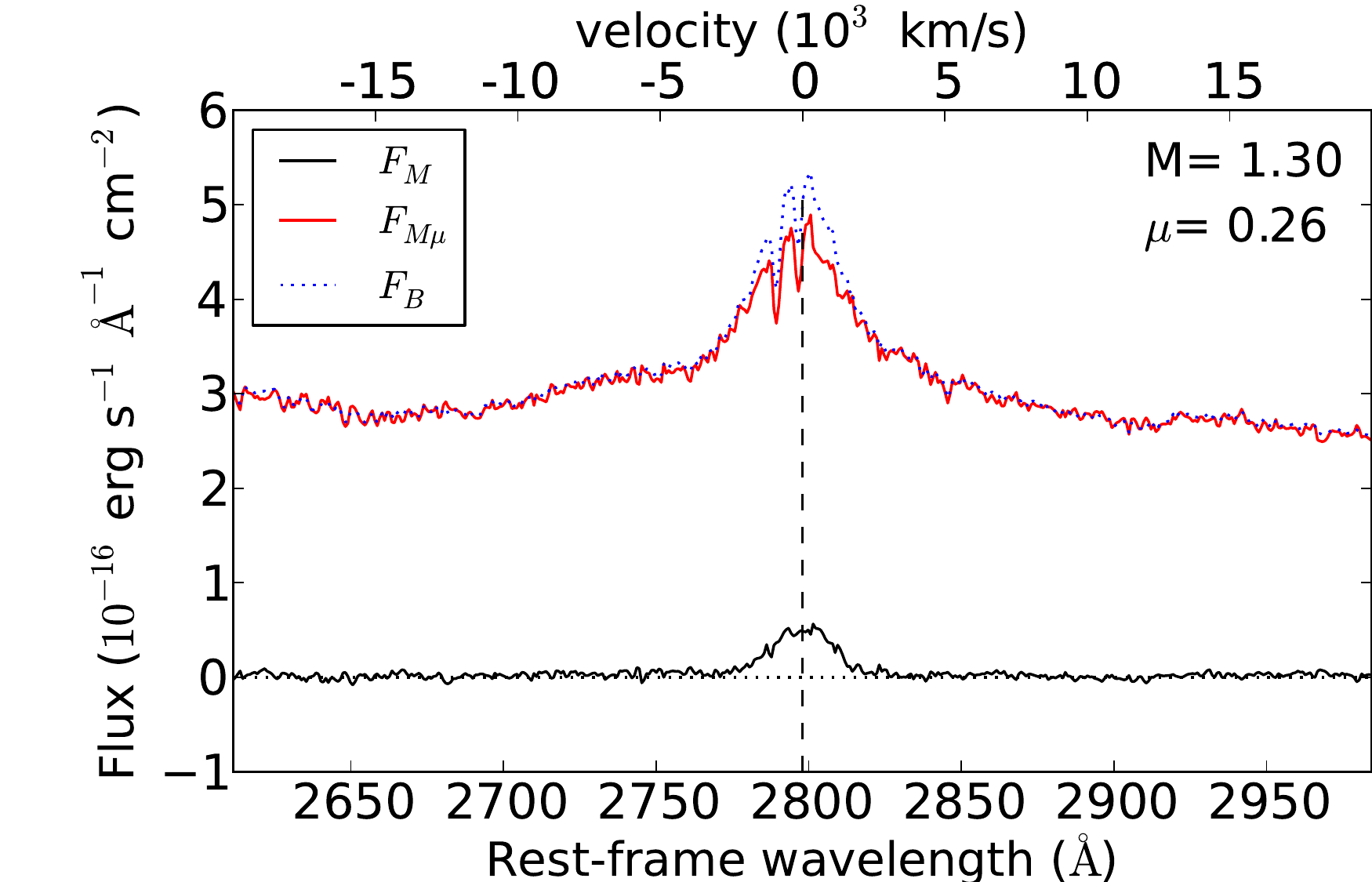}}   
\setcounter{subfigure}{14}
\renewcommand{\thesubfigure}{(\alph{subfigure}2)}
\subfigure[J1131-1231, image pair B-C (\Hbeta+\OIII)]{\includegraphics[scale=0.44]{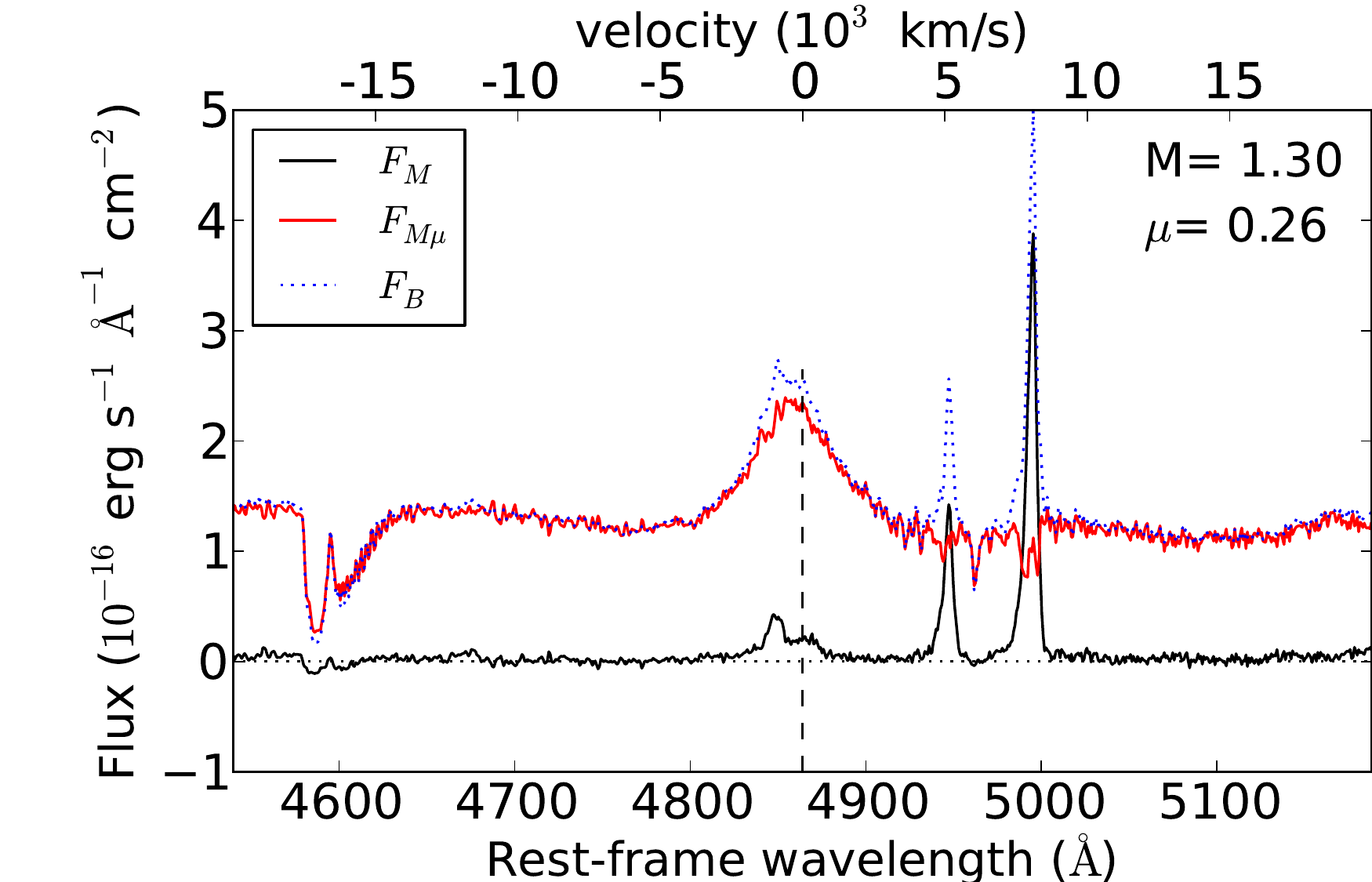}}

\setcounter{figure}{3}
\caption[]{continued}
\end{figure*}

\begin{figure*} 
\centering

\setcounter{subfigure}{14}
\renewcommand{\thesubfigure}{(\alph{subfigure}3)}
  \subfigure[J1131-1231, image pair A-B (\MgII)]{\includegraphics[scale=0.44]{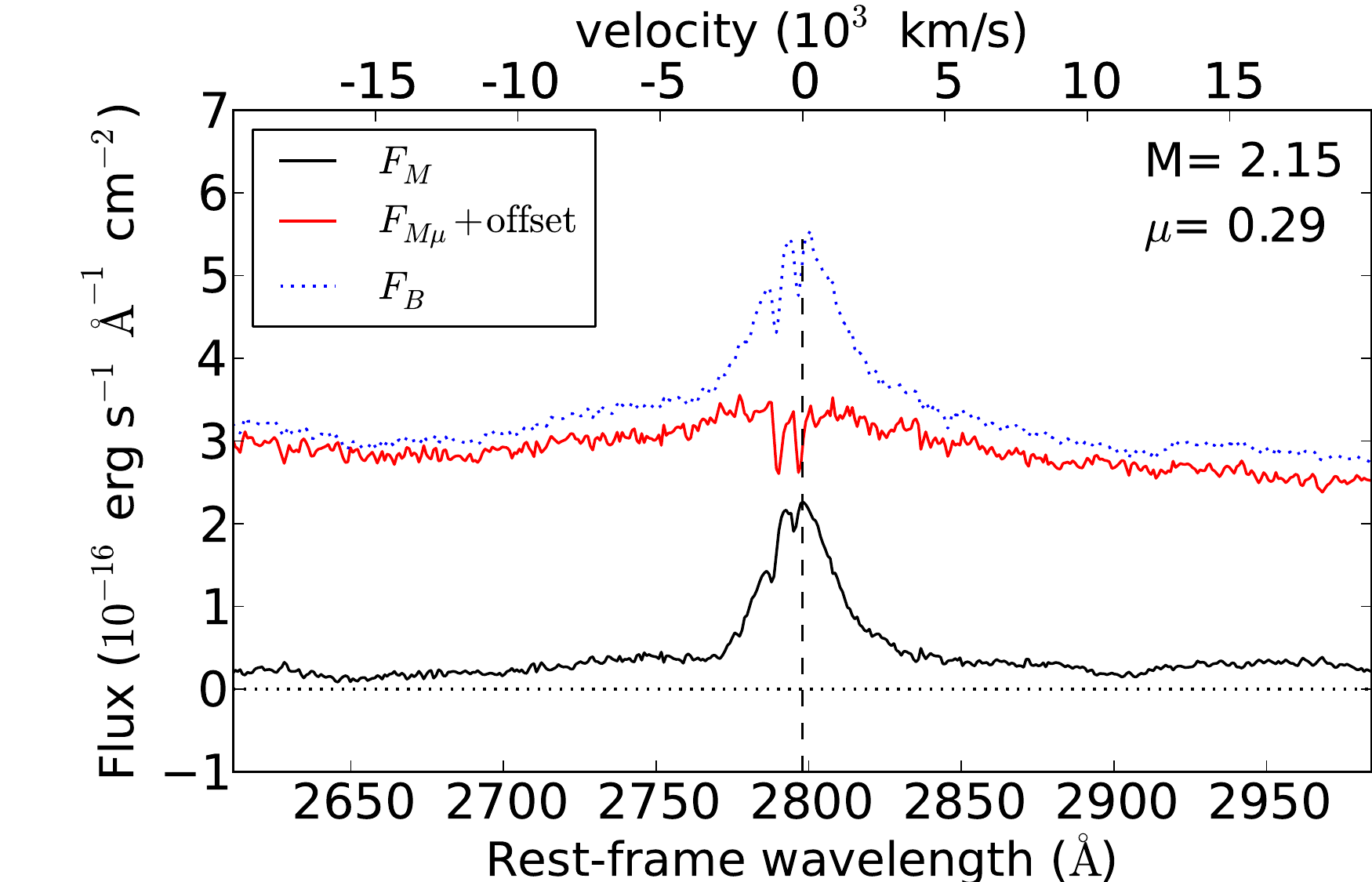}}
\setcounter{subfigure}{14}
\renewcommand{\thesubfigure}{(\alph{subfigure}4)}
\subfigure[J1131-1231, image pair A-B (\Hbeta+\OIII)]{\includegraphics[scale=0.44]{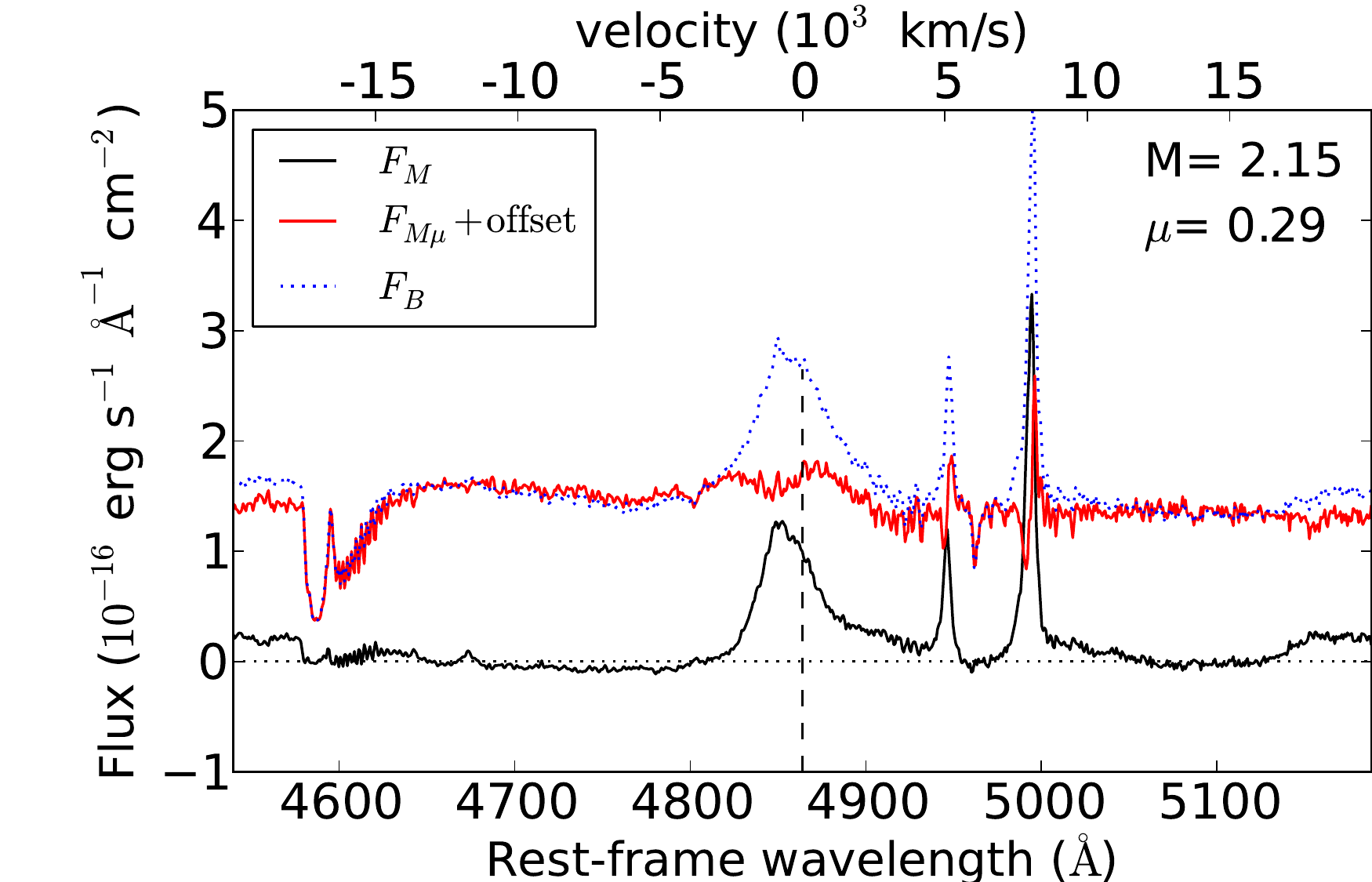}} 

\setcounter{subfigure}{15}
\renewcommand{\thesubfigure}{(\alph{subfigure}1)}
  \subfigure[H1413+117 (\CIV)]{\includegraphics[scale=0.44]{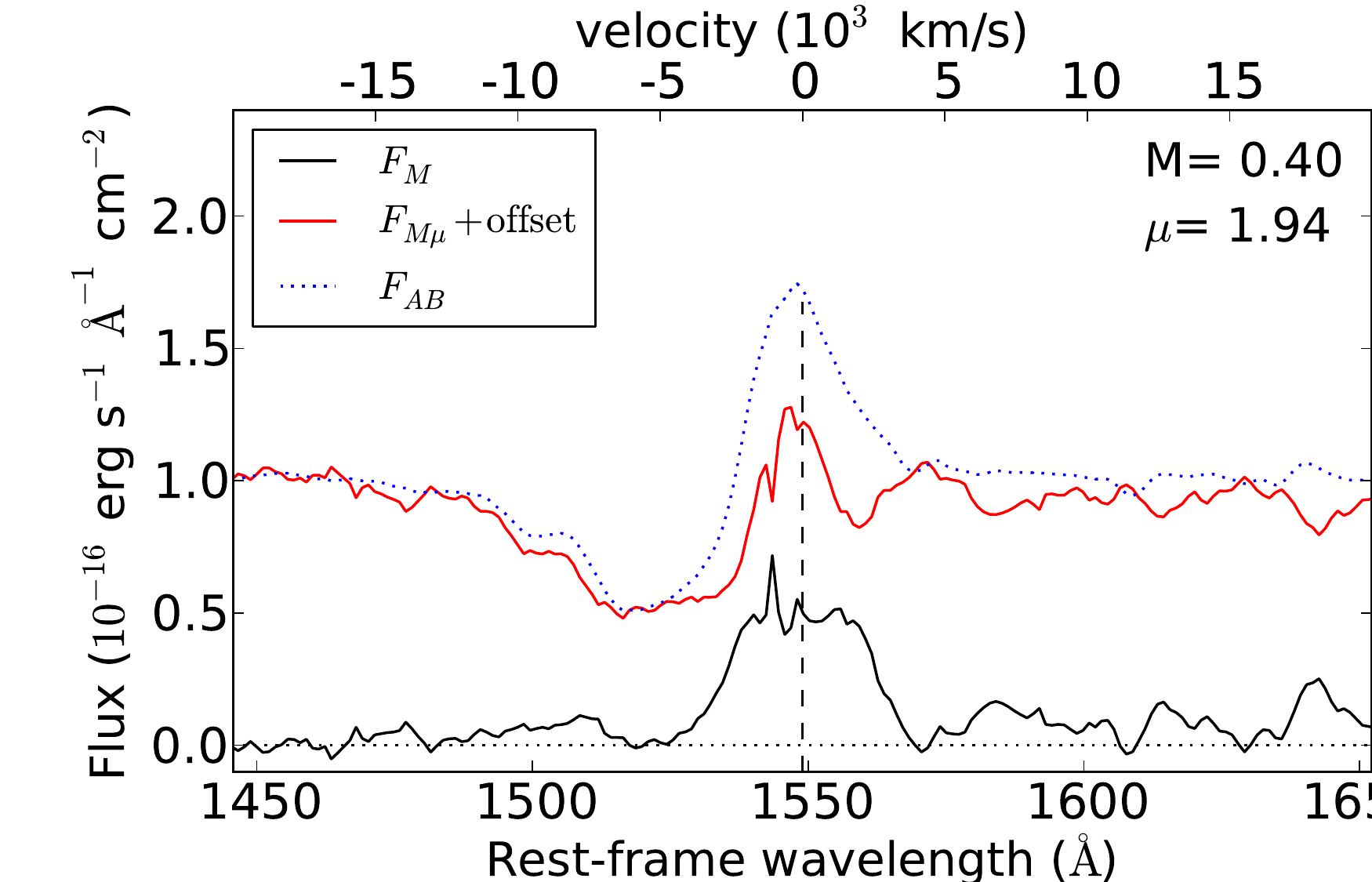}} 
\setcounter{subfigure}{15}
\renewcommand{\thesubfigure}{(\alph{subfigure}2)}
\subfigure[H1413+117 (\Hbeta+\OIII)]{\includegraphics[scale=0.44]{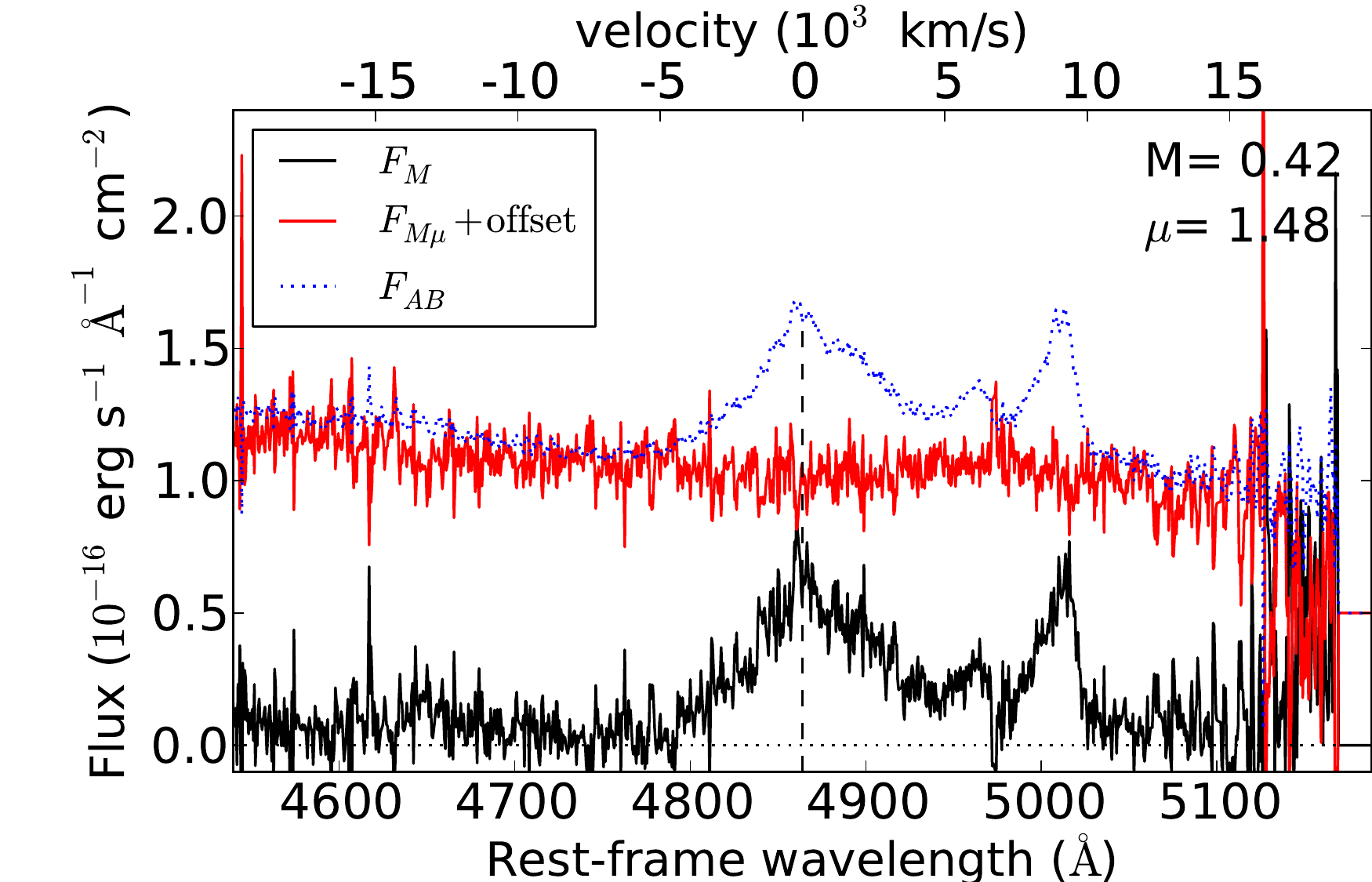}} 

 \setcounter{subfigure}{16}
\renewcommand{\thesubfigure}{(\alph{subfigure})}
\subfigure[HE 2149-2745 (\CIII)]{\includegraphics[scale=0.44]{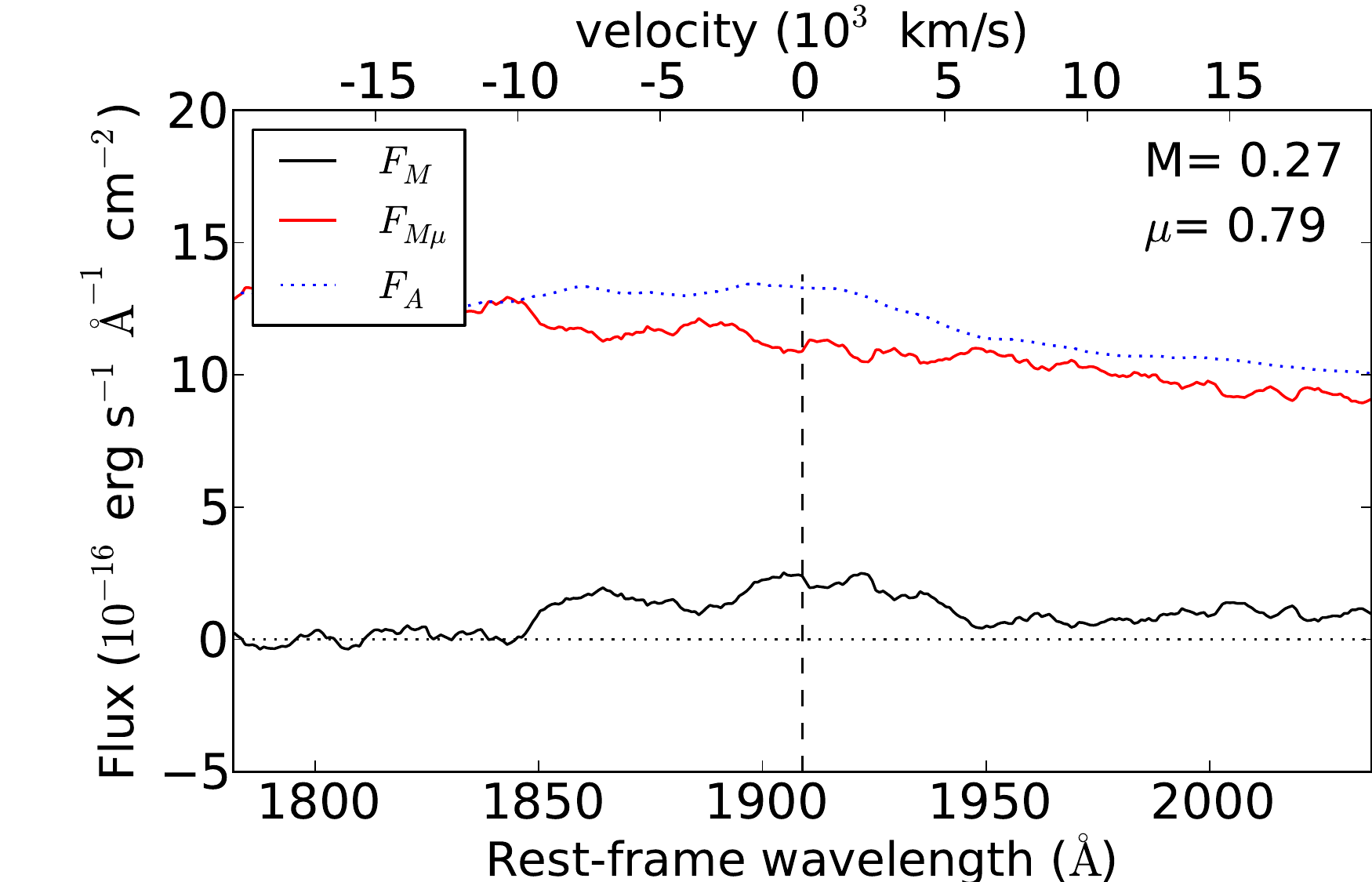}}
\setcounter{subfigure}{17}
\renewcommand{\thesubfigure}{(\alph{subfigure}1)}
  \subfigure[Q2237+0305 (\CIV)]{\includegraphics[scale=0.44]{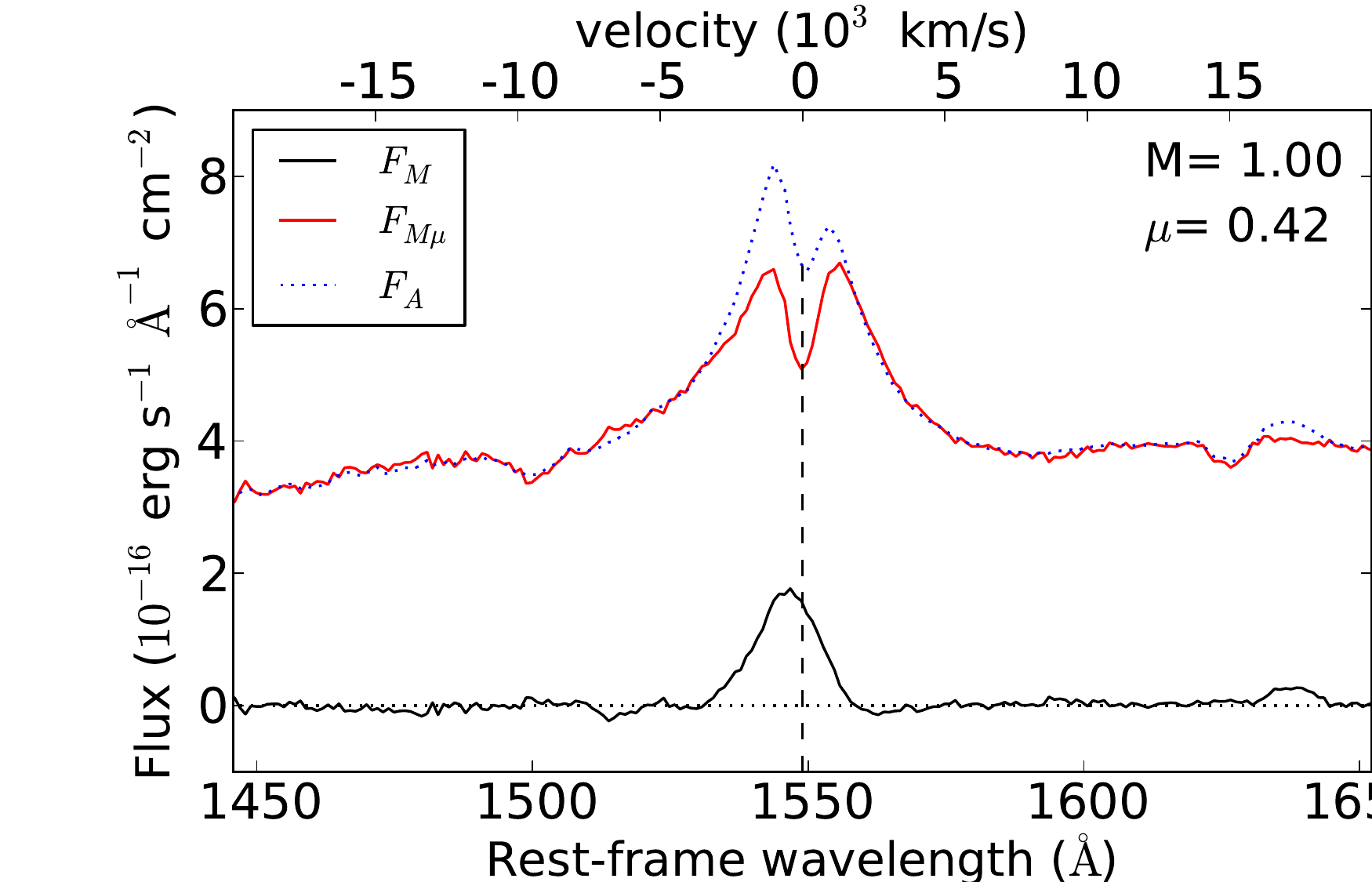}}

\setcounter{subfigure}{17}
\renewcommand{\thesubfigure}{(\alph{subfigure}2)}
\subfigure[Q2237+0305 (\CIII)]{\includegraphics[scale=0.44]{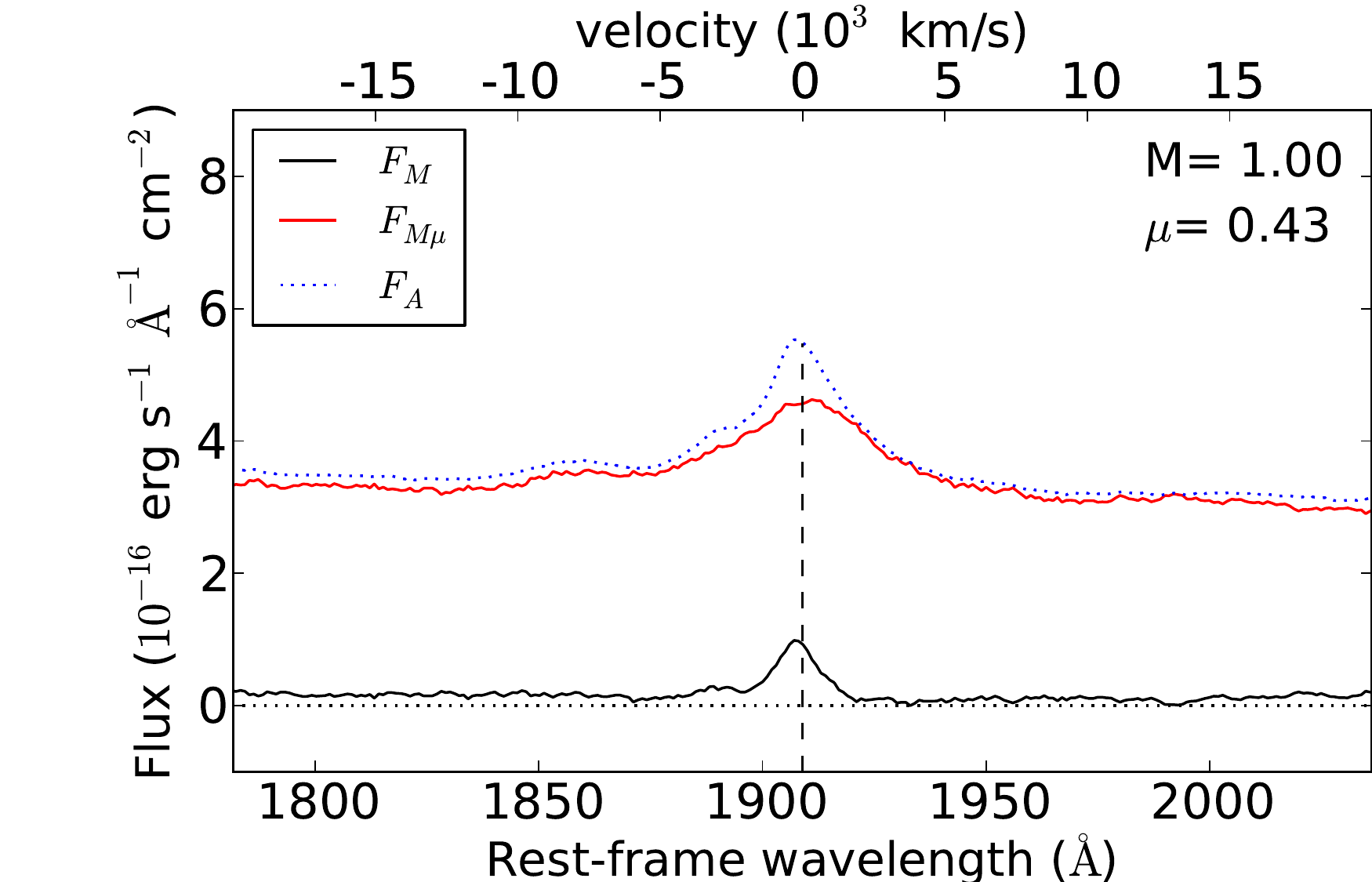}} 
\setcounter{subfigure}{17}
\renewcommand{\thesubfigure}{(\alph{subfigure}3)}
  \subfigure[Q2237+0305 (\MgII+\,atm)]{\includegraphics[scale=0.44]{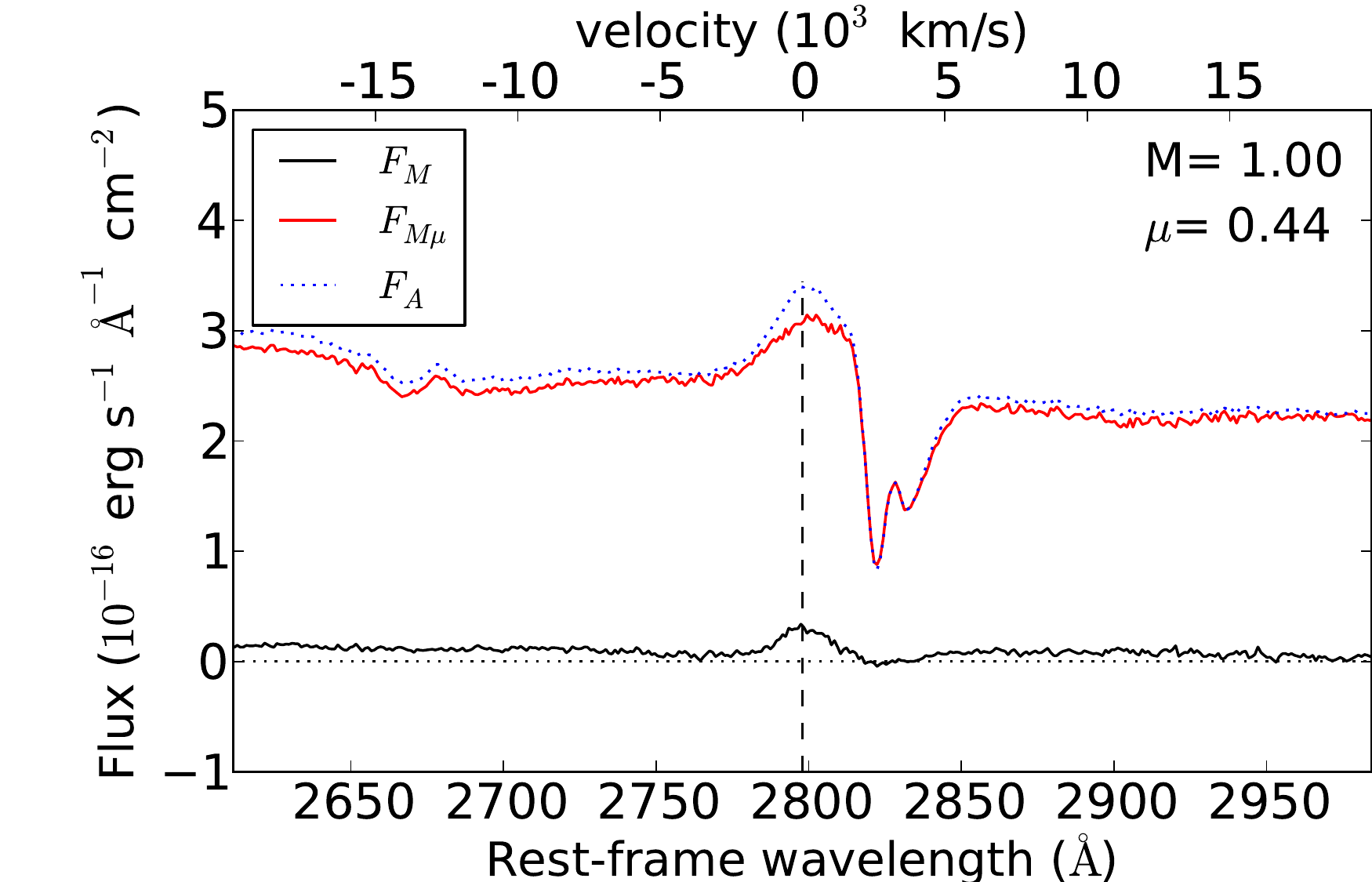}}

\setcounter{figure}{3}
\caption[]{continued}
\end{figure*}

\begin{table*}[ht!]
\begin{center}
\begin{tabular}{l|ccc|ccclccc}
\hline
Object & $F_1/F_2$ & $\lambda$ (\AA) & Line & $A$ & $M$ & $\mu_{\rm cont}$ & Chromaticity & BW & LC & RW \\
\hline
(a) HE~0047-1756     & B/A & 5100 & \CIII$^{(a)}$ & 0.277$\pm$0.003 & 0.231$\pm$0.007 &   1.199$\pm$0.039 &  DE, CML & (1) & 0 & 0 \\
                 & B/A & 7500 & \MgII & 0.258$\pm$0.003 & 0.220$\pm$0.005 &    1.173$\pm$0.030 & - & 1 & 0 & 1 \\ 
(b) Q0142-100        & B/A & 5200 & \OVI  & 0.126$\pm$0.001 & 0.126$\pm$0.001 & 1.00        & DE, IV & - & - & - \\
                 & B/A & 7100 & \CIII & 0.131$\pm$0.003 & 0.131$\pm$0.001 & 1.00        & - &  - & - & - \\
(c) SDSS~J0246-0825  & B/A & 5100 & \CIII & 0.249$\pm$0.003 & 0.315$\pm$0.015  &   0.790$\pm$0.039 & DE, CML & 1 & 0 & 1 \\
                 & B/A & 7500 & \MgII & 0.260$\pm$0.002 & 0.340$\pm$0.010  &   0.765$\pm$0.023 & - & 1 & 0 & 1 \\
(d) HE~0435-1223     & B/D & 5150 & \CIII & 1.093$\pm$0.003 & 1.370$\pm$0.030  &   0.798$\pm$0.018 & DE, CML & 0 & 0 & 1 \\
                 & B/D & 7500 & \MgII & 1.210$\pm$0.002 & 1.470$\pm$0.050  &   0.823$\pm$0.028 & - & 0 & 1 & 1 \\ 
(e) SDSS~J0806+2006  & B/A & 4850 & \CIII$^{(a)}$ & 0.780$\pm$0.046 & 0.485$\pm$0.065  &    1.608$\pm$0.235 & - & 0 & 0 & 0\\ 
                 & B/A & 7100 & \MgII & 0.755$\pm$0.046 &   0.420$\pm$0.040  &    1.797$\pm$0.203 &  - & 0 & 0 & 0 \\ 
(f) FBQ~0951+2635    & B/A & 6300 & \MgII & 0.295$\pm$0.010 &   0.230$\pm$0.020  &    1.283$\pm$0.120  & - & 0 & (0) & (0) \\ 
(g) BRI~0952-0115    & B/A & 6650 & \Lyalpha & 0.295$\pm$0.008 &    0.440$\pm$0.040 &   0.670$\pm$0.064 & - & 0 & 1 & 0  \\ 
(h) SDSS~J1138+0314  & C/B & 5300 & \CIV  & 0.779$\pm$0.035 &    1.100$\pm$0.050 &   0.708$\pm$0.045 & - & 1 & 1 & 1 \\
                 & C/B & 6550 & \CIII & 0.806$\pm$0.042 &   1.100$\pm$0.050 &    0.733$\pm$0.051 & - & 1 & 1 & 1 \\ 
(i) J1226-0006       & B/A & 5950 & \MgII$^{(b)}$ & 0.456$\pm$0.007 &    0.850$\pm$0.050 &  0.536$\pm$0.033  & DE, LC & (1) & (1) & 0? \\  
                 & B/A & 7930 & \OII  & 0.410$\pm$0.006 &    0.875$\pm$0.075 &   0.464$\pm$0.041 & - & 0 & 0 & 0 \\  
(j) SDSS~J1335+0118  & B/A & 4890 & \CIII$^{(c)}$ & 0.295$\pm$0.003 &   0.250$\pm$0.010 &    1.180$\pm$0.049 & DE, IV & 0 & 0 & 1 \\
                 & B/A & 7200 & \MgII & 0.321$\pm$0.004 &   0.230$\pm$0.020 &    1.396$\pm$0.123 & - & 1 & 1 & 1 \\ 
(k) Q1355-2257     & B/A & 4500 & \CIII$^{(a)}$ & 0.183$\pm$0.002 &    0.330$\pm$0.020 &   0.550$\pm$0.034 & CML, DE, LC & 0 & 0 & (0) \\
                 & B/A & 6630 & \MgII & 0.215$\pm$0.004 &    0.340$\pm$0.030 &   0.632$\pm$0.057 & - & 0 & 1 & 1  \\
(l) WFI~2033-4723    & C/B & 5070 & \CIII$^{(a)}$ & 0.709$\pm$0.022 &    0.900$\pm$0.020&   0.788$\pm$0.030 & - & 1 & 0 & 0 \\ 
                 & C/B & 7450 & \MgII & 0.728$\pm$0.040 & 0.900$\pm$0.020 &  0.809$\pm$0.048 & - & 0 & 0 & 0 \\  
(m) HE~2149-2745     & B/A & 4700 & \CIV$^{(a)}$  & 0.242$\pm$0.001 & 0.242$\pm$0.001 & 1.00 & DE, LC, IV & - & - & - \\
                 & B/A & 5750 & \CIII & 0.245$\pm$0.002 & 0.245$\pm$0.002 & 1.00 & - & - & - & - \\
(n) SDSS~J0924+0219$^\dagger$ & A/B & 4800 & \CIII & 0.304$\pm$0.019 & 0.39$\pm$0.01 & 0.779$\pm$0.053 & - & 0 & 0 & 0 \\
      &    & 7050 & \MgII & 0.296$\pm$0.023 & 0.38$\pm$0.01 & 0.779$\pm$0.063 & - & 0 & 0 & 0 \\
(o) J1131-1231$^\dagger$ & B/C & 4650 & \MgII & 0.334$\pm$0.012 & 1.30$\pm$0.07 & 0.257$\pm$0.016 & $(d)$ & 1 & 1 & 1 \\
      &    & 8050 & \Hbeta& 0.344$\pm$0.030 & 1.30$\pm$0.07 & 0.265$\pm$0.026 & -  & 1 & 1  & 1 \\
      & A/B & 4650 & \MgII & 0.630$\pm$0.050 & 2.15$\pm$0.05 & 0.293$\pm$0.024 & $(d)$ & 0 & 1 & 0 \\
      &    & 8050 & \Hbeta & 0.630$\pm$0.050 & 2.15$\pm$0.05 & 0.293$\pm$0.024& - & 1 & 0 & 1 \\
(p) H1413+117$^\dagger$ & AB/D& 5500 & \CIV & 0.775$\pm$0.010& 0.40$\pm$0.04& 1.940$\pm$0.220 & CML & 0 & 1 & 0 \\
      &     & 17200  & \Hbeta& 0.630$\pm$0.010& 0.43$\pm$0.02& 1.480$\pm$0.080 & - & 0 & 0 & 0 \\
(q) HE~2149-2745$^\dagger$ & A/B & 5750  & \CIII & 0.214$\pm$0.003& 0.27$\pm$0.02& 0.785$\pm$0.025 & CML, (DE, IV) & 0 & (1) & 0 \\ 
(r) Q2237+0305$^\dagger$ & A/D  & 4150  & \CIV & 0.425$\pm$0.002 & 1.00$^{0.20}_{-0.13}$ & 0.425$\pm$0.085 & $(e)$& 1 & 1 & 1 \\
      &     & 5150  & \CIII& 0.429$\pm$0.002 & 1.00$^{0.20}_{-0.13}$ & 0.429$\pm$0.085 & - & 1 & 1 & 1 \\
      &     & 7550  & \MgII& 0.438$\pm$0.003 & 1.00$^{0.20}_{-0.13}$ & 0.438$\pm$0.088 & - & 1 & 1 & 1  \\
\hline
                 
\end{tabular}
\end{center}
\vspace{0.2cm}
{\tiny{Notes: $(a)$ Uncertain due to the proximity to the edge. $(b)$ Likely cross contamination between the lens and the QSO images. $(c)$ The noisy decomposition around \CIII~might hide a faint broad component similar to the one observed in \MgII. There is a narrow component in the red wing superimposed to the broad one. $(d)$ Spectra already corrected for chromatic effects induced by the host galaxy \citep[see][]{Sluse2007}. $(e)$ Spectra already corrected for differential extinction between A \& D \citep[see][]{Sluse2011a}. $\dagger$ Objects from the extended sample. }}
\caption{Results of the MmD technique (see Sect.~\ref{subsec:results} and Appendix~\ref{appendixA} \& ~\ref{appendixB} for details). Cols. \#1 and \#2 give the object name and the image pair used for the decomposition, cols. \#5, \#6, and \#7 give the continuum flux ratio $A$, the macro-magnification ratio $M$, and the micro-magnification $\mu_{\rm cont}$ of the continuum. These quantities are given at the approximate wavelength (observed frame) given in col.\#3, which corresponds to the center of the emission line given in col.~\#4 (see Sect.~\ref{subsec:FMFMmu}). Column~\#8 lists the possible origins of the chromatic changes of $A$, namely  differential extinction (DE), chromatic microlensing (CML), contamination by the lens (LC), or intrinsic variability (IV). The last 3 columns indicate if microlensing of the broad line is seen in the blue wing (BW), line core (LC; [-500:+500] km/s), or red wing (RW). The value 0 is used when there is no microlensing and 1 when it is present. The value is in parentheses when the signal is weak or depends strongly on the exact value of $M$.
}
\label{tab:analysis}
\end{table*}

The MmD at the location of the main emission lines is shown in Fig.~\ref{fig:MmD}. The spectra of the systems Q0142-100 and HE~2149-2745 from our main sample are not shown because of the absence of microlensing. A detailed description of the microlensing signal is provided in Appendix~\ref{appendixA} (resp. Appendix~\ref{appendixB}) for the main (resp. extended) sample. In Table~\ref{tab:analysis}, we summarize the results of the MmD. For each object we provide $A$, $M$, $\mu$, we list the possible sources of chromatic variations of $A$ and the part of the broad emission line affected by microlensing, i.e. the line core ({\it {LC}}) which corresponds to the velocity range [-500:+500] km/s, the blue wing ({\it {BW}}), or the red wing ({\it {RW}}).  When the signal is uncertain (because it is weak or depends on the exact value of $M$), we put the value in parentheses. When we do not detect microlensing of the continuum, we do not fill these three columns. The definition of these three regions does not rely on a strict definition based on the line width and velocity ranges because of the variety of lines studied (sometimes blended with other lines) and of the uncertainty of the systemic redshift. We found, however, that this qualitative rating gives a fairly good synthetic view of the observed signal, with the drawback that it does not reflect the relative amplitude of each microlensed component. A more quantitative description, based on multi-component line fit and simulations, will be presented in a forthcoming paper.  

\subsection{Accuracy of the intrinsic flux ratios}
\label{subsec:intrinsic}

\begin{figure*}[ht!]
\begin{tabular}{cc}
\centering
 {\includegraphics[scale=0.45]{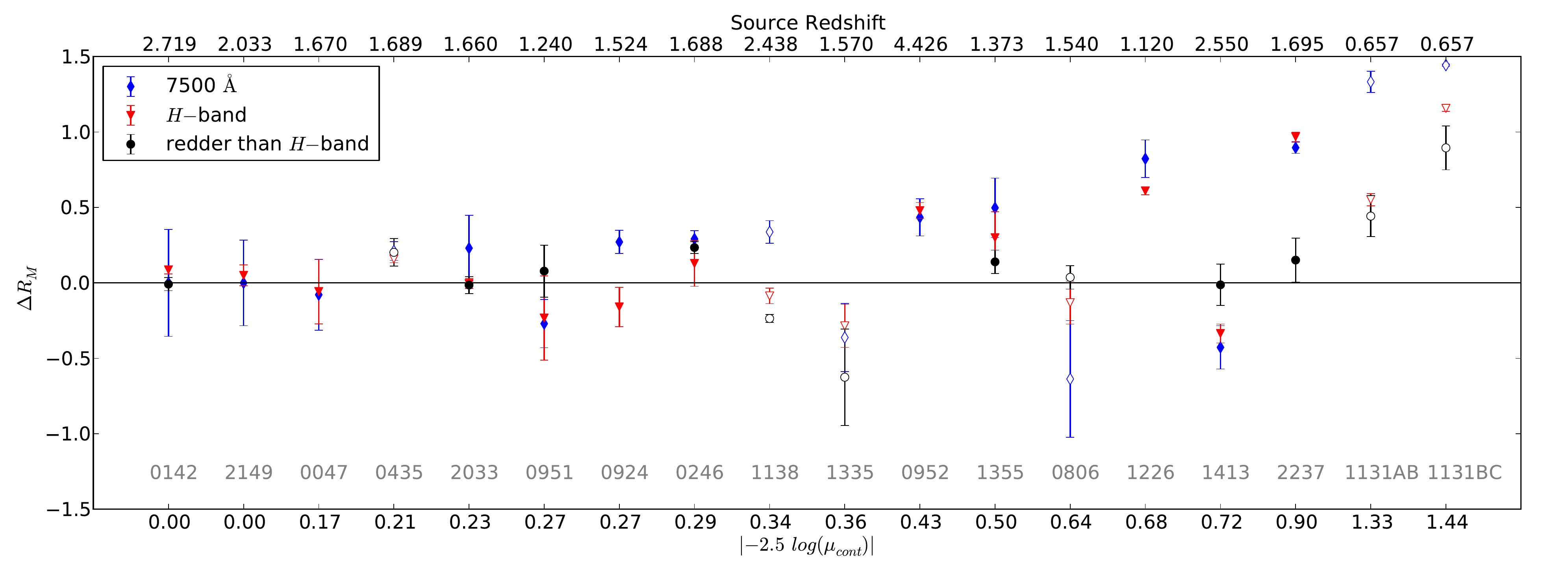}} 
\end{tabular}

\caption{Pseudo color $\Delta R_M = -2.5\,\log(M_w/M)$, where $M$ is the  intrinsic flux ratio  derived from our spectra using the MmD and $M_w$ is the flux ratio measured in 3 different wavelength ranges:  (1) the region around 7500\,\AA~(blue diamonds), (2) the  $H-$band (red triangles), and (3) wavelengths redder than $H-$band (black circle). Open symbols are used for the systems where our spectroscopic estimate of $M$ disagrees with the value at wavelengths redder than $H-$band (see Sect.~\ref{subsec:intrinsic}). The objects are ordered from left to right as a function of increasing microlensing of the continuum (note that the steps are not linear). The gray labels provide the name of the systems (shortened to the first 4 digits of the RA) and the upper axis indicates the redshift of the source. 
}
\label{fig:Hband}
\end{figure*}

We check that our measurement of the flux ratio $M$ is a good estimate of the macro-model flux ratio. For this purpose, we compare it with flux ratios collected from literature, measured in the near-infrared or at larger wavelengths. Near-infrared flux ratios are believed to be less affected by micro-lensing because of the larger size of the emitting region \citep{Wambsganss1991}. They should therefore be a good proxy of the macro-model flux ratio. However, we may not exclude that in some cases, the size of the emission region is small enough to be affected by microlensing \citep{Agol2009, Kishimoto2009, Kishimoto2011}. Table~\ref{tab:macro} summarizes measurements of the flux ratios from our spectra (col.\#3) and at larger wavelengths (col.\#4, from literature). In that table, the value of $M$ in col.~\#3 is an average between the values derived in the blue and the red part of the spectrum when we cannot disentangle between chromatic ML and differential extinction (the spread is then quoted in parentheses and becomes our  formal uncertainty). Otherwise, the values are measured at our reddest wavelength. Figure~\ref{fig:Hband} shows the pseudo-color $\Delta R_M = -2.5\,\log(M_w/M)$, where $M$ is the flux ratio derived from the MmD and $M_w$ a broad band measurement in various spectral window (Table~\ref{tab:macro}). Three spectral windows are considered. First the region around 7500 $\AA$ (which corresponds to the reddest part of our spectra). Second, the $H-$band ($\sim$ 1.6\,$\mu $m). Third, the reddest available wavelength ($K-$band = 2.2\,$\mu $m, $L-$band = 3.8\,$\mu $m, 11.5\,$\mu $m or radio 8.4 Ghz; see Table~\ref{tab:macro}). There are five objects which are lacking flux measurements at a wavelength redder than $H-$band. In that figure, we have quadratically added the expected variability over the time delay period (see Table~\ref{tab:variability} in Appendix) to the error bar of the continuum flux ratio. This error in principle affects $M$, and should therefore be propagated to all the other measurements. However, to ease legibility and identify more clearly discrepancies, we have included the variability error budget only to the spectroscopic flux ratios.

Figure~\ref{fig:Hband} shows that there is a general trend towards $\Delta R_M =$ 0 with increasing wavelength (black circle symbols), confirming that our estimate of $M$ is a good proxy to intrinsic flux ratios (see Appendix~\ref{appendixA} \& ~\ref{appendixB} for a discussion of each object individually). Small discrepancies of up to 0.1 mag are however observed because of intrinsic variability, and possibly of differential extinction. The effect of the intrinsic variability is in general smaller than the prediction from the structure function (Table~\ref{tab:variability}). Among the 12 systems for which $K$, $L$, MIR or radio flux ratios were available, five objects (depicted with open symbols) have flux ratios in the red which disagree significantly with our estimate of $M$. In one case (J1131-1231), it is likely that the flux ratio in $K-$band is significantly affected by microlensing. Two other systems (HE~0435-1223 and SDSS~J0806-2006) are already flagged as anomalous because of the discrepancy between their $K-$ and $L-$band flux ratios{\footnote{The case of HE~0435-1223 is a bit puzzling since  two discrepant measurements of the $K-$band flux ratios have been published by \cite{Fadely2011a} and \cite{Blackburne2011a}.}} \citep{Fadely2011a, Fadely2011b}. This anomaly might be produced by a massive substructure in the lens \citep{Fadely2011b}, and deserves further investigation. The origin of the disagreement between our estimate of $M$ and the $K-$band flux ratios for the last two objects (SDSS~J1138+0314, SDSS~J1335+0118) is less clear. The most simple explanation is intrinsic variability larger than our predictions and/or significant differential extinction{\footnote{For SDSS~J1138+0314, we cannot exclude a systematic error due to the slit centering because we could not cross check the spectroscopic flux ratio with broad band simultaneous data.}}. 

Figure~\ref{fig:Hband} also allows us to test whether ML and differential extinction are negligible in $H-$band, as often stated in literature. We clearly observe in the figure that the $H-$band value of $\Delta R_M$ is systematically closer to 0 (no ML) than the value at 7500\,\AA. This trend supports the idea that intrinsic variability and differential extinction do not add much noise in the estimation of $M$ (otherwise the effect would not be systematic). In addition, $\Delta R_M$ is in general less close to 0 in $H-$band than at redder wavelength. This confirms that ML is often detectable in $H-$band. We would expect the source at larger redshift (i.e. for which $H-$band corresponds to the bluest and therefore smallest emitting regions) to be more prone to ML in the $H-$band than the low redshift ones, but $H-$band ML is also observed in the low redshift sources. There is neither a clear correlation with the black-hole mass, which would be naively expected because the source size scales with the black hole mass. The absence of such correlations is likely related to ML variability, because the $H-$band data and optical spectra are not simultaneous, and to the variety of microlensing events which may not all lead to significant ML in the $H-$band.  

The above discussion confirms the efficiency of the MmD to derive intrinsic flux ratios, with a typical inaccuracy of 0.1 mag associated to intrinsic variability and, in a few cases, systematic errors associated to differential extinction and microlensing.

\begin{table}[h]
\begin{center}
\begin{tabular}{l|c|ccl}
\hline
Object & $F_1/F_2$ & $M$(this work) & $M_w$(literature) & Filter/line  \\
\hline
(a) 0047    & A/B & 0.220$\pm$0.05 & 0.240$\pm0.020$ & \CIII$^1$ \\ 	       	  
    &  & & 0.253$\pm$0.002 & $H-$band$^9$ \\ 	       	  
(b) 0142     & A/B & 0.131$\pm$0.003 & 0.121$\pm$0.004 & $H$-band$^{11}$\\
          &     &       & 0.128$\pm$0.004 & $K$-band$^{11}$\\
          &     &       & 0.132$\pm$0.060 & $L'$-band$^2$ \\  	  
(c) 0246  & A/B & 0.340$\pm$0.010 & 0.302$\pm$0.062 & $H-$band$^{12}$ \\
 & &  & 0.258$\pm$0.015 & $K-$band$^{2}$ \\
 & &  & 0.290$\pm$0.019 & $K'-$band$^{12, \dagger}$ \\
 & &  & 0.331$\pm$0.016 & $L'-$band$^{2}$ \\
(d) 0435     & B/D  &  1.470$\pm$0.050 & 1.270$\pm$0.016 &  $H$-band$^8$ \\
     & &  &    1.493$\pm$0.120 &  $K$-band$^2$ \\
     & &  &    1.270$\pm$0.040 &  $K$-band$^{16}$ \\
     & & &  1.220$\pm$0.170 &  $L'$-band$^2$ \\
(e) 0806 & A/B & 0.420$\pm$0.040  & 0.474$\pm$0.035 & $H-$band$^{13}$\\
            & & & 0.406$\pm$0.030 & $K-$band$^2$\\
	    & & & $<$0.164 & $L'-$band$^2$\\
(f) 0951    & A/B & 0.230 $\pm$ 0.020 & 0.285$\pm$0.005 & $H-$band$^{14}$\\
                 & & & 0.214$\pm$0.042 & 8.4Ghz$^{4, \ddagger}$ \\	  
(g) 0952   & A/B &  0.440$\pm$ 0.040 & 0.283$\pm$0.03 & $H-$band$^8$ \\  
(h) 1138  & C/B & 1.100$\pm$0.050  & 1.191$\pm$0.015 & $H-$band$^9$\\ 
       &  &    & 1.367 $\pm$ 0.135 & $Ks-$band$^{16}$\\
(i) 1226       & A-B  & 0.850$\pm$0.050 & 0.499$\pm$0.006 & $H-$band$^9$ \\ 
(j) 1335  & A/B  &0.230$\pm$0.020 & 0.299$\pm$0.008 & $H-$band$^8$\\
       &  &    & 0.409$\pm$0.027 &  $K-$band$^{5, \dagger}$ \\
(k) 1355  & A/B & 0.340$\pm$0.030 &0.258 $\pm$ 0.08& $H-$band$^8$\\  
& & & 0.299$\pm$0.019 & $K-$band$^{6, \dagger}$ \\	  
(l) 2033    & B/C  &0.900$\pm$0.020 &  0.904$\pm$0.025  & $H-$band$^{7}$ \\
& & &  0.912$\pm$0.050 &  $Ks-$band$^{16}$\\
(m) 2149     & A/B  & 0.245$\pm$0.002 &  0.245$\pm$0.005 & \CIII$^{15}$ \\
& & &  0.234$\pm$0.001 & $H-$band$^{10}$ \\
(n) 0924 & A/B & 0.38$\pm$0.01 & 0.44$\pm$ 0.04 & $H-$band$^{17}$\\
(o) 1131  & B/A &  2.15$\pm$0.05 & 1.294$\pm$0.156 & $H-$band$^{18}$ \\
  &  &   & 1.430$\pm$ 0.026 & $K-$band$^{18}$ \\
   & C/B &  1.30$\pm$0.07 & 0.449$\pm$0.079 & $H-$band$^{18}$\\
  &  &   & 0.570$\pm$0.026 & $K-$band$^{18}$\\
(p) 1413  & AB/D & 0.425$\pm$0.015 & 0.579$\pm$0.005 & $H-$band$^{20}$ \\
       &     &                  & 0.430$\pm$0.060 & 11.2\,$\mu $m$^{21}$ \\
(r) 2237 & A/D & 1.0$\pm$0.200 & 0.410$\pm$0.010 & $H-$band$^{19}$ \\
 &  &  & 0.87$\pm$0.05 & 11.67\,$\mu $m$^{19}$ \\
\hline 

\end{tabular}
\end{center}
\vspace{0.2cm}
{\tiny {Notes: $\dagger$: We assumed 0.05 mag on the photometry of the individual images, $\ddagger$: We arbitrarily assumed 20\% uncertainty on the flux ratio.}}
\tablebib{
(1) \cite{Wisotzki2004}; (2) \cite{Fadely2011a}; (3) \cite{Courbin2010}; (4) \cite{Schechter1998}; (5) \cite{Oguri2004a}; (6) \cite{Morgan2004}; (7) \cite{Vuissoz2008}; (8) CASTLES; (9) \cite{Chantry2010}; (10) \cite{Sluse2011b}; (11) \cite{Lehar2000}; (12) \cite{Inada2005}; (13) \cite{Sluse2008}; (14) \cite{Falco1999}; (15) \cite{Burud2002a}; (16) \cite{Blackburne2011a}; (17) \cite{Eigenbrod2006a}; (18) \cite{Sluse2006}; (19) \cite{Falco1996}; (20) \cite{Chantry2007}; (21) \cite{McLeod2009}
}

\caption{Comparison of $M$ from this paper (col.\,\#3) with values derived in the literature (col.\,\#4). The object name (col.\,\#1) has been shortened to the first 4 digits of the RA. The last column indicates the filter/line for which the literature measurement has been performed. See Sect.~\ref{subsec:intrinsic} for discussion.}
\label{tab:macro}
\end{table}

\section{Discussion}
\label{sec:discussion}

The objects of our main sample (Sect.~\ref{subsec:obs}) were targeted with the goal of measuring the redshift of the lensing galaxy and independently of any known detection of microlensing. Therefore, we can make some basic statistics regarding the chance of detection of microlensing in a lensed quasar based on these systems. Because observations were not performed exactly at the same time but over a two-years period, and because of possible secondary selection biases (e.g. selection of the brighter targets, large source redshift range, bright lensing galaxy, mix of doubles and quads), this estimate should not be very accurate. On the other hand, we can use the complete sample to discuss the variety of microlensing deformations of the broad lines. 
We first address the occurrence of microlensing of the continuum and its chromatic variations in Sect.~\ref{subsec:mucont}. Afterwards, we discuss the occurrence of microlensing of the broad emission lines and try to identify some trends as a function of the object physical and spectral properties (Sect.~\ref{subsec:muBLR}). We finally discuss the consequences of our observations for our understanding of the BLR in Sect.~\ref{subsec:BLRgeom}

\subsection{Microlensing of the continuum} 
\label{subsec:mucont}

Our observations confirm that microlensing of the continuum is observed at a level $> 0.05$\,mag in 11 out of 13 systems, i.e. $\sim$ 85\% of the systems. Because Q0142-100 and HE2149-2745 are the 2 systems with the largest black hole mass, we may be tempted to associate the absence of microlensing in these systems to a large size of the continuum emitting region. However, simple estimates of the latter based on the accretion disc theory lead to continuum sizes at least 10 times smaller than the microlens Einstein radius \citep{Mosquera2011}. On the other hand, the past detection of microlensing in HE2149-2745 \citep{Burud2002a} teaches us that microlensing of the continuum can be observed in these massive quasars. 

Despite the large occurrence of microlensing of the continuum, significant chromatic variations over the optical wavelength range, and more specifically those possibly caused by microlensing, are less common. Only four out of twelve systems show changes possibly associated to microlensing (HE~0047-1756, SDSS~J0246-0825, HE~0435-1223, Q1355-2257). This probably reflects that the source often lies far enough from a microlensing-caustic such that chromatic ML in the optical range is only weak. We should however notice that microlensing-induced chromatic changes are expected for all the microlensed sources if the wavelength coverage is large enough \citep[i.e. extending up to 2.5\,$\mu$m, cf. ][]{Bate2008, Yonehara2008, Floyd2009a, Blackburne2011a}.

\subsection{Microlensing of the BLR}
\label{subsec:muBLR}

\begin{figure*}[ht!]
\begin{tabular}{cc}
\centering
 \includegraphics[scale=0.45]{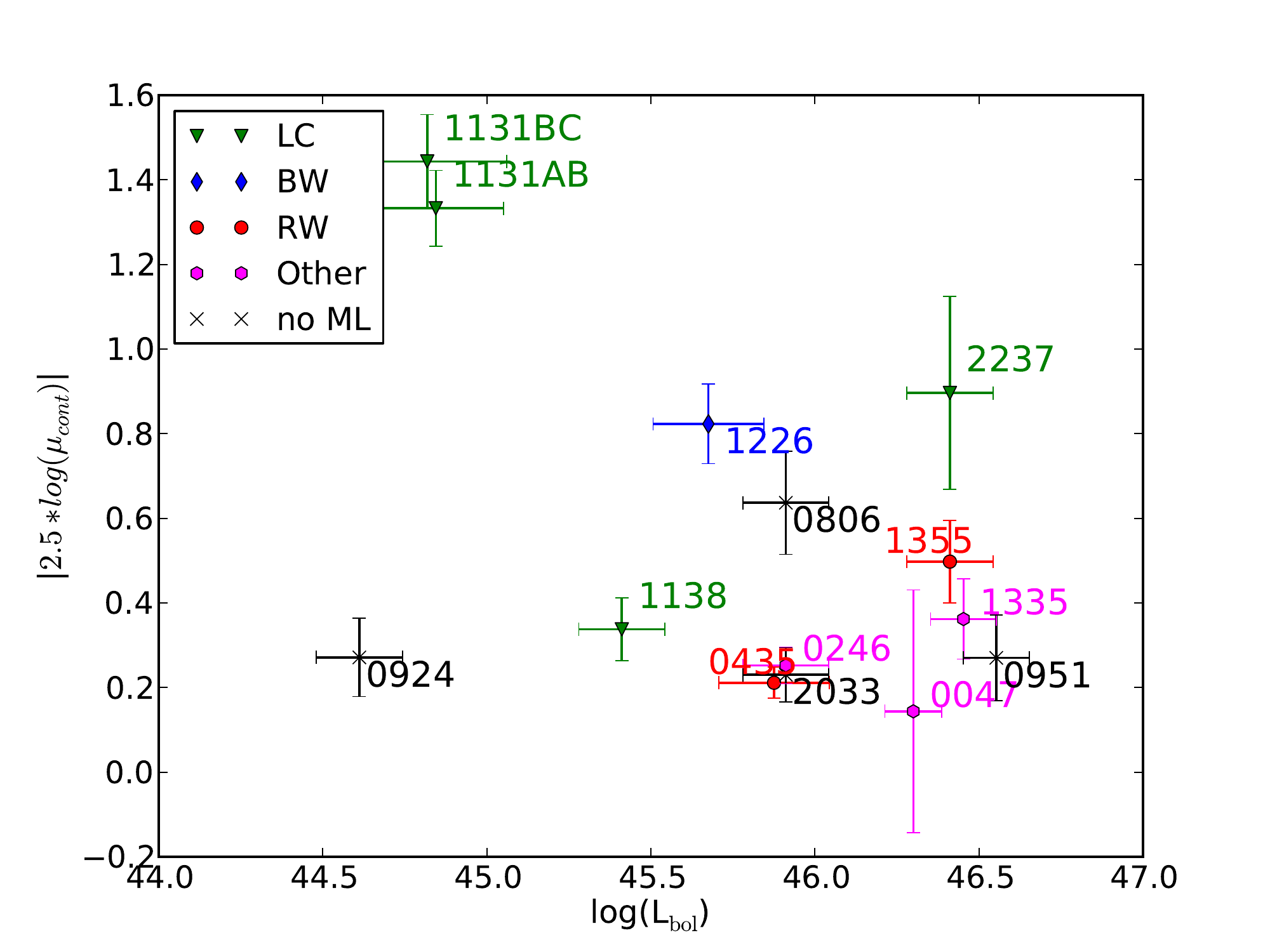}  & \includegraphics[scale=0.45]{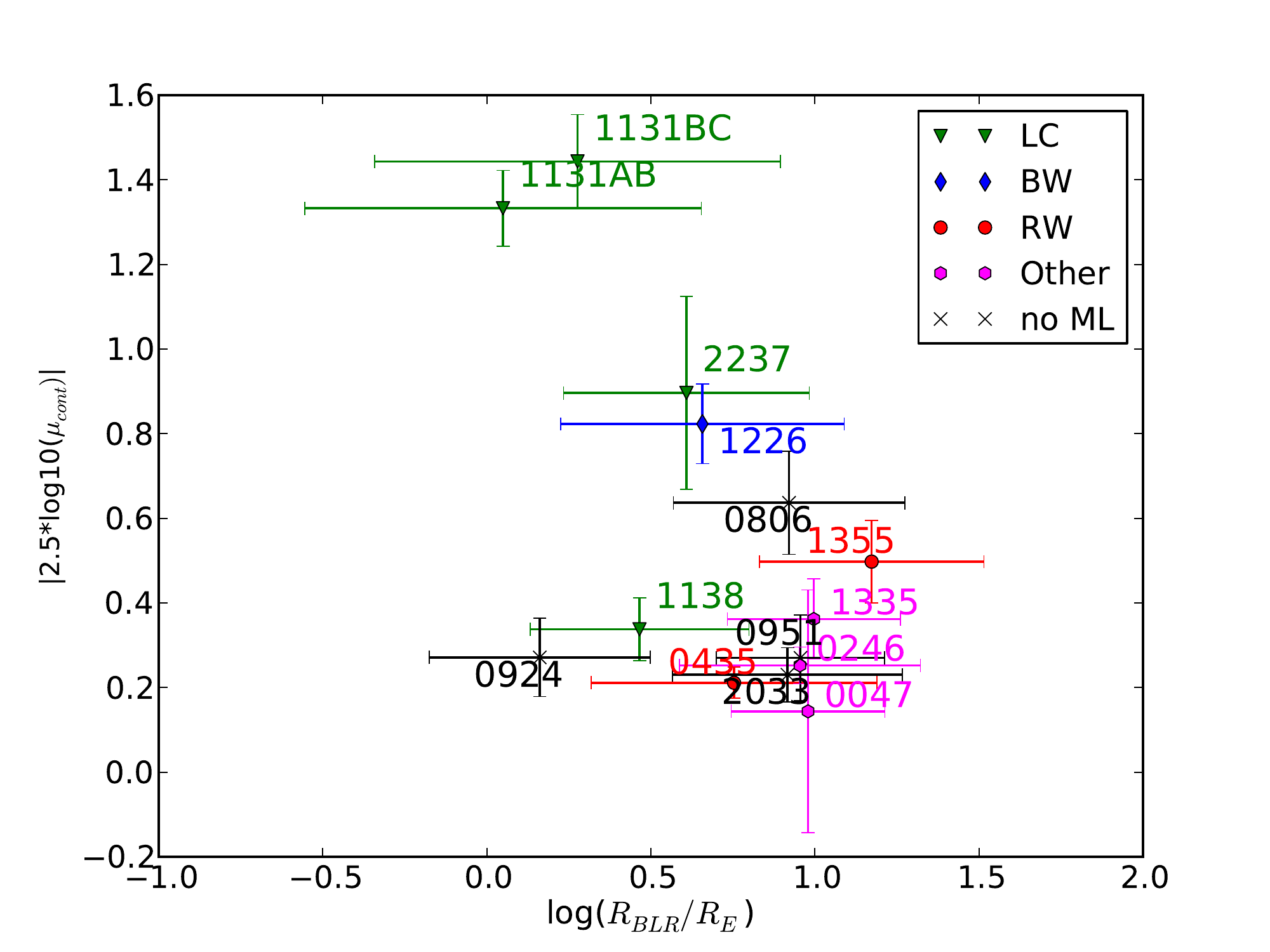} \\
\end{tabular}
\caption{Amplitude of microlensing in the continuum as a function of $\log(L_{\rm {bol}})$ (left) and of $\log(R_{\rm {BLR}} / R_E)$ (right). Different symbols are used to indicate the characteristics of the ML observed in the \MgII~or \CIII~ emission (Table~\ref{tab:analysis}). A green triangle corresponds to ML of the line core (LC), a blue losange to ML of the blue wing (BW), a red circle to ML of the red wing (RW), a magenta hexagon to a more complex signal (other), and a black cross is used for object where no ML of the line is detected. }
\label{fig:physprop}
\end{figure*}

The most common broad emission lines (BEL) detected in our sample are \CIII~ and \MgII. Therefore we focus our discussion on these two lines. Considering only the objects showing microlensing of the continuum in our main sample, we found microlensing of one of these two lines in $\sim$ 80\% of the systems (8 out of 10). We now extend the discussion to the whole sample and address the question of the variety of line deformations. Looking to the results of the MmD we found: 

\begin{itemize}
\item The \CIII~ and \MgII~emission lines are simultaneously observed in nine systems. For two of them, microlensing affects none of the lines and for six others, both lines are affected (but the signal in \CIII~ is uncertain in two cases). Microlensing of \CIII~ only, is seen in one system (WFI~2033-4723), but in this particular case the signal might also be due to microlensing of the \AlIII~ or \SiIII~ lines blended with the blue wing of \CIII.   
\item In most cases either the red wing or the blue wing is microlensed, with roughly equal occurrence. In addition, the microlensing signal always starts relatively close to the systemic redshift{\footnote{This is not the case for J1226-0006 but the signal is flagged as possibly contaminated by the flux from the lensing galaxy.}} (typically within 1000 km/s). 
\item Two of the six systems where microlensing of \MgII~and \CIII~is simultaneously observed display $F_{M\mu}$ with a different shape in \MgII~and \CIII. These systems (SDSS~J1335+0118 and Q1355-2257) are also those ones where a significant blueshift ($\sim$ 1000\,km/s) of \CIII~ is visible. 
\item For one system (SDSS~J1335+0118), the microlensing signal suggests two velocity components in the BLR, one which gives rise to a broad symmetric profile and the other one to a narrow component centered at v$\sim$ 1500\,km/s. These components are detected in both \CIII~and \MgII~ profiles despite their different shapes. Note also that the \MgII~line in this object is more asymmetric than in any other system of our sample.

\item The strongest microlensing effects in the BEL are in general associated to a symmetric signal affecting both positive (red) and negative (blue) velocities (Q2237+0305, J1131-1231, BRI0952-115 and SDSS~J1138+0314). This effect is expected for most of the BLR geometries studied in the literature when the source lies in a large demagnification valley of the micro-caustic network or when a large fraction of the source crosses a micro-caustic \citep[e.g.][]{Abajas2007}. In the first case the symmetry should last several crossing time-scale{\footnote{The crossing time scale is the time it takes the source to cross a distance equivalent to its own radius}} while in the second case, the effect should be more transient.  
\item Figure~\ref{fig:physprop} shows the properties of the microlensing signal in \MgII~or \CIII~as a function of the bolometric luminosity $L_{\rm {bol}}$ and of $R_{\rm {BLR}} / R_E$. Almost any kind of microlensing signal is observed at a given $L_{\rm {bol}}$ or $R_{\rm {BLR}} / R_E$. The objects with smaller $R_{\rm {BLR}} / R_E$ seem more prone to stronger deformation of the lines, as expected. Possibly because of the large error bars, we did not detect any correlation of the ML signal with the other physical properties calculated in Table~\ref{tab:MBH}.

\end{itemize}

\subsection{Consequences for the BLR}
\label{subsec:BLRgeom}

The current analysis has identified various signatures of microlensing of the BLR of quasars. A confrontation of the observed signal to different geometries of the BLR is beyond the scope of the present paper. However, the data presented here allow us to derive several qualitative results.
\begin{enumerate}[(a)]
\item  The typical amount of microlensing we observe suggests that the BLR cannot be much more extended than the microlensing Einstein radius $R_E$. Assuming that 10\% of the emission line flux is microlensed (see Fig.~\ref{fig:MmD}), the volume of the region magnified by a microlens of $R_E$ should roughly constitute 10\% of the total emitting volume. For a spherical BLR, this means $\pi\,R_E^2 \times 2\,R_{\rm {BLR}} \sim 0.1\times4/3\,\pi\times R_{\rm {BLR}}^3$, i.e. $R_{\rm {BLR}} \sim 4 R_E$. A similar value is obtained for a face-on disk. Although this is a very simplified treatment of the problem, this is in agreement with the independent estimates of $R_{\rm {BLR}}$ given in Table~\ref{tab:MBH}. 
\item The frequent observation of asymmetric blue/red deformations of the emission lines demonstrates that the BLR does not have, in general, a spherically symmetric geometry and velocity field. As shown by the simulation of \cite{Schneider1990a, Abajas2002, Lewis2004a}, asymmetric deformation of the broad lines can only be produced when the BLR has an anisotropic geometry or velocity field (i.e. axially symmetric, biconic, ...). Note that the symmetric microlensing of the broad line observed in some systems (e.g. Q2237+0305, J1131-1231, SDSS~1138+0314) may still occur with an anisotropic BLR. In this case the signal could become asymmetric and revealed by a dedicated monitoring campaign. 
\item The comparison of the microlensing signal in \MgII~and \CIII~ suggests that differences observed in emission lines of different ionisation degrees are more pronounced when one of the two lines is not centered at the systemic redshift. This may be of particular importance to disentangle blueshift caused by obscuration of the redward part of the line and blueshift associated to the dynamics of the BLR \citep[e.g.][]{Richards2002, Richards2011}. 

\item For one system (SDSS~J0246-0825), a symmetric microlensing of the emission line is observed with a dip in the line core. This is strikingly similar to the signal expected for a biconical disk \citep{Abajas2002, Lewis2004a, Abajas2007}. However, the solution may not be unique. Microlensing of keplerian disks (modified disks like those introduced in \citealt{Abajas2002} or relativistic discs as discussed in \citealt{Popovic2001}) sometimes lead to a dip in the microlensed signal at zero velocity. Only one other system (HE~0047-1756) shows ML similar to the one observed in SDSS~J0246-0825 but with a larger signal in one wing than in the other, a case which should in principle be more likely \citep{Abajas2002, Abajas2007}. This might be an observational bias (microlensing of the other wing is too small to be detected), or not. In this case, it seems, according to published simulations, that microlensing of a keplerian disc qualitatively better reproduces the observed signal.

\end{enumerate}

\section{Conclusions}
\label{sec:conclusions}

In this paper we have searched for the presence of microlensing, in the continuum and in the broad emission lines, among strongly lensed quasars based on optical spectroscopy of a sample of 13 systems (3 quadruply and 10 doubly-imaged systems). The spectra of these systems, originally targeted for detecting and measuring the redshift of their lensing galaxy \citep{Eigenbrod2006b, Eigenbrod2007}, have been re-reduced and deconvolved in order to accurately deblend the flux of the lensed images and of the lensing galaxy. For the three systems BRI~0952-0115, SDSS~J1138+0314, SDSS~J1226-0006, these are the first published  spectra of the individual lensed images. In order to get a more complete overview of the variety of microlensing signals which can be detected, we have complemented our main sample with previously analysed objects, i.e. SDSS~J0924+0219, HE~2149-2745, J1131-1231, Q2237+0305 and H1413+117, extending our sample to 17 objects. 

We derive robust estimates of the intrinsic flux ratios $M$ between the lensed images and of the amplitude $\mu$ of microlensing of the continuum. Based on optical spectra only, we have shown that we can retrieve $M$ to typically 0.1 mag accuracy. Higher accuracy may be expected by using spectra separated by the time delay, over a wavelength range extending from UV to NIR. Our ability to derive accurate $M$ and $\mu$ is particularly relevant for the study of the so called ``flux ratio anomalies'' produced by dark matter substructures or microlensing \citep{Mao1998, Dalal2002}. Indeed, current studies of this effect are challenged by the limited number of reliable flux ratios \citep{Keeton2005, Metcalf2011, Xu2011}. Therefore, the use of our technique should help in increasing the number of interesting systems. Our results may also be used to derive the fraction of smooth matter in lensing galaxies. A study of the probability of deriving the observed values of $\mu$ in sample of lensed quasars, varying the fraction of compact/smooth dark matter along the line-of-sight of the lensed images, makes possible to constrain the fraction of matter in compact form in these galaxies \citep{Schechter2004, Pooley2009, Mediavilla2009, Bate2011, Pooley2012}.   

We find in our sample, which contains sources with bolometric luminosities in the range [$10^{44.7}, 10^{47.4}$]\,erg/s and black hole masses in the range [$10^{7.6}, 10^{9.8}$]$\,M_{\sun}$, that 85\% of the sources show microlensing of the continuum. The microlensing of the continuum is not systematically associated to significant chromatic changes in the optical range. These chromatic changes are not necessarily absent but relatively weak. Because we observe microlensing of the continuum in our sample in the range 0.2-0.8 mag in $R-$band (observed frame), microlensing may in general not be neglected in the $H-$band (1.6 $\mu$m). This implies that studies of the accretion disc based on microlensing will strongly benefit of a large wavelength coverage, extending at least up to the $K-$band. Another consequence is that microlensing may often exceed differential extinction in the NIR, leading to possible biases in studies of extragalactic extinction curves in lensed quasars.

Using our MmD decomposition technique, we have been able to unveil microlensing-induced deformation of the emission lines independently of a detailed modeling of the quasar spectrum. Among the systems with a microlensed continuum, 80\% show deformations of at least one broad emission line. The two major characteristics of the signal are its relatively low amplitude (typically 10 \% of the line is affected) and its variety. We searched for correlation between the observed microlensing signal and the luminosity, black hole mass and Eddington ratio of the microlensed quasars, but we did not find any clear correlation. Contrary to most previous observations, we frequently detect microlensing of either the blue or the red wing of the broad emission lines instead of a signal affecting roughly symmetrically both components. This simple observation implies that the BLR does not have a spherically symmetric geometry and velocity field, at least in those objects. Microlensing of only one wing of the line has been observed recurrently in the past for SDSS~J1004+4112 \citep{Richards2004a, Lamer2006, Gomez2006}. \cite{Abajas2007} has shown that the signal observed in that system is compatible with microlensing of a biconical wind but that in general the two wings should be affected by microlensing, with different amplitudes. Such a signal is detected in only one system of our sample. This suggests that the simple biconical model is not adequate for the BLR of the lines we studied (\CIII~and \MgII). A modified keplerian disc \citep{Abajas2002} seems to be a promising alternative as a ``generic'' BLR model. 

\begin{acknowledgements}
We are grateful to the referee for his (her) constructive comments. We thank Markus Demleitner for setting-up the GAVO spectral repository where the spectra presented in this paper are available. DS acknowledges partial support from the Alexander von Humboldt fundation, from the German Virtual Observatory (GAVO) and from the Deutsche Forschungsgemeinschaft, reference SL172/1-1. This project is partially supported by the Swiss National Science Foundation (SNSF).
\end{acknowledgements}

\bibliographystyle{aa}
\bibliography{microsluse.bib}

\begin{thebibliography}{141}
\expandafter\ifx\csname natexlab\endcsname\relax\def\natexlab#1{#1}\fi

\bibitem[{{Abajas} {et~al.}(2002){Abajas}, {Mediavilla}, {Mu{\~ n}oz},
  {Popovi{\' c}}, \& {Oscoz}}]{Abajas2002}
{Abajas}, C., {Mediavilla}, E., {Mu{\~ n}oz}, J.~A., {Popovi{\' c}}, L.~{\v
  C}., \& {Oscoz}, A. 2002, \apj, 576, 640

\bibitem[{{Abajas} {et~al.}(2007){Abajas}, {Mediavilla}, {Mu{\~n}oz},
  {G{\'o}mez-{\'A}lvarez}, \& {Gil-Merino}}]{Abajas2007}
{Abajas}, C., {Mediavilla}, E., {Mu{\~n}oz}, J.~A., {G{\'o}mez-{\'A}lvarez},
  P., \& {Gil-Merino}, R. 2007, \apj, 658, 748

\bibitem[{{Agol} {et~al.}(2009){Agol}, {Gogarten}, {Gorjian}, \&
  {Kimball}}]{Agol2009}
{Agol}, E., {Gogarten}, S., {Gorjian}, V., \& {Kimball}, A. 2009, \apj, 697,
  1010

\bibitem[{{Agol} {et~al.}(2001){Agol}, {Wyithe}, {Jones}, \&
  {Fluke}}]{Agol2001}
{Agol}, E., {Wyithe}, S., {Jones}, B., \& {Fluke}, C. 2001, Publications of the
  Astronomical Society of Australia, 18, 166

\bibitem[{{Angonin} {et~al.}(1990){Angonin}, {Vanderriest}, {Remy}, \&
  {Surdej}}]{Angonin1990}
{Angonin}, M.-C., {Vanderriest}, C., {Remy}, M., \& {Surdej}, J. 1990, \aap,
  233, L5

\bibitem[{{Anguita} {et~al.}(2008{\natexlab{a}}){Anguita}, {Faure}, {Yonehara},
  {Wambsganss}, {Kneib}, {Covone}, \& {Alloin}}]{Anguita2008b}
{Anguita}, T., {Faure}, C., {Yonehara}, A., {et~al.} 2008{\natexlab{a}}, \aap,
  481, 615

\bibitem[{{Anguita} {et~al.}(2008{\natexlab{b}}){Anguita}, {Schmidt}, {Turner},
  {Wambsganss}, {Webster}, {Loomis}, {Long}, \& {McMillan}}]{Anguita2008a}
{Anguita}, T., {Schmidt}, R.~W., {Turner}, E.~L., {et~al.} 2008{\natexlab{b}},
  \aap, 480, 327

\bibitem[{{Assef} {et~al.}(2011){Assef}, {Denney}, {Kochanek}, {Peterson},
  {Kozlowski}, {Ageorges}, {Buschkamp}, {Falco}, {Feiz}, {Gemperlein},
  {Germeroth}, {Grier}, {Hofmann}, {Juette}, {Khan}, {Kilic}, {Knierim},
  {Laun}, {Lederer}, {Lehmitz}, {Lenzen}, {Mall}, {Mandel}, {Martini},
  {Mueller}, {Naranjo}, {Pasquali}, {Polsterer}, {Pogge}, {Quirrenbach},
  {Seifert}, {Shappee}, {Storz}, {Van Saders}, {Weiser}, \&
  {Zhang}}]{Assef2011}
{Assef}, R.~J., {Denney}, K.~D., {Kochanek}, C.~S., {et~al.} 2011, \apj, 742,
  93

\bibitem[{{Bachev} {et~al.}(2004){Bachev}, {Marziani}, {Sulentic}, {Zamanov},
  {Calvani}, \& {Dultzin-Hacyan}}]{Bachev2004}
{Bachev}, R., {Marziani}, P., {Sulentic}, J.~W., {et~al.} 2004, \apj, 617, 171

\bibitem[{{Bate} {et~al.}(2008){Bate}, {Floyd}, {Webster}, \&
  {Wyithe}}]{Bate2008}
{Bate}, N.~F., {Floyd}, D.~J.~E., {Webster}, R.~L., \& {Wyithe}, J.~S.~B. 2008,
  \mnras, 391, 1955

\bibitem[{{Bate} {et~al.}(2011){Bate}, {Floyd}, {Webster}, \&
  {Wyithe}}]{Bate2011}
{Bate}, N.~F., {Floyd}, D.~J.~E., {Webster}, R.~L., \& {Wyithe}, J.~S.~B. 2011,
  \apj, 731, 71

\bibitem[{{Bate} \& {Fluke}(2012)}]{Bate2012a}
{Bate}, N.~F. \& {Fluke}, C.~J. 2012, \apj, 744, 90

\bibitem[{{Bate} {et~al.}(2010){Bate}, {Fluke}, {Barsdell}, {Garsden}, \&
  {Lewis}}]{Bate2010a}
{Bate}, N.~F., {Fluke}, C.~J., {Barsdell}, B.~R., {Garsden}, H., \& {Lewis},
  G.~F. 2010, \na, 15, 726

\bibitem[{{Belle} \& {Lewis}(2000)}]{Belle2000}
{Belle}, K.~E. \& {Lewis}, G.~F. 2000, \pasp, 112, 320

\bibitem[{{Blackburne} {et~al.}(2011){Blackburne}, {Pooley}, {Rappaport}, \&
  {Schechter}}]{Blackburne2011a}
{Blackburne}, J.~A., {Pooley}, D., {Rappaport}, S., \& {Schechter}, P.~L. 2011,
  \apj, 729, 34

\bibitem[{{Burud} {et~al.}(2002){Burud}, {Courbin}, {Magain}, {Lidman},
  {Hutsem{\'e}kers}, {Kneib}, {Hjorth}, {Brewer}, {Pompei}, {Germany},
  {Pritchard}, {Jaunsen}, {Letawe}, \& {Meylan}}]{Burud2002a}
{Burud}, I., {Courbin}, F., {Magain}, P., {et~al.} 2002, \aap, 383, 71

\bibitem[{{Chang} \& {Refsdal}(1979)}]{Chang1979}
{Chang}, K. \& {Refsdal}, S. 1979, \nat, 282, 561

\bibitem[{{Chantry} \& {Magain}(2007)}]{Chantry2007}
{Chantry}, V. \& {Magain}, P. 2007, \aap, 470, 467

\bibitem[{{Chantry} {et~al.}(2010){Chantry}, {Sluse}, \&
  {Magain}}]{Chantry2010}
{Chantry}, V., {Sluse}, D., \& {Magain}, P. 2010, \aap, 522, A95+

\bibitem[{{Chartas} {et~al.}(2009){Chartas}, {Kochanek}, {Dai}, {Poindexter},
  \& {Garmire}}]{Chartas2009}
{Chartas}, G., {Kochanek}, C.~S., {Dai}, X., {Poindexter}, S., \& {Garmire}, G.
  2009, \apj, 693, 174

\bibitem[{{Claeskens} {et~al.}(2001){Claeskens}, {Khmil}, {Lee}, {Sluse}, \&
  {Surdej}}]{Claeskens2001}
{Claeskens}, J.-F., {Khmil}, S.~V., {Lee}, D.~W., {Sluse}, D., \& {Surdej}, J.
  2001, \aap, 367, 748

\bibitem[{{Collin} {et~al.}(2006){Collin}, {Kawaguchi}, {Peterson}, \&
  {Vestergaard}}]{Collin2006}
{Collin}, S., {Kawaguchi}, T., {Peterson}, B.~M., \& {Vestergaard}, M. 2006,
  \aap, 456, 75

\bibitem[{{Courbin} {et~al.}(2011){Courbin}, {Chantry}, {Revaz}, {Sluse},
  {Faure}, {Tewes}, {Eulaers}, {Koleva}, {Asfandiyarov}, {Dye}, {Magain}, {van
  Winckel}, {Coles}, {Saha}, {Ibrahimov}, \& {Meylan}}]{Courbin2010}
{Courbin}, F., {Chantry}, V., {Revaz}, Y., {et~al.} 2011, \aap, 536, A53

\bibitem[{{Courbin} {et~al.}(2000){Courbin}, {Magain}, {Kirkove}, \&
  {Sohy}}]{COU00}
{Courbin}, F., {Magain}, P., {Kirkove}, M., \& {Sohy}, S. 2000, \apj, 529, 1136

\bibitem[{{Dai} {et~al.}(2003){Dai}, {Chartas}, {Agol}, {Bautz}, \&
  {Garmire}}]{Dai2003a}
{Dai}, X., {Chartas}, G., {Agol}, E., {Bautz}, M.~W., \& {Garmire}, G.~P. 2003,
  \apj, 589, 100

\bibitem[{{Dai} {et~al.}(2010){Dai}, {Kochanek}, {Chartas}, {Kozlowski},
  {Morgan}, {Garmire}, \& {Agol}}]{Dai2010a}
{Dai}, X., {Kochanek}, C.~S., {Chartas}, G., {et~al.} 2010, \apj, 709, 278

\bibitem[{{Dalal} \& {Kochanek}(2002)}]{Dalal2002}
{Dalal}, N. \& {Kochanek}, C.~S. 2002, \apj, 572, 25

\bibitem[{{Decarli} {et~al.}(2008){Decarli}, {Labita}, {Treves}, \&
  {Falomo}}]{Decarli2008}
{Decarli}, R., {Labita}, M., {Treves}, A., \& {Falomo}, R. 2008, \mnras, 387,
  1237

\bibitem[{{Eigenbrod} {et~al.}(2006{\natexlab{a}}){Eigenbrod}, {Courbin},
  {Dye}, {Meylan}, {Sluse}, {Vuissoz}, \& {Magain}}]{Eigenbrod2006a}
{Eigenbrod}, A., {Courbin}, F., {Dye}, S., {et~al.} 2006{\natexlab{a}}, \aap,
  451, 747

\bibitem[{{Eigenbrod} {et~al.}(2007){Eigenbrod}, {Courbin}, \&
  {Meylan}}]{Eigenbrod2007}
{Eigenbrod}, A., {Courbin}, F., \& {Meylan}, G. 2007, \aap, 465, 51

\bibitem[{{Eigenbrod} {et~al.}(2006{\natexlab{b}}){Eigenbrod}, {Courbin},
  {Meylan}, {Vuissoz}, \& {Magain}}]{Eigenbrod2006b}
{Eigenbrod}, A., {Courbin}, F., {Meylan}, G., {Vuissoz}, C., \& {Magain}, P.
  2006{\natexlab{b}}, \aap, 451, 759

\bibitem[{{Eigenbrod} {et~al.}(2008){Eigenbrod}, {Courbin}, {Sluse}, {Meylan},
  \& {Agol}}]{Eigenbrod2008a}
{Eigenbrod}, A., {Courbin}, F., {Sluse}, D., {Meylan}, G., \& {Agol}, E. 2008,
  \aap, 480, 647

\bibitem[{{El{\'{\i}}asd{\'o}ttir} {et~al.}(2006){El{\'{\i}}asd{\'o}ttir},
  {Hjorth}, {Toft}, {Burud}, \& {Paraficz}}]{Elliasdotir2006}
{El{\'{\i}}asd{\'o}ttir}, {\'A}., {Hjorth}, J., {Toft}, S., {Burud}, I., \&
  {Paraficz}, D. 2006, \apjs, 166, 443

\bibitem[{{Fadely} \& {Keeton}(2011)}]{Fadely2011a}
{Fadely}, R. \& {Keeton}, C.~R. 2011, \aj, 141, 101

\bibitem[{{Fadely} \& {Keeton}(2012)}]{Fadely2011b}
{Fadely}, R. \& {Keeton}, C.~R. 2012, \mnras, 419, 936

\bibitem[{{Falco} {et~al.}(1999){Falco}, {Impey}, {Kochanek}, {Leh{\'a}r},
  {McLeod}, {Rix}, {Keeton}, {Mu{\~n}oz}, \& {Peng}}]{Falco1999}
{Falco}, E.~E., {Impey}, C.~D., {Kochanek}, C.~S., {et~al.} 1999, \apj, 523,
  617

\bibitem[{{Falco} {et~al.}(1996){Falco}, {Lehar}, {Perley}, {Wambsganss}, \&
  {Gorenstein}}]{Falco1996}
{Falco}, E.~E., {Lehar}, J., {Perley}, R.~A., {Wambsganss}, J., \&
  {Gorenstein}, M.~V. 1996, \aj, 112, 897

\bibitem[{{Faure} {et~al.}(2011){Faure}, {Sluse}, {Cantale}, {Tewes},
  {Courbin}, {Durrer}, \& {Meylan}}]{Faure2011}
{Faure}, C., {Sluse}, D., {Cantale}, N., {et~al.} 2011, \aap, 536, A29

\bibitem[{{Filippenko}(1989)}]{Fillipenko1989a}
{Filippenko}, A.~V. 1989, \apjl, 338, L49

\bibitem[{{Fine} {et~al.}(2008){Fine}, {Croom}, {Hopkins}, {Hernquist},
  {Bland-Hawthorn}, {Colless}, {Hall}, {Miller}, {Myers}, {Nichol}, {Pimbblet},
  {Ross}, {Schneider}, {Shanks}, \& {Sharp}}]{Fine2008}
{Fine}, S., {Croom}, S.~M., {Hopkins}, P.~F., {et~al.} 2008, \mnras, 390, 1413

\bibitem[{{Fine} {et~al.}(2011){Fine}, {Jarvis}, \& {Mauch}}]{Fine2011}
{Fine}, S., {Jarvis}, M., \& {Mauch}, T. 2011, \mnras, 412, 213

\bibitem[{{Floyd} {et~al.}(2009){Floyd}, {Bate}, \& {Webster}}]{Floyd2009a}
{Floyd}, D.~J.~E., {Bate}, N.~F., \& {Webster}, R.~L. 2009, \mnras, 398, 233

\bibitem[{{Garsden} {et~al.}(2011){Garsden}, {Bate}, \& {Lewis}}]{Garsden2011b}
{Garsden}, H., {Bate}, N.~F., \& {Lewis}, G.~F. 2011, \mnras, 418, 1012

\bibitem[{{Goicoechea} \& {Shalyapin}(2010)}]{Goicoechea2010}
{Goicoechea}, L.~J. \& {Shalyapin}, V.~N. 2010, \apj, 708, 995

\bibitem[{{G{\'o}mez-{\'A}lvarez} {et~al.}(2006){G{\'o}mez-{\'A}lvarez},
  {Mediavilla}, {Mu{\~n}oz}, {Arribas}, {S{\'a}nchez}, {Oscoz}, {Prada}, \&
  {Serra-Ricart}}]{Gomez2006}
{G{\'o}mez-{\'A}lvarez}, P., {Mediavilla}, E., {Mu{\~n}oz}, J.~A., {et~al.}
  2006, \apjl, 645, L5

\bibitem[{{Graham} {et~al.}(2011){Graham}, {Onken}, {Athanassoula}, \&
  {Combes}}]{Graham2011}
{Graham}, A.~W., {Onken}, C.~A., {Athanassoula}, E., \& {Combes}, F. 2011,
  \mnras, 412, 2211

\bibitem[{{Grieger} {et~al.}(1988){Grieger}, {Kayser}, \&
  {Refsdal}}]{Grieger1988}
{Grieger}, B., {Kayser}, R., \& {Refsdal}, S. 1988, \aap, 194, 54

\bibitem[{{Hutsem{\'e}kers}(1993)}]{Hutsemekers1993}
{Hutsem{\'e}kers}, D. 1993, \aap, 280, 435

\bibitem[{{Hutsem{\'e}kers} {et~al.}(2010){Hutsem{\'e}kers}, {Borguet},
  {Sluse}, {Riaud}, \& {Anguita}}]{Hutsemekers2010}
{Hutsem{\'e}kers}, D., {Borguet}, B., {Sluse}, D., {Riaud}, P., \& {Anguita},
  T. 2010, \aap, 519, A103+

\bibitem[{{Hutsem{\'e}kers} {et~al.}(1994){Hutsem{\'e}kers}, {Surdej}, \& {van
  Drom}}]{Hutsemekers1994}
{Hutsem{\'e}kers}, D., {Surdej}, J., \& {van Drom}, E. 1994, \apss, 216, 361

\bibitem[{{Inada} {et~al.}(2005){Inada}, {Burles}, {Gregg}, {Becker},
  {Schechter}, {Eisenstein}, {Oguri}, {Castander}, {Hall}, {Johnston},
  {Pindor}, {Richards}, {Schneider}, {White}, {Brinkmann}, {Szalay}, \&
  {York}}]{Inada2005}
{Inada}, N., {Burles}, S., {Gregg}, M.~D., {et~al.} 2005, \aj, 130, 1967

\bibitem[{{Irwin} {et~al.}(1989){Irwin}, {Webster}, {Hewett}, {Corrigan}, \&
  {Jedrzejewski}}]{Irwin1989}
{Irwin}, M.~J., {Webster}, R.~L., {Hewett}, P.~C., {Corrigan}, R.~T., \&
  {Jedrzejewski}, R.~I. 1989, \aj, 98, 1989

\bibitem[{{Jakobsson} {et~al.}(2005){Jakobsson}, {Hjorth}, {Burud}, {Letawe},
  {Lidman}, \& {Courbin}}]{Jakobsson2005}
{Jakobsson}, P., {Hjorth}, J., {Burud}, I., {et~al.} 2005, \aap, 431, 103

\bibitem[{{Jin} {et~al.}(2011){Jin}, {Ward}, {Done}, \& {Gelbord}}]{Jin2011}
{Jin}, C., {Ward}, M., {Done}, C., \& {Gelbord}, J.~M. 2011, arXiv1109.2069

\bibitem[{{Kaspi} {et~al.}(2000){Kaspi}, {Smith}, {Netzer}, {Maoz}, {Jannuzi},
  \& {Giveon}}]{Kaspi2000}
{Kaspi}, S., {Smith}, P.~S., {Netzer}, H., {et~al.} 2000, \apj, 533, 631

\bibitem[{{Kayser} {et~al.}(1986){Kayser}, {Refsdal}, \&
  {Stabell}}]{Kayser1986}
{Kayser}, R., {Refsdal}, S., \& {Stabell}, R. 1986, \aap, 166, 36

\bibitem[{{Keeton}(2001)}]{Keeton2001}
{Keeton}, C.~R. 2001, astro-ph/0102340

\bibitem[{{Keeton} {et~al.}(2006){Keeton}, {Burles}, {Schechter}, \&
  {Wambsganss}}]{Keeton2006a}
{Keeton}, C.~R., {Burles}, S., {Schechter}, P.~L., \& {Wambsganss}, J. 2006,
  \apjs, 639, 1

\bibitem[{{Keeton} {et~al.}(2003){Keeton}, {Gaudi}, \& {Petters}}]{Keeton2003}
{Keeton}, C.~R., {Gaudi}, B.~S., \& {Petters}, A.~O. 2003, \apj, 598, 138

\bibitem[{{Keeton} {et~al.}(2005){Keeton}, {Gaudi}, \& {Petters}}]{Keeton2005}
{Keeton}, C.~R., {Gaudi}, B.~S., \& {Petters}, A.~O. 2005, \apj, 635, 35

\bibitem[{{Kishimoto} {et~al.}(2009){Kishimoto}, {Hoenig}, {Antonucci},
  {Kotani}, {Barvainis}, {Tristram}, \& {Weigelt}}]{Kishimoto2009}
{Kishimoto}, M., {Hoenig}, S.~F., {Antonucci}, R., {et~al.} 2009, \aap, 507,
  L57

\bibitem[{{Kishimoto} {et~al.}(2011){Kishimoto}, {Hoenig}, {Antonucci},
  {Millour}, {Tristram}, \& {Weigelt}}]{Kishimoto2011}
{Kishimoto}, M., {Hoenig}, S.~F., {Antonucci}, R., {et~al.} 2011, \aap, 536,
  A78

\bibitem[{{Kneib} {et~al.}(1998){Kneib}, {Alloin}, \& {Pello}}]{Kneib1998}
{Kneib}, J.-P., {Alloin}, D., \& {Pello}, R. 1998, \aap, 339, L65

\bibitem[{{Kochanek}(2004)}]{Kochanek2004a}
{Kochanek}, C.~S. 2004, \apj, 605, 58

\bibitem[{{Koptelova} {et~al.}(2010){Koptelova}, {Oknyanskij}, {Artamonov}, \&
  {Burkhonov}}]{Koptelova2010}
{Koptelova}, E., {Oknyanskij}, V.~L., {Artamonov}, B.~P., \& {Burkhonov}, O.
  2010, \mnras, 401, 2805

\bibitem[{{La Mura} {et~al.}(2009){La Mura}, {Di Mille}, {Popovi{\'c}},
  {Ciroi}, {Rafanelli}, \& {Ili{\'c}}}]{Lamura2009}
{La Mura}, G., {Di Mille}, F., {Popovi{\'c}}, L.~{\v C}., {et~al.} 2009, \nar,
  53, 162

\bibitem[{{Lamer} {et~al.}(2006){Lamer}, {Schwope}, {Wisotzki}, \&
  {Christensen}}]{Lamer2006}
{Lamer}, G., {Schwope}, A., {Wisotzki}, L., \& {Christensen}, L. 2006, \aap,
  454, 493

\bibitem[{{Leh{\' a}r} {et~al.}(2000){Leh{\' a}r}, {Falco}, {Kochanek},
  {McLeod}, {Mu{\~ n}oz}, {Impey}, {Rix}, {Keeton}, \& {Peng}}]{Lehar2000}
{Leh{\' a}r}, J., {Falco}, E.~E., {Kochanek}, C.~S., {et~al.} 2000, \apj, 536,
  584

\bibitem[{{Lewis} \& {Belle}(1998)}]{Lewis1998b}
{Lewis}, G.~F. \& {Belle}, K.~E. 1998, \mnras, 297, 69

\bibitem[{{Lewis} \& {Ibata}(2004)}]{Lewis2004a}
{Lewis}, G.~F. \& {Ibata}, R.~A. 2004, \mnras, 348, 24

\bibitem[{{Lewis} {et~al.}(1998){Lewis}, {Irwin}, {Hewett}, \&
  {Foltz}}]{Lewis1998}
{Lewis}, G.~F., {Irwin}, M.~J., {Hewett}, P.~C., \& {Foltz}, C.~B. 1998,
  \mnras, 295, 573

\bibitem[{{MacLeod} {et~al.}(2009){MacLeod}, {Kochanek}, \&
  {Agol}}]{McLeod2009}
{MacLeod}, C.~L., {Kochanek}, C.~S., \& {Agol}, E. 2009, \apj, 699, 1578

\bibitem[{{Mao} \& {Schneider}(1998)}]{Mao1998}
{Mao}, S. \& {Schneider}, P. 1998, \mnras, 295, 587

\bibitem[{{Marziani} {et~al.}(2006){Marziani}, {Dultzin-Hacyan}, \&
  {Sulentic}}]{Marziani2006}
{Marziani}, P., {Dultzin-Hacyan}, D., \& {Sulentic}, J.~W. 2006, {Accretion
  onto Supermassive Black Holes in Quasars: Learning from Optical/UV
  Observations}, ed. {Kreitler, P.~V.} (Nova Science Publishers), 123--+

\bibitem[{{Marziani} \& {Sulentic}(2012)}]{Marziani2012a}
{Marziani}, P. \& {Sulentic}, J.~W. 2012, \nar, 56, 49

\bibitem[{{McLure} \& {Dunlop}(2004)}]{McLure2004}
{McLure}, R.~J. \& {Dunlop}, J.~S. 2004, \mnras, 352, 1390

\bibitem[{{Mediavilla} {et~al.}(2011){Mediavilla}, {Mediavilla}, {Mu{\~n}oz},
  {Ariza}, {Lopez}, {Gonzalez-Morcillo}, \&
  {Jimenez-Vicente}}]{Mediavilla2011b}
{Mediavilla}, E., {Mediavilla}, T., {Mu{\~n}oz}, J.~A., {et~al.} 2011, \apj,
  741, 42

\bibitem[{{Mediavilla} {et~al.}(2009){Mediavilla}, {Munoz}, {Falco}, {Motta},
  {Guerras}, {Canovas}, {Jean}, {Oscoz}, \& {Mosquera}}]{Mediavilla2009}
{Mediavilla}, E., {Munoz}, J.~A., {Falco}, E., {et~al.} 2009, \apj, 706, 1451

\bibitem[{{Metcalf} \& {Amara}(2012)}]{Metcalf2011}
{Metcalf}, R.~B. \& {Amara}, A. 2012, \mnras, 419, 3414

\bibitem[{{Metcalf} \& {Madau}(2001)}]{Metcalf2001}
{Metcalf}, R.~B. \& {Madau}, P. 2001, \apj, 563, 9

\bibitem[{{Minezaki} {et~al.}(2009){Minezaki}, {Chiba}, {Kashikawa}, {Inoue},
  \& {Kataza}}]{Minezaki2009}
{Minezaki}, T., {Chiba}, M., {Kashikawa}, N., {Inoue}, K.~T., \& {Kataza}, H.
  2009, \apj, 697, 610

\bibitem[{{Morgan} {et~al.}(2004){Morgan}, {Caldwell}, {Schechter}, {Dressler},
  {Egami}, \& {Rix}}]{Morgan2004}
{Morgan}, N.~D., {Caldwell}, J.~A.~R., {Schechter}, P.~L., {et~al.} 2004, \aj,
  127, 2617

\bibitem[{{Morgan} {et~al.}(2003){Morgan}, {Gregg}, {Wisotzki}, {Becker},
  {Maza}, {Schechter}, \& {White}}]{Morgan2003}
{Morgan}, N.~D., {Gregg}, M.~D., {Wisotzki}, L., {et~al.} 2003, \aj, 126, 696

\bibitem[{{Morgan} {et~al.}(2006){Morgan}, {Kochanek}, {Falco}, \&
  {Dai}}]{Morgan2006}
{Morgan}, N.~D., {Kochanek}, C.~S., {Falco}, E.~E., \& {Dai}, X. 2006,
  astro-ph/0605321

\bibitem[{{Mosquera} \& {Kochanek}(2011)}]{Mosquera2011}
{Mosquera}, A.~M. \& {Kochanek}, C.~S. 2011, \apj, 738, 96

\bibitem[{{Mu{\~n}oz} {et~al.}(2011){Mu{\~n}oz}, {Mediavilla}, {Kochanek},
  {Falco}, \& {Mar{\'{\i}}a Mosquera}}]{Munoz2011}
{Mu{\~n}oz}, J.~A., {Mediavilla}, E., {Kochanek}, C.~S., {Falco}, E., \&
  {Mar{\'{\i}}a Mosquera}, A. 2011, \apj, 742, 67

\bibitem[{{Nemiroff}(1988)}]{Nemiroff1988}
{Nemiroff}, R.~J. 1988, \apj, 335, 593

\bibitem[{{O'Dowd} {et~al.}(2011){O'Dowd}, {Bate}, {Webster}, {Wayth}, \&
  {Labrie}}]{Odowd2011}
{O'Dowd}, M., {Bate}, N.~F., {Webster}, R.~L., {Wayth}, R., \& {Labrie}, K.
  2011, \mnras, 415, 1985

\bibitem[{{Oguri} {et~al.}(2004){Oguri}, {Inada}, {Castander}, {Gregg},
  {Becker}, {Ichikawa}, {Pindor}, {Brinkmann}, {Eisenstein}, {Frieman}, {Hall},
  {Johnston}, {Richards}, {Schechter}, {Schneider}, \& {Szalay}}]{Oguri2004a}
{Oguri}, M., {Inada}, N., {Castander}, F.~J., {et~al.} 2004, \pasj, 56, 399

\bibitem[{{{\O}stman} {et~al.}(2008){{\O}stman}, {Goobar}, \&
  {M{\"o}rtsell}}]{Orstman2008}
{{\O}stman}, L., {Goobar}, A., \& {M{\"o}rtsell}, E. 2008, \aap, 485, 403

\bibitem[{{Paczynski}(1986)}]{Paczynski1986}
{Paczynski}, B. 1986, \apjs, 301, 503

\bibitem[{{Patat} {et~al.}(2011){Patat}, {Moehler}, {O'Brien}, {Pompei},
  {Bensby}, {Carraro}, {de Ugarte Postigo}, {Fox}, {Gavignaud}, {James},
  {Korhonen}, {Ledoux}, {Randall}, {Sana}, {Smoker}, {Stefl}, \&
  {Szeifert}}]{Patat2011}
{Patat}, F., {Moehler}, S., {O'Brien}, K., {et~al.} 2011, \aap, 527, A91+

\bibitem[{{Peng} {et~al.}(2006){Peng}, {Impey}, {Rix}, {Kochanek}, {Keeton},
  {Falco}, {Leh{\'a}r}, \& {McLeod}}]{Peng2006b}
{Peng}, C.~Y., {Impey}, C.~D., {Rix}, H., {et~al.} 2006, \apj, 649, 616

\bibitem[{{Poindexter} \& {Kochanek}(2010{\natexlab{a}})}]{Poindexter2010a}
{Poindexter}, S. \& {Kochanek}, C.~S. 2010{\natexlab{a}}, \apj, 712, 668

\bibitem[{{Poindexter} \& {Kochanek}(2010{\natexlab{b}})}]{Poindexter2010b}
{Poindexter}, S. \& {Kochanek}, C.~S. 2010{\natexlab{b}}, \apj, 712, 658

\bibitem[{{Pooley} {et~al.}(2007){Pooley}, {Blackburne}, {Rappaport}, \&
  {Schechter}}]{Pooley2007a}
{Pooley}, D., {Blackburne}, J.~A., {Rappaport}, S., \& {Schechter}, P.~L. 2007,
  \apj, 661, 19

\bibitem[{{Pooley} {et~al.}(2009){Pooley}, {Rappaport}, {Blackburne},
  {Schechter}, {Schwab}, \& {Wambsganss}}]{Pooley2009}
{Pooley}, D., {Rappaport}, S., {Blackburne}, J., {et~al.} 2009, \apj, 697, 1892

\bibitem[{{Pooley} {et~al.}(2012){Pooley}, {Rappaport}, {Blackburne},
  {Schechter}, \& {Wambsganss}}]{Pooley2012}
{Pooley}, D., {Rappaport}, S., {Blackburne}, J.~A., {Schechter}, P.~L., \&
  {Wambsganss}, J. 2012, \apj, 744, 111

\bibitem[{{Popovi{\'c}} \& {Chartas}(2005)}]{Popovic2005}
{Popovi{\'c}}, L.~{\v C}. \& {Chartas}, G. 2005, \mnras, 357, 135

\bibitem[{{Popovi{\'c}} {et~al.}(2001){Popovi{\'c}}, {Mediavilla}, \&
  {Mu{\~n}oz}}]{Popovic2001}
{Popovi{\'c}}, L.~{\v C}., {Mediavilla}, E.~G., \& {Mu{\~n}oz}, J.~A. 2001,
  \aap, 378, 295

\bibitem[{{Richards} {et~al.}(2004){Richards}, {Keeton}, {Pindor}, {Hennawi},
  {Hall}, {Turner}, {Inada}, {Oguri}, {Ichikawa}, {Becker}, {Gregg}, {White},
  {Wyithe}, {Schneider}, {Johnston}, {Frieman}, \& {Brinkmann}}]{Richards2004a}
{Richards}, G.~T., {Keeton}, C.~R., {Pindor}, B., {et~al.} 2004, \apj, 610, 679

\bibitem[{{Richards} {et~al.}(2011){Richards}, {Kruczek}, {Gallagher}, {Hall},
  {Hewett}, {Leighly}, {Deo}, {Kratzer}, \& {Shen}}]{Richards2011}
{Richards}, G.~T., {Kruczek}, N.~E., {Gallagher}, S.~C., {et~al.} 2011, \aj,
  141, 167

\bibitem[{{Richards} {et~al.}(2002){Richards}, {Vanden Berk}, {Reichard},
  {Hall}, {Schneider}, {SubbaRao}, {Thakar}, \& {York}}]{Richards2002}
{Richards}, G.~T., {Vanden Berk}, D.~E., {Reichard}, T.~A., {et~al.} 2002, \aj,
  124, 1

\bibitem[{{Robinson}(1995)}]{Robinson1995}
{Robinson}, A. 1995, \mnras, 272, 647

\bibitem[{{Schechter} {et~al.}(1998){Schechter}, {Gregg}, {Becker}, {Helfand},
  \& {White}}]{Schechter1998}
{Schechter}, P.~L., {Gregg}, M.~D., {Becker}, R.~H., {Helfand}, D.~J., \&
  {White}, R.~L. 1998, \aj, 115, 1371

\bibitem[{{Schechter} \& {Wambsganss}(2004)}]{Schechter2004}
{Schechter}, P.~L. \& {Wambsganss}, J. 2004, 220, 103

\bibitem[{{Schmidt} \& {Wambsganss}(2010)}]{Schmidt2010}
{Schmidt}, R.~W. \& {Wambsganss}, J. 2010, General Relativity and Gravitation,
  42, 2127

\bibitem[{{Schneider} \& {Wambsganss}(1990)}]{Schneider1990a}
{Schneider}, P. \& {Wambsganss}, J. 1990, \aap, 237, 42

\bibitem[{{Shen} {et~al.}(2008){Shen}, {Greene}, {Strauss}, {Richards}, \&
  {Schneider}}]{Shen2008}
{Shen}, Y., {Greene}, J.~E., {Strauss}, M.~A., {Richards}, G.~T., \&
  {Schneider}, D.~P. 2008, \apj, 680, 169

\bibitem[{{Sluse} {et~al.}(2012){Sluse}, {Chantry}, {Magain}, {Courbin}, \&
  {Meylan}}]{Sluse2011b}
{Sluse}, D., {Chantry}, V., {Magain}, P., {Courbin}, F., \& {Meylan}, G. 2012,
  \aap, 538, A99

\bibitem[{{Sluse} {et~al.}(2006){Sluse}, {Claeskens}, {Altieri}, {Cabanac},
  {Garcet}, {Hutsem{\'e}kers}, {Jean}, {Smette}, \& {Surdej}}]{Sluse2006}
{Sluse}, D., {Claeskens}, J.-F., {Altieri}, B., {et~al.} 2006, \aap, 449, 539

\bibitem[{{Sluse} {et~al.}(2007){Sluse}, {Claeskens}, {Hutsem{\'e}kers}, \&
  {Surdej}}]{Sluse2007}
{Sluse}, D., {Claeskens}, J.-F., {Hutsem{\'e}kers}, D., \& {Surdej}, J. 2007,
  \aap, 468, 885

\bibitem[{{Sluse} {et~al.}(2008{\natexlab{a}}){Sluse}, {Claeskens},
  {Hutsem{\'e}kers}, \& {Surdej}}]{Sluse2008b}
{Sluse}, D., {Claeskens}, J.-F., {Hutsem{\'e}kers}, D., \& {Surdej}, J.
  2008{\natexlab{a}}, in Revista Mexicana de Astronomia y Astrofisica, vol. 27,
  Vol.~32, Revista Mexicana de Astronomia y Astrofisica Conference Series,
  83--85

\bibitem[{{Sluse} {et~al.}(2008{\natexlab{b}}){Sluse}, {Courbin}, {Eigenbrod},
  \& {Meylan}}]{Sluse2008}
{Sluse}, D., {Courbin}, F., {Eigenbrod}, A., \& {Meylan}, G.
  2008{\natexlab{b}}, \aap, 492, L39

\bibitem[{{Sluse} {et~al.}(2011){Sluse}, {Schmidt}, {Courbin},
  {Hutsem{\'e}kers}, {Meylan}, {Eigenbrod}, {Anguita}, {Agol}, \&
  {Wambsganss}}]{Sluse2011a}
{Sluse}, D., {Schmidt}, R., {Courbin}, F., {et~al.} 2011, \aap, 528, A100+

\bibitem[{{Small} {et~al.}(1997){Small}, {Sargent}, \& {Steidel}}]{Small1997}
{Small}, T.~A., {Sargent}, W.~L.~W., \& {Steidel}, C.~C. 1997, \aj, 114, 2254

\bibitem[{{Steidel} \& {Sargent}(1991)}]{Steidel1991}
{Steidel}, C.~C. \& {Sargent}, W.~L.~W. 1991, \aj, 102, 1610

\bibitem[{{Sugai} {et~al.}(2007){Sugai}, {Kawai}, {Shimono}, {Hattori},
  {Kosugi}, {Kashikawa}, {Inoue}, \& {Chiba}}]{Sugai2007}
{Sugai}, H., {Kawai}, A., {Shimono}, A., {et~al.} 2007, \apj, 660, 1016

\bibitem[{{Tsuzuki} {et~al.}(2006){Tsuzuki}, {Kawara}, {Yoshii}, {Oyabu},
  {Tanab{\'e}}, \& {Matsuoka}}]{Tsuzuki2006}
{Tsuzuki}, Y., {Kawara}, K., {Yoshii}, Y., {et~al.} 2006, \apj, 650, 57

\bibitem[{{Vestergaard} \& {Peterson}(2006)}]{Vestergaard2006}
{Vestergaard}, M. \& {Peterson}, B.~M. 2006, \apj, 641, 689

\bibitem[{{Vestergaard} \& {Wilkes}(2001)}]{Vestergaard2001}
{Vestergaard}, M. \& {Wilkes}, B.~J. 2001, \apjs, 134, 1

\bibitem[{{Vuissoz} {et~al.}(2008){Vuissoz}, {Courbin}, {Sluse}, {Meylan},
  {Chantry}, {Eulaers}, {Morgan}, {Eyler}, {Kochanek}, {Coles}, {Saha},
  {Magain}, \& {Falco}}]{Vuissoz2008}
{Vuissoz}, C., {Courbin}, F., {Sluse}, D., {et~al.} 2008, \aap, 488, 481

\bibitem[{{Wambsganss}(1998)}]{Wambsganss1998}
{Wambsganss}, J. 1998, Living Reviews in Relativity, 1, 12

\bibitem[{{Wambsganss}(2006)}]{Wambsganss2006}
{Wambsganss}, J. 2006, in Saas-Fee Advanced Course 33: Gravitational Lensing:
  Strong, Weak and Micro, ed. {G.~Meylan, P.~Jetzer, P.~North, P.~Schneider,
  C.~S.~Kochanek, \& J.~Wambsganss}, 453--540

\bibitem[{{Wambsganss} \& {Paczy{\' n}ski}(1991)}]{Wambsganss1991}
{Wambsganss}, J. \& {Paczy{\' n}ski}, B. 1991, \aj, 102, 864

\bibitem[{{Wandel} {et~al.}(1999){Wandel}, {Peterson}, \&
  {Malkan}}]{Wandel1999}
{Wandel}, A., {Peterson}, B.~M., \& {Malkan}, M.~A. 1999, \apj, 526, 579

\bibitem[{{Wang} {et~al.}(2011){Wang}, {Wang}, {Zhou}, {Liu}, {Wang}, {Yuan},
  \& {Dong}}]{Wang2011}
{Wang}, H., {Wang}, T., {Zhou}, H., {et~al.} 2011, \apj, 738, 85

\bibitem[{{Wang} {et~al.}(2009){Wang}, {Dong}, {Wang}, {Ho}, {Yuan}, {Wang},
  {Zhang}, {Zhang}, \& {Zhou}}]{Wang2009}
{Wang}, J., {Dong}, X., {Wang}, T., {et~al.} 2009, \apj, 707, 1334

\bibitem[{{Wayth} {et~al.}(2005){Wayth}, {O'Dowd}, \& {Webster}}]{Wayth2005}
{Wayth}, R.~B., {O'Dowd}, M., \& {Webster}, R.~L. 2005, \mnras, 359, 561

\bibitem[{{Wilhite} {et~al.}(2008){Wilhite}, {Brunner}, {Grier}, {Schneider},
  \& {vanden Berk}}]{Wilhite2008}
{Wilhite}, B.~C., {Brunner}, R.~J., {Grier}, C.~J., {Schneider}, D.~P., \&
  {vanden Berk}, D.~E. 2008, \mnras, 383, 1232

\bibitem[{{Wisotzki} {et~al.}(1993){Wisotzki}, {Koehler}, {Kayser}, \&
  {Reimers}}]{Wisotzki1993}
{Wisotzki}, L., {Koehler}, T., {Kayser}, R., \& {Reimers}, D. 1993, \aap, 278,
  L15

\bibitem[{{Wisotzki} {et~al.}(2004){Wisotzki}, {Schechter}, {Chen},
  {Richstone}, {Jahnke}, {S{\'a}nchez}, \& {Reimers}}]{Wisotzki2004}
{Wisotzki}, L., {Schechter}, P.~L., {Chen}, H.-W., {et~al.} 2004, \aap, 419,
  L31

\bibitem[{{Wo{\'z}niak} {et~al.}(2000){Wo{\'z}niak}, {Alard}, {Udalski},
  {Szyma{\'n}ski}, {Kubiak}, {Pietrzy{\'n}ski}, \& {Zebru{\'n}}}]{Wozniak2000}
{Wo{\'z}niak}, P.~R., {Alard}, C., {Udalski}, A., {et~al.} 2000, \apj, 529, 88

\bibitem[{{Wucknitz} {et~al.}(2003){Wucknitz}, {Wisotzki}, {Lopez}, \&
  {Gregg}}]{Wucknitz2003}
{Wucknitz}, O., {Wisotzki}, L., {Lopez}, S., \& {Gregg}, M.~D. 2003, \aap, 405,
  445

\bibitem[{{Wyithe} {et~al.}(2000{\natexlab{a}}){Wyithe}, {Webster}, \&
  {Turner}}]{Wyithe2000b}
{Wyithe}, J.~S.~B., {Webster}, R.~L., \& {Turner}, E.~L. 2000{\natexlab{a}},
  \mnras, 318, 762

\bibitem[{{Wyithe} {et~al.}(2000{\natexlab{b}}){Wyithe}, {Webster}, {Turner},
  \& {Mortlock}}]{Wyithe2000a}
{Wyithe}, J.~S.~B., {Webster}, R.~L., {Turner}, E.~L., \& {Mortlock}, D.~J.
  2000{\natexlab{b}}, \mnras, 315, 62

\bibitem[{{Xu} {et~al.}(2012){Xu}, {Mao}, {Cooper}, {Gao}, {Frenk}, {Angulo},
  \& {Helly}}]{Xu2011}
{Xu}, D.~D., {Mao}, S., {Cooper}, A., {et~al.} 2012, \mnras, 421, 2553

\bibitem[{{Yonehara}(2001)}]{Yonehara2000}
{Yonehara}, A. 2001, \apjl, 548, L127

\bibitem[{{Yonehara} {et~al.}(2008){Yonehara}, {Hirashita}, \&
  {Richter}}]{Yonehara2008}
{Yonehara}, A., {Hirashita}, H., \& {Richter}, P. 2008, \aap, 478, 95

\bibitem[{{Zackrisson} \& {Riehm}(2010)}]{Zackrisson2010}
{Zackrisson}, E. \& {Riehm}, T. 2010, Advances in Astronomy, 2010

\bibitem[{{Zimmer} {et~al.}(2011){Zimmer}, {Schmidt}, \&
  {Wambsganss}}]{Zimmer2011}
{Zimmer}, F., {Schmidt}, R.~W., \& {Wambsganss}, J. 2011, \mnras, 413, 1099

\end{thebibliography}

\newpage

\appendix

\section{Effect of intrinsic variability on the MmD}
\label{appendixC}
\begin{figure*}[htb!]
\begin{center}
\includegraphics[scale=0.75]{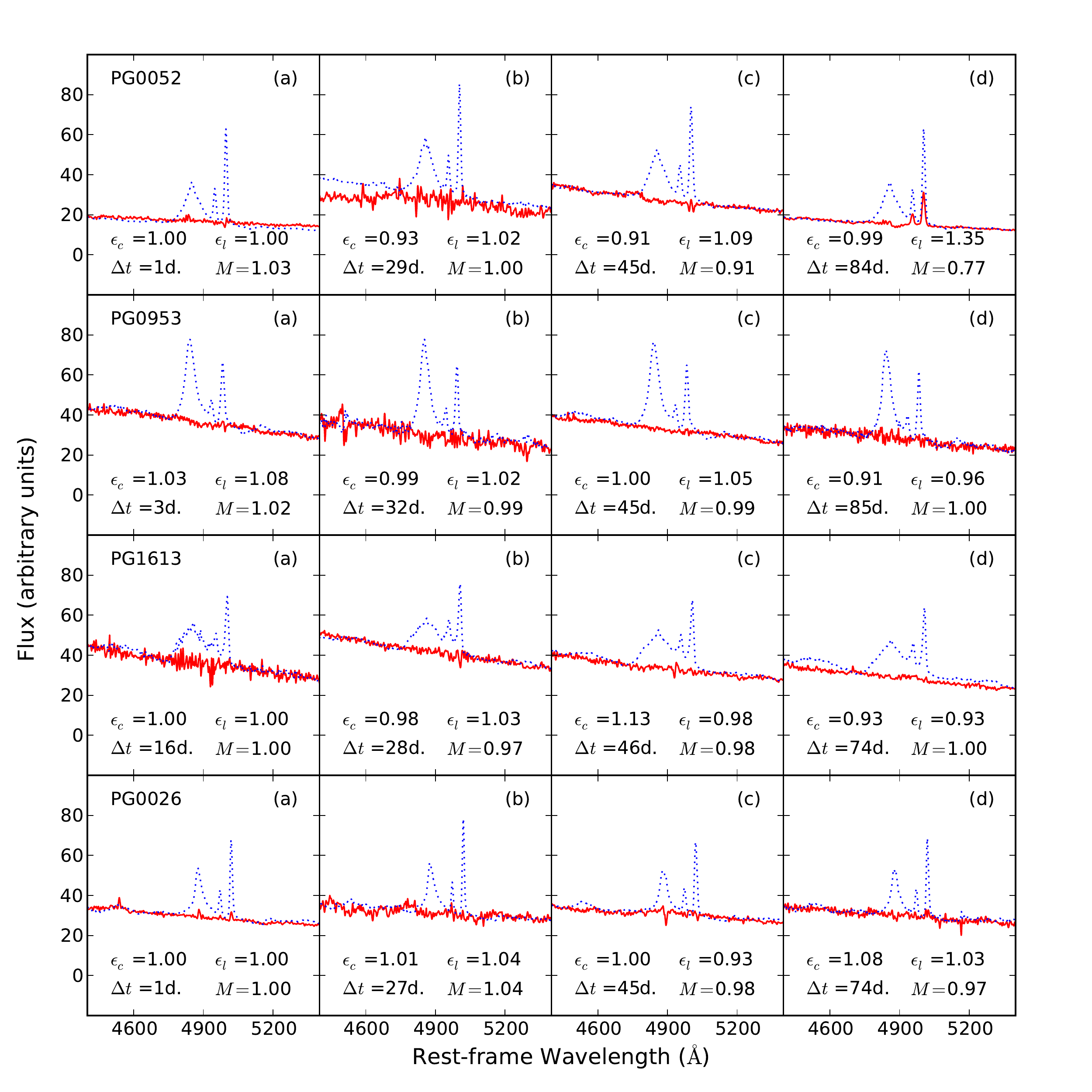}
\end{center}
\caption{Application of the MmD method for \Hbeta~\& \OIII~ for mock lensed quasars simulated based on spectro-photometric monitoring data of Palomar-Green quasars. For each pair of spectra separated by a delay $\Delta t$, the continuum of the one of the spectra has been artificially microlensed by $\mu=$1.5 and $M=$1 has been assumed. For legibility, only the fraction $F_{M\mu}$ (red solid line) and the reference spectrum (blue dotted line) are shown. Each row corresponds to a different object. The time-delay between the pairs of spectra is increasing from pannel (a) to (d). For each pannel, we provide $M$ retrieved with the decomposition, the variability in the continuum $\epsilon_c$ and in the line $\epsilon_l$. }
\label{fig:MmDrevmap}
\end{figure*}

The MmD technique described in Sect.~\ref{subsec:FMFMmu} should ideally be applied to pairs of spectra separated by the time-delay in order correct for any effect introduced by intrinsic variability. Hereafter, we investigate how intrinsic variability may affect the MmD technique applied to spectra obtained at a single epoch. For this purpose, we have decided to create mock lensed systems based on existing spectra of quasars observed at several epochs. The principle of our simulation is to pick pairs of spectra of an object at two different epochs $t_1, t_2$. A pair of spectra simulates the single-epoch spectra of two images of a macro-lensed quasar with a time delay $\Delta t = t_1-t_2$ and $M=$1. Then, we amplify the continuum of one image to simulate microlensing and apply the MmD. Since only the continuum is microlensed, we do not expect emission lines or part of them in $F_{M\mu}$, except possible contamination due to intrinsic variability. 

Specifically, we proceeded as follows. First, we used publicly available reverberation mapping data{\footnote{\url{http://wise-obs.tau.ac.il/~shai/PG/}}} of Palomar-Green quasars. In this database, we choose pairs of spectra of the same object separated by a delay $\Delta t$ in the ranges (a) 1-20 days, (2) 20-40 days, (3) 40-60 days, (4) 60-100 days. Second, we artificially microlensed the continuum of the first spectrum by a factor $\mu=$1.5. Third, we applied the MmD, estimating automatically $A$ and choosing $M$ to minimize the flux in $F_{M\mu}$ at the position of the \Hbeta~line{\footnote{For those spectra, we are able to derive $M$ directly from \OIII~which does not vary, but we did not choose this option to follow the same procedure as in Sect.~\ref{subsec:FMFMmu}}}. Note that we had to restrict ourselves to pairs of spectra obtained with the same instrumental setup to avoid spurious line deformation introduced by variable spectral resolution. We show in Fig.~\ref{fig:MmDrevmap} the result of this procedure at the position of the \Hbeta~line for PG0052 ($R_{\rm {BLR}}\sim$ 134\,light days), PG0953 ($R_{\rm {BLR}}\sim$ 151\,light days), PG1613 ($R_{\rm {BLR}}\sim$ 39\,light days), PG0026 ($R_{\rm {BLR}}\sim$ 113\,light days). We also report the measured fractional variation of the continuum ($\epsilon_c$) and of the line ($\epsilon_l$) during $\Delta t$. This figure illustrates that in general the deformations of the emission lines caused by intrinsic variability are too weak to mimic microlensing and introduce a significant signal above the continuum in $F_{M\mu}$ at the location of \Hbeta. When the delay becomes large (typically $>$ 40 days), it happens that a weak signal is detected in the emission lines  (e.g. pannel (c) and (d) for PG0052, pannel (c) for PG0026). This happens when $\epsilon_l$ is large and when it differs significantly from $\epsilon_c$. From this figure, it seems that differences of $\epsilon_l$ and $\epsilon_c$ by more than 10\% are needed to introduce noticeable line deformations in the decomposition. In order to derive how frequent this situation appears, we have calculated $\epsilon_l/\epsilon_c$  as a function of $\Delta t$ for the objects of the sample of \cite{Kaspi2000}. We did not find a clear change of $\epsilon_l$/$\epsilon_c$ nor of the standard deviation $\sigma_\epsilon$ with the size of the BLR. Therefore, we report in Table~\ref{tab:epsilon} the average value of  $\epsilon_l$/$\epsilon_c$ together with the average value of $\sigma_\epsilon$. We see in Table~\ref{tab:epsilon} that $\epsilon_l$/$\epsilon_c$ is in average equal to 1 with a scatter $< 10\%$ on periods corresponding to a time delay $< 50 $ days. Since most of our targets have $\Delta t < 50$ days (13 out of 17 targets), we may safely conclude that statistically, intrinsic variability is unlikely to mimic microlensing of the broad lines for such time-delays. The situation might be less favourable for the objects with $\Delta t > 50$ days, but only two out of four of these systems (Q1355-2257 and WFI~2033-4723) show possible microlensing of the emission lines.

\begin{table}[t!]
\begin{center}
\begin{tabular}{l|cccc}
\hline
$\Delta t$ & $\epsilon_l$/$\epsilon_c$(\Hbeta) & $\sigma_\epsilon$(\Hbeta) \\  
\hline
5  & 1.00 & 0.02 \\ 
10  & 1.00  & 0.03 \\
20  & 1.00  & 0.05  \\
30  & 1.00  & 0.06 \\
40  & 1.00  & 0.07 \\
50  & 1.01  & 0.08 \\
100 & 1.01  & 0.10  \\
150 &1.01  & 0.11 \\

\hline
\end{tabular}
\end{center}
\caption{Average value of fractional variation of the \Hbeta~line and of the continuum ($\epsilon_l$/$\epsilon_c$) on periods $\Delta t$ (col. 1), for the sample of reverberation mapped quasars of published in \cite{Kaspi2000}. The average scatter $\sigma_\epsilon$ is given in col. 3. }

\label{tab:epsilon}
\end{table}

\section{MmD applied to a simulated spectrum}
\label{appendixD}

\begin{figure}[htb!]
\includegraphics[scale=0.45]{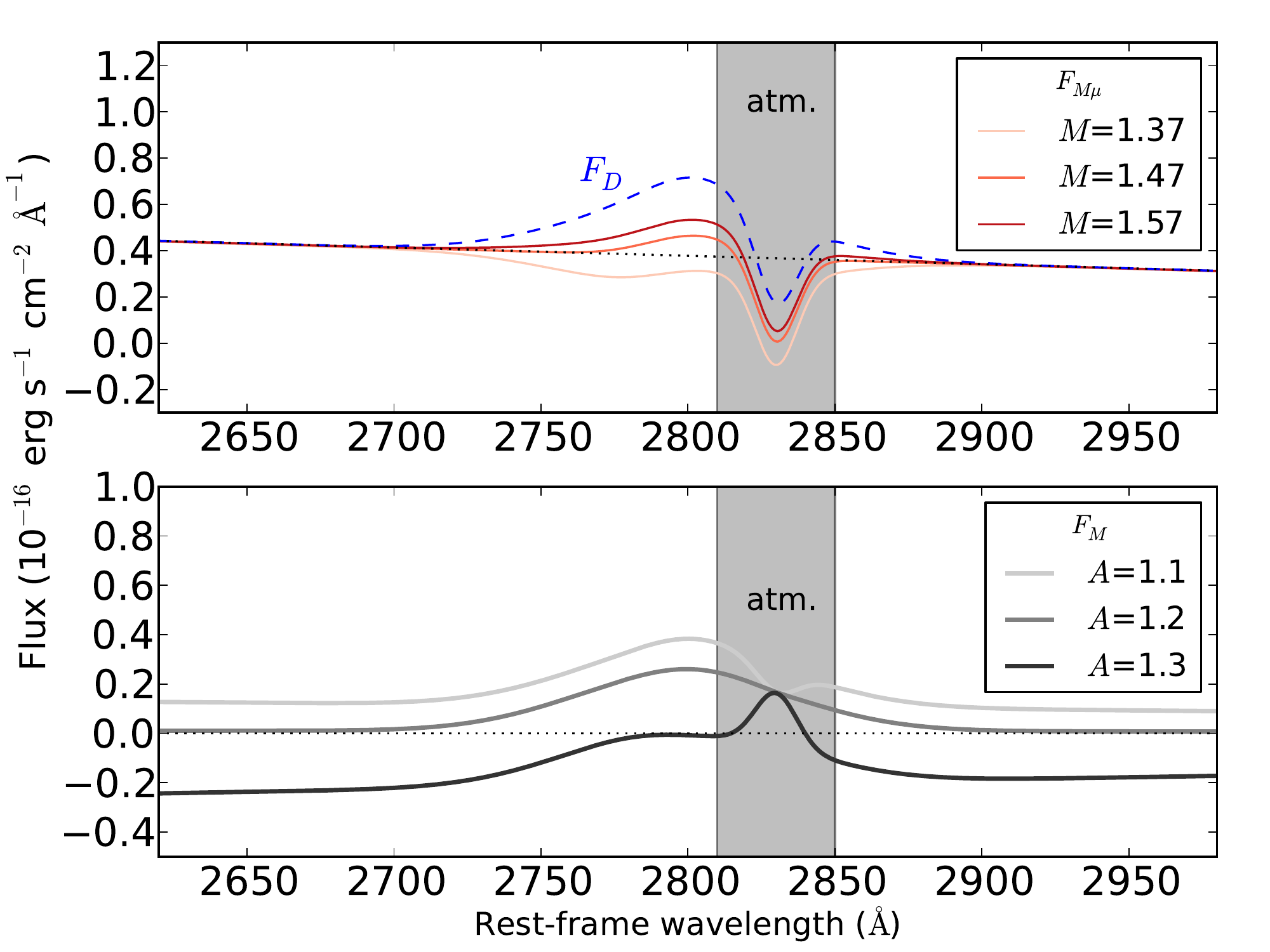}

\caption{Macro-micro decomposition (MmD) applied to simulated spectra of HE~0435-1223 mimicking the observed data. The decomposition is similar to the one showed in Fig.~\ref{fig:MmDexample}. The bottom pannel shows $F_M$ for three different values of $A$ and an arbitrary value of $M$. The best value of $A$ is $A\sim$1.2 because it leads to $F_M=0$ in the continuum regions blueward and redward of the emission line. The upper pannel shows the decomposition for 3 different values of $M$. The best value is $M=$1.47 because it minimizes the emission {\it {in the line}}, keeping the flux above the apparent local continuum depicted as a dotted black line. }
\label{fig:MmDexamplesimu}
\end{figure}

Similarly to the example of MmD applied to HE~0435-1223 in Sect.~\ref{subsec:example}, we show in Fig.~\ref{fig:MmDexamplesimu}, the MmD applied to mock spectra roughly mimicking our spectra of HE~0435-1223. The mock spectra of HE~0435-1223 are defined in the following way:
\begin {equation}
\begin{array}{l}
F_D^{\rm{mock}}= \mathcal{M}_D\,(\mu_D\,F_c + \mu^l_{D}E_a + E_b)  \\ 
\vspace{2mm} F_B^{\rm{mock}}= \mathcal{M}_B\,(\mu_B\,F_c +\mu^l_{B}\,E_a+E_b ) \\ 
\end{array}
\label{equ:mock}
\end {equation}
\noindent where $F_c$ is the continuum emission, ($E_b$, $E_a$) are gaussian emission profiles centered on ($\lambda_{E_b}$, $\lambda_{E_a}) = $(2798, 2803) $\AA$ and with $(FWHM _{E_b}, FWHM_{E_a})$ = (5700, 2500) km/s. In this equation, the macro model magnification $\mathcal{M}$ and the micro-magnification $\mu$ (for the continuum) and $\mu^l$ (for the line)  of individual images have been written explicitly. These quantities have been chosen arbitrarily such that $M=\mathcal{M}_B/\mathcal{M}_D=6.46/4.39=1.47$ (matching the macro-model), $\mu=\mu_B/\mu_D=0.74/0.9=0.82$, and $\mu^l=\mu^l_B/\mu^l_D=0.85/1.0=0.85$. We used a different micro-magnification factor in the line and in the continuum to account for the fact that lines are emitted in a region larger than the continuum (i.e. $\mu_l$ closer to 1 than $\mu_c$). We have also added a fake atmospheric absorption to the spectra of B and D in order to increase the similarity with the observed spectrum of H~0435-1223. Despite the similarity with the data, the model of the emission line of Eq.~\ref{equ:mock} should be considered only for illustration purpose. The MmD applied to these spectra is the same as the one discussed in Sect.~\ref{subsec:example}. In absence of noise, the value $M=1.47$ and $\mu=0.82$ are retrieved. The component $F_{M\mu}$ shows that a red fraction of the emission line (corresponding to our input component $E_a$) is retrieved as microlensed, in agreement with our input model. 

\section{Characteristics of  the main sample}
\label{appendixA}

\begin{figure*} 
\centering
\setcounter{subfigure}{0}
\renewcommand{\thesubfigure}{(\alph{subfigure})}
  \subfigure[HE~0047-1756 (B/A)]{\includegraphics[scale=0.35]{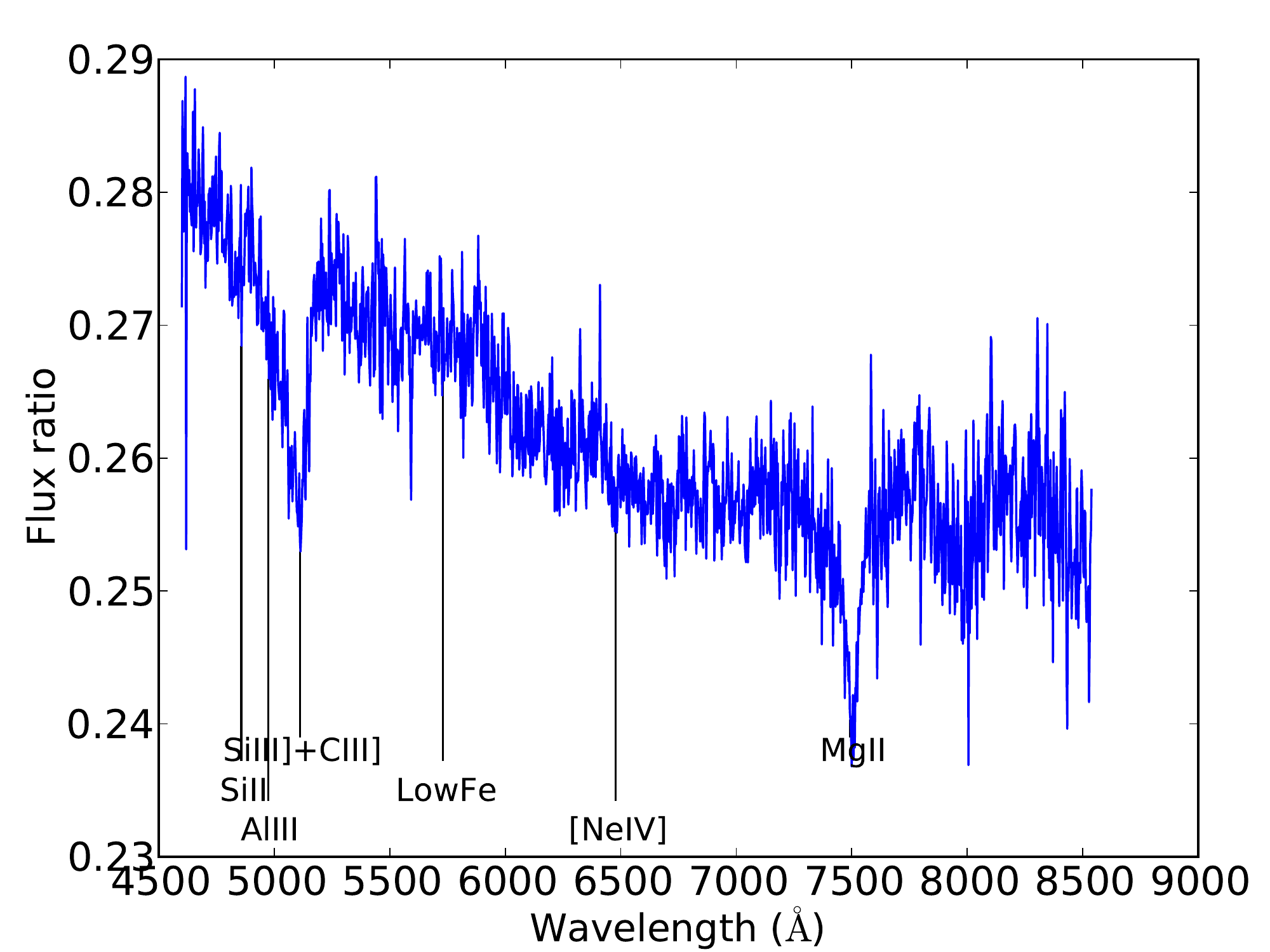}} 
\setcounter{subfigure}{1}
\renewcommand{\thesubfigure}{(\alph{subfigure})}
  \subfigure[Q0142-100 (B/A)]{\includegraphics[scale=0.35]{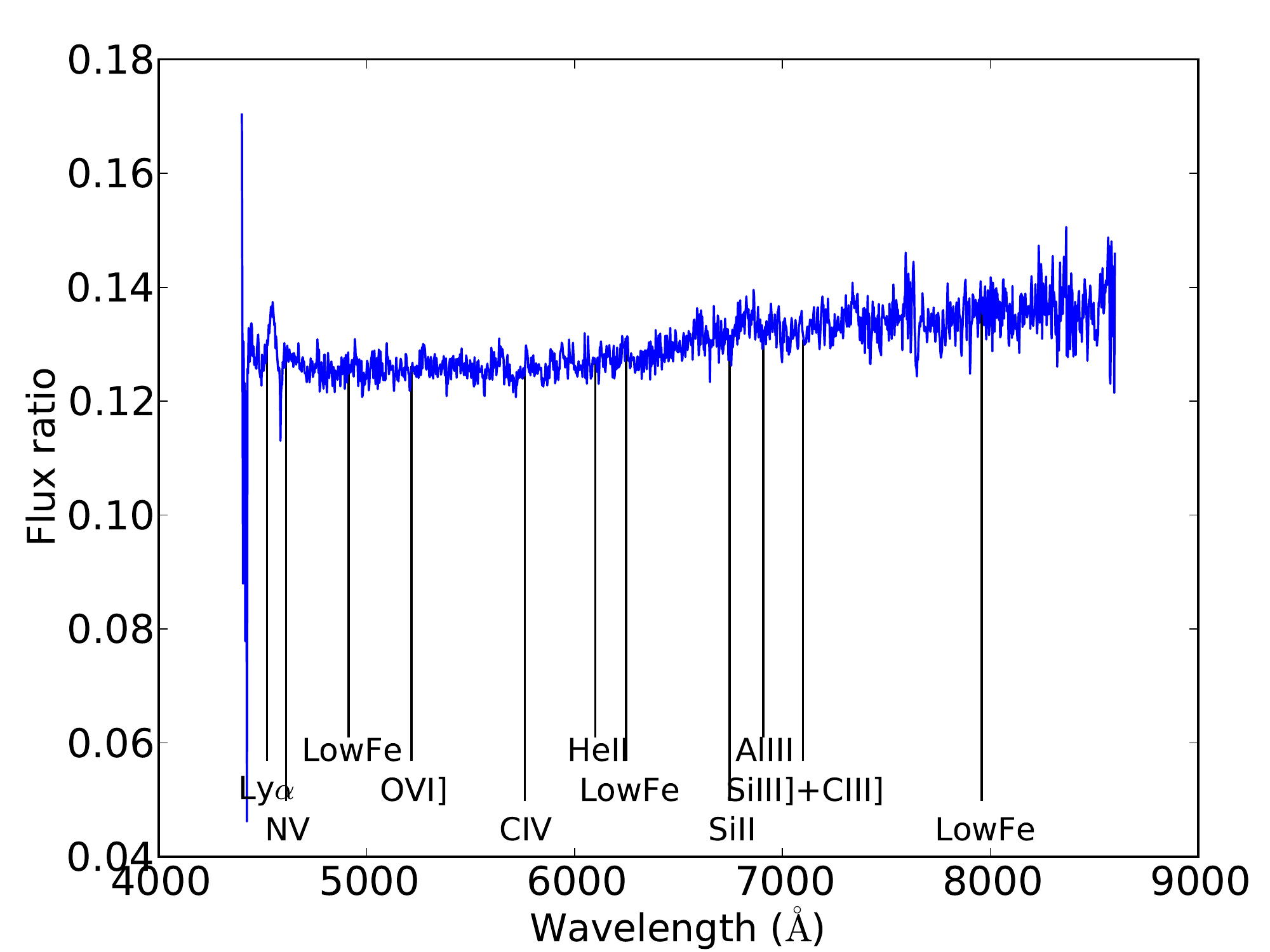}} 

\setcounter{subfigure}{2}
\renewcommand{\thesubfigure}{(\alph{subfigure})}
  \subfigure[SDSS~J0246-082 (B/A)]{\includegraphics[scale=0.35]{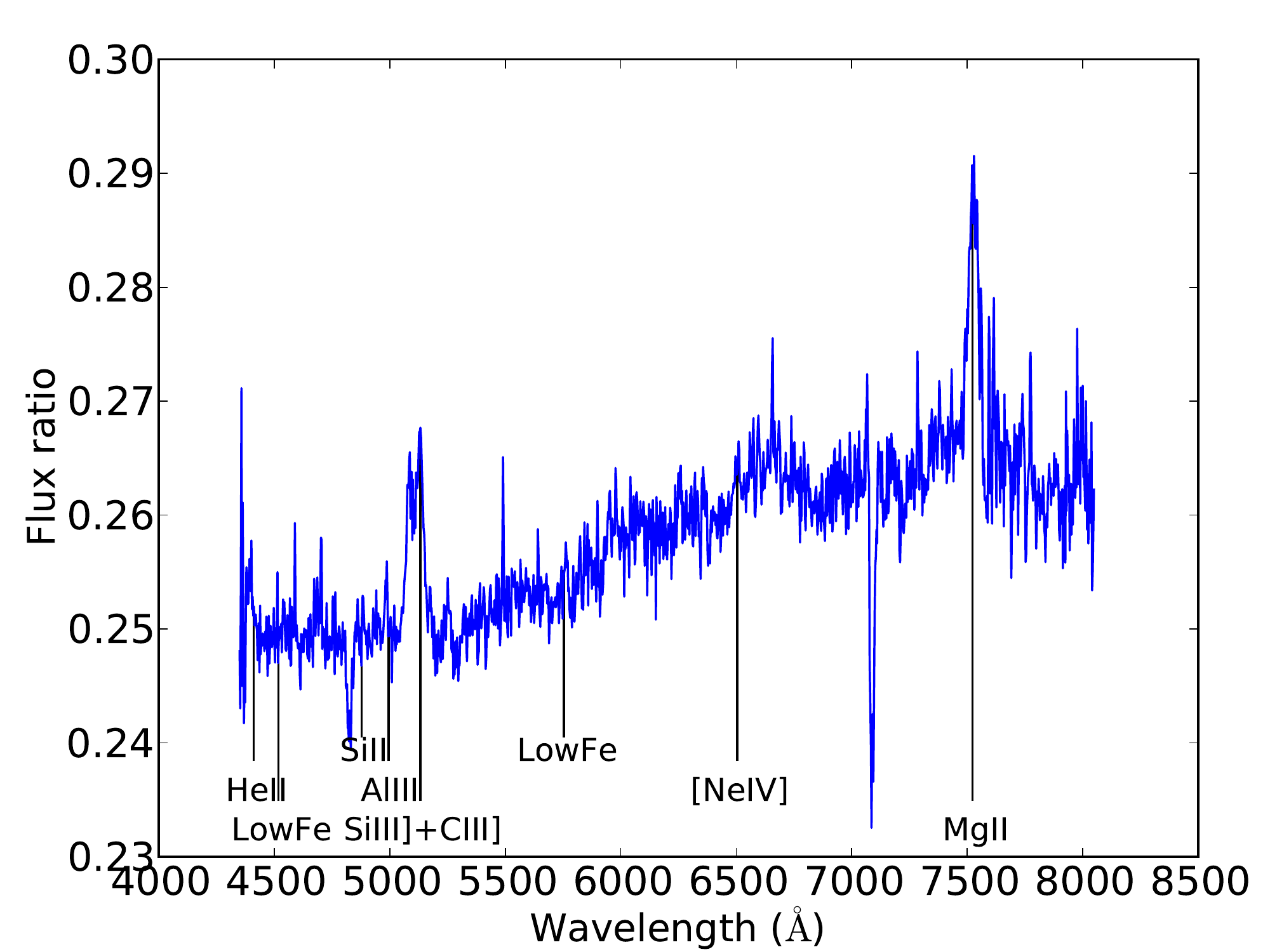}} 
\setcounter{subfigure}{3}
\renewcommand{\thesubfigure}{(\alph{subfigure})}
  \subfigure[HE~0435-1223 (B/D)]{\includegraphics[scale=0.35]{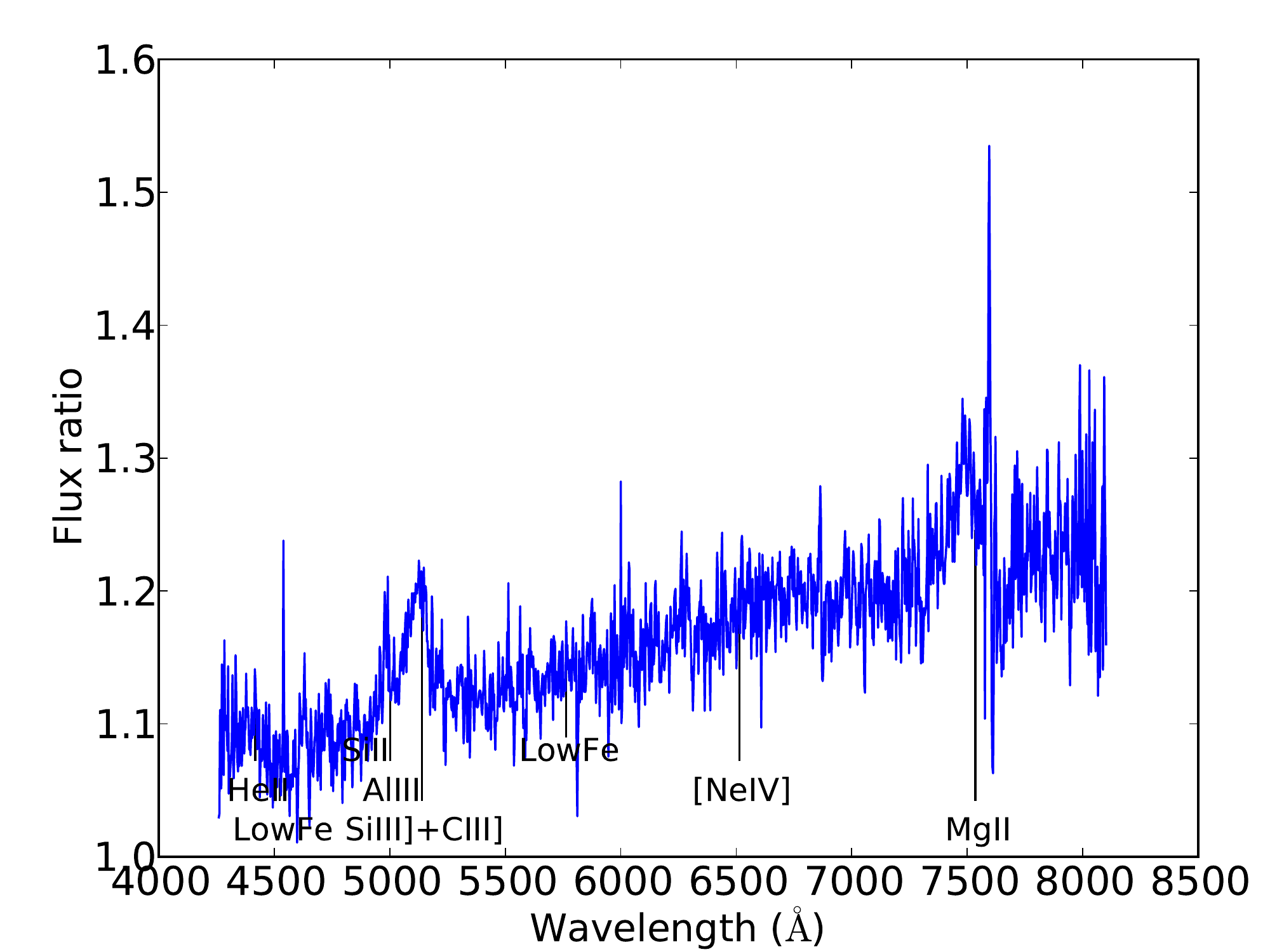}}              

\setcounter{subfigure}{4}
\renewcommand{\thesubfigure}{(\alph{subfigure})}
  \subfigure[SDSS~J0806+2006 (B/A)]{\includegraphics[scale=0.35]{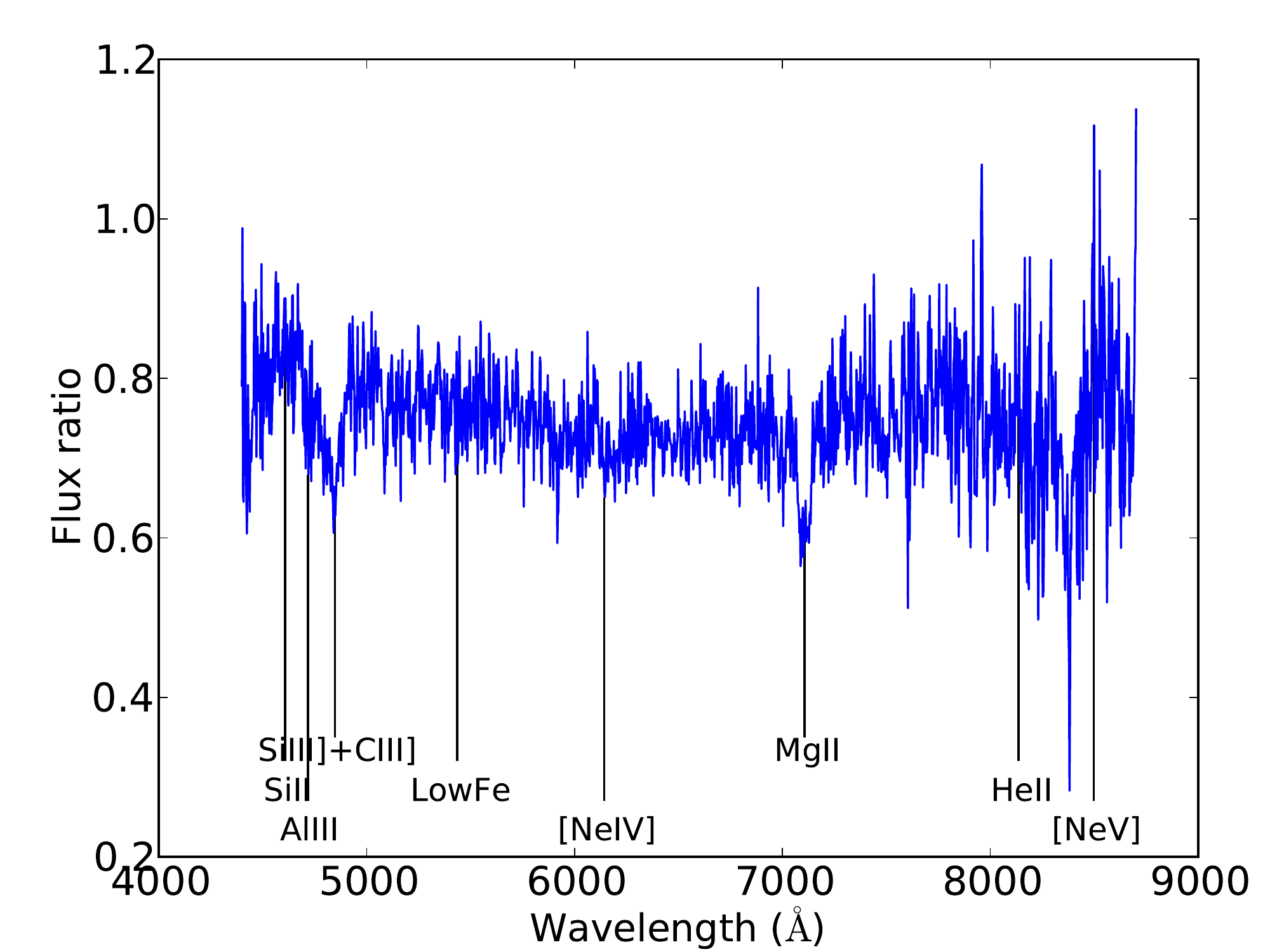}} 
\setcounter{subfigure}{5}
\renewcommand{\thesubfigure}{(\alph{subfigure})}
  \subfigure[FBQ~0951+2635 (B/A)]{\includegraphics[scale=0.35]{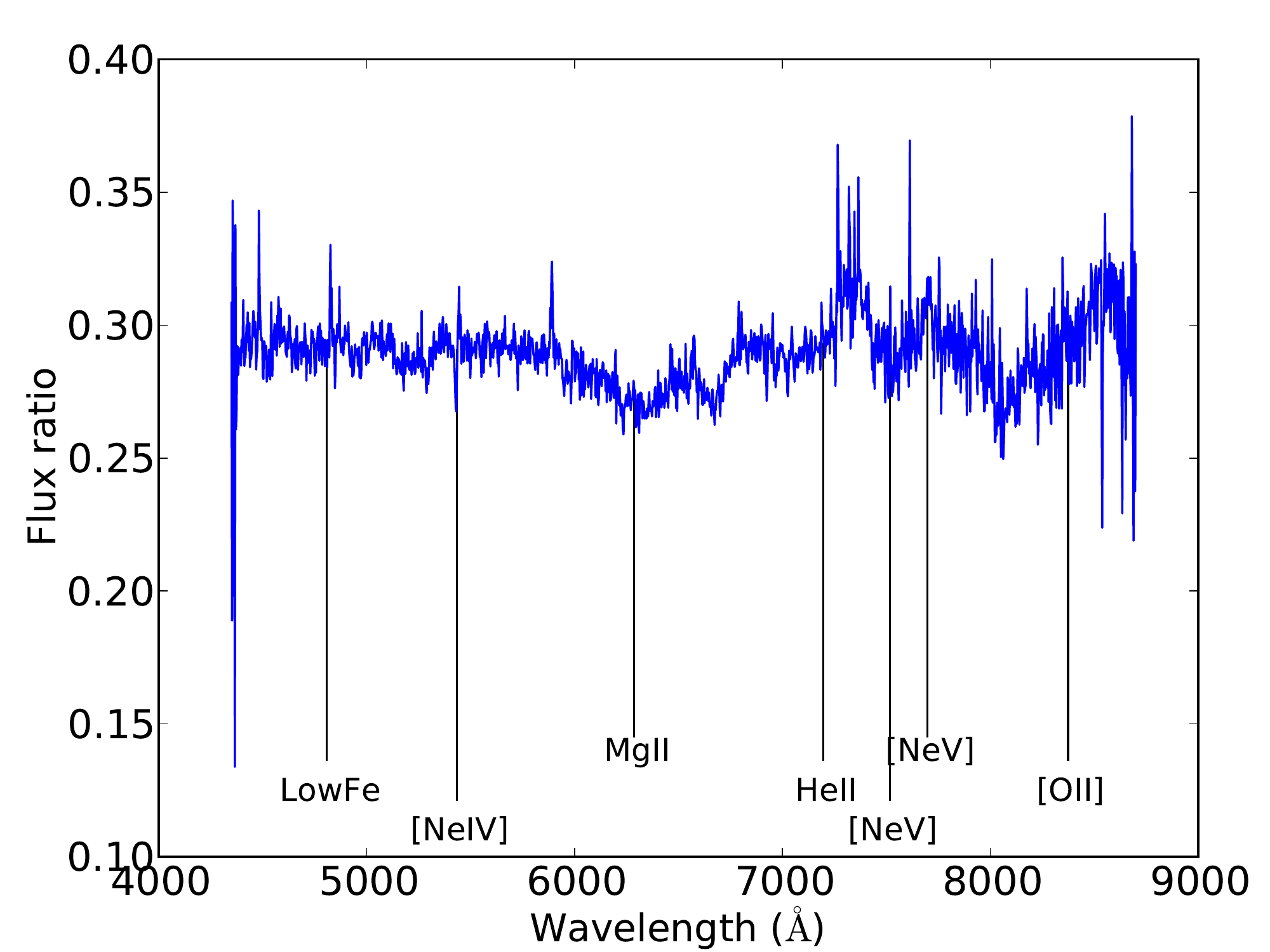}}              

\setcounter{subfigure}{6}
\renewcommand{\thesubfigure}{(\alph{subfigure})}
  \subfigure[BRI~0952-0115 (B/A)]{\includegraphics[scale=0.35]{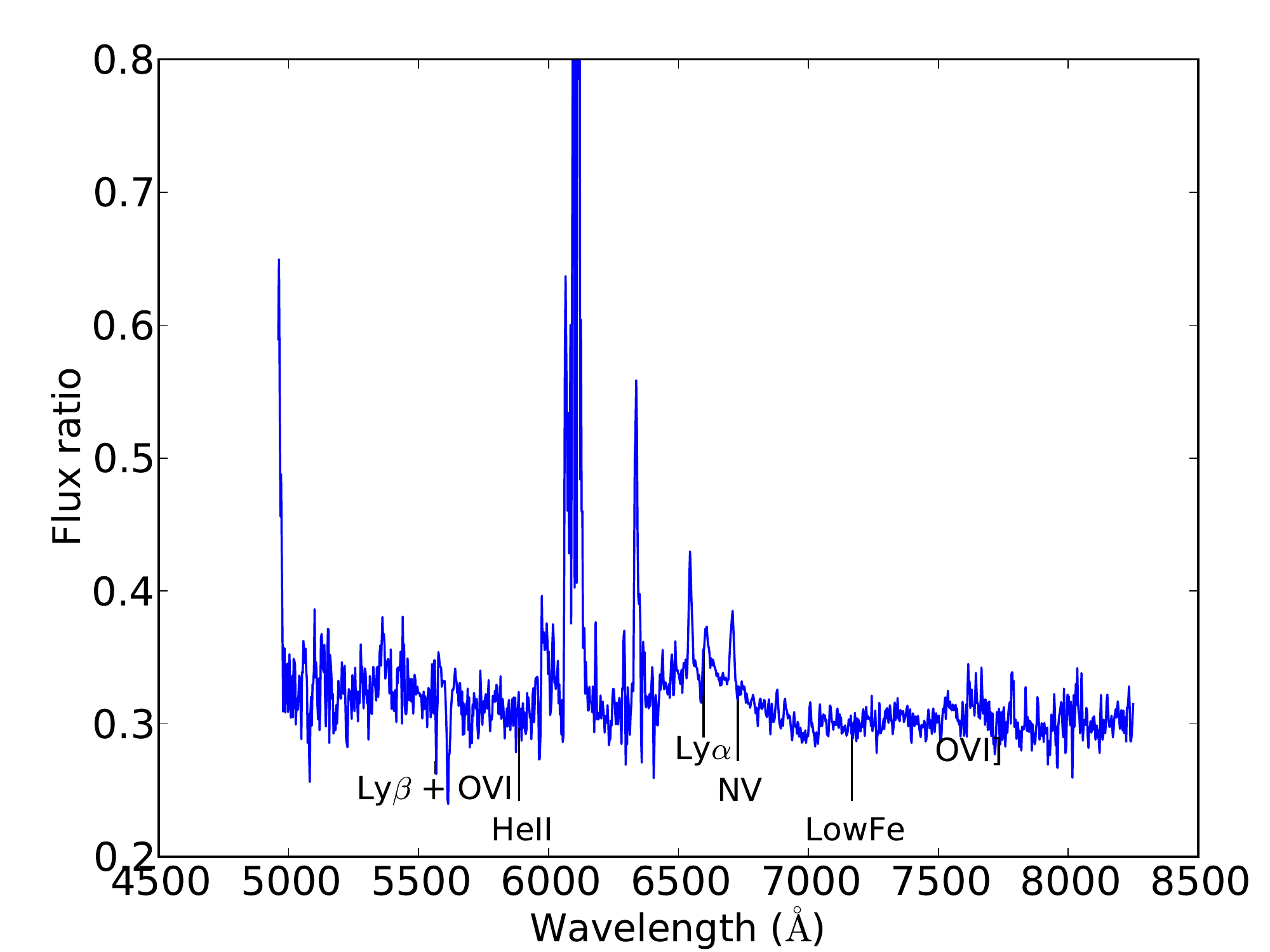}} 
\setcounter{subfigure}{7}
\renewcommand{\thesubfigure}{(\alph{subfigure})}
  \subfigure[SDSS~J1138+0314 (C/B)]{\includegraphics[scale=0.35]{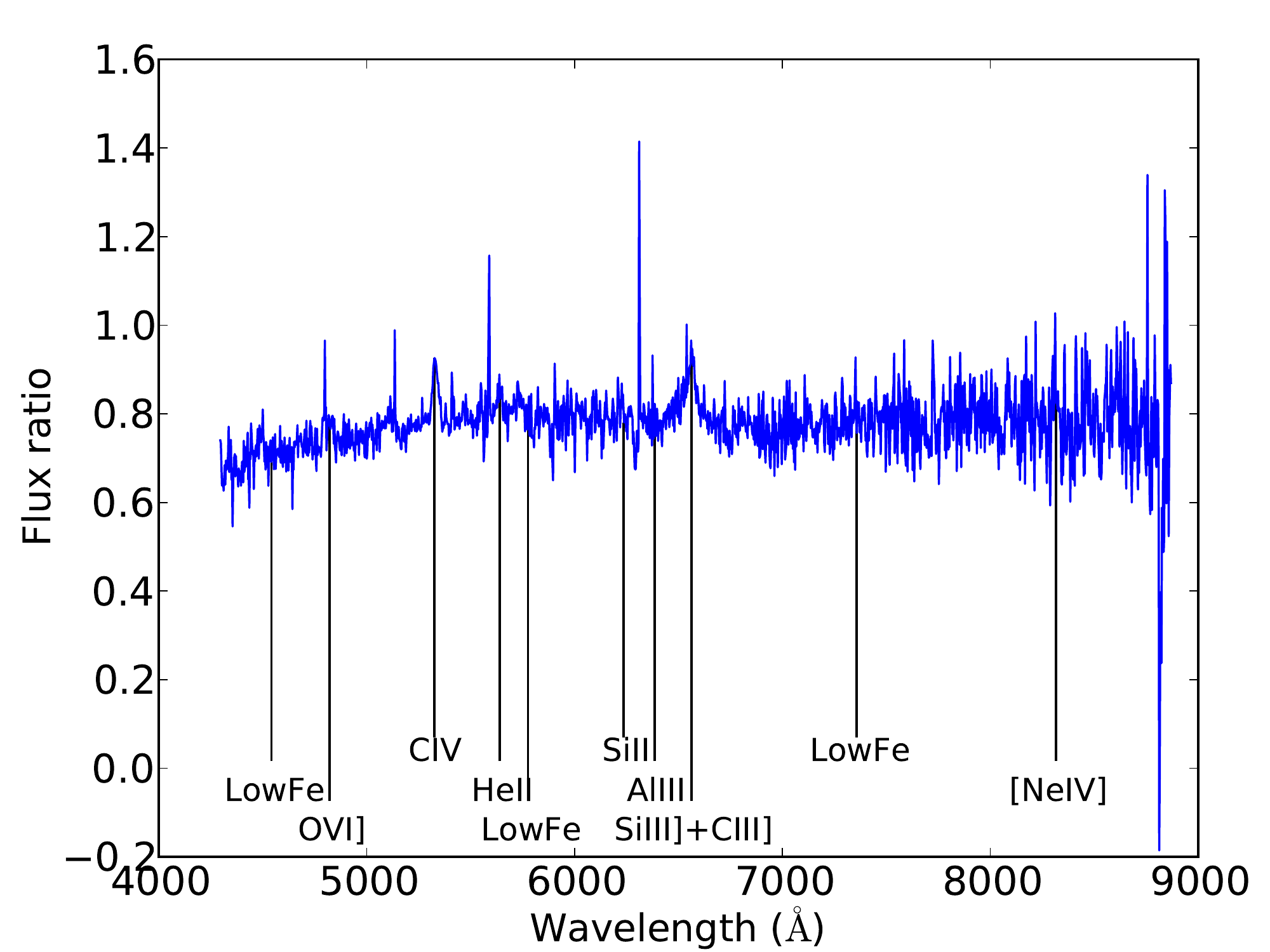}}    

\caption{Flux ratio for all the image pairs of our main sample.}
\label{fig:Ratiomain}
\end{figure*}

\begin{figure*} 
\centering
\setcounter{subfigure}{8}
\renewcommand{\thesubfigure}{(\alph{subfigure})}
  \subfigure[J1226-006 (B/A)]{\includegraphics[scale=0.35]{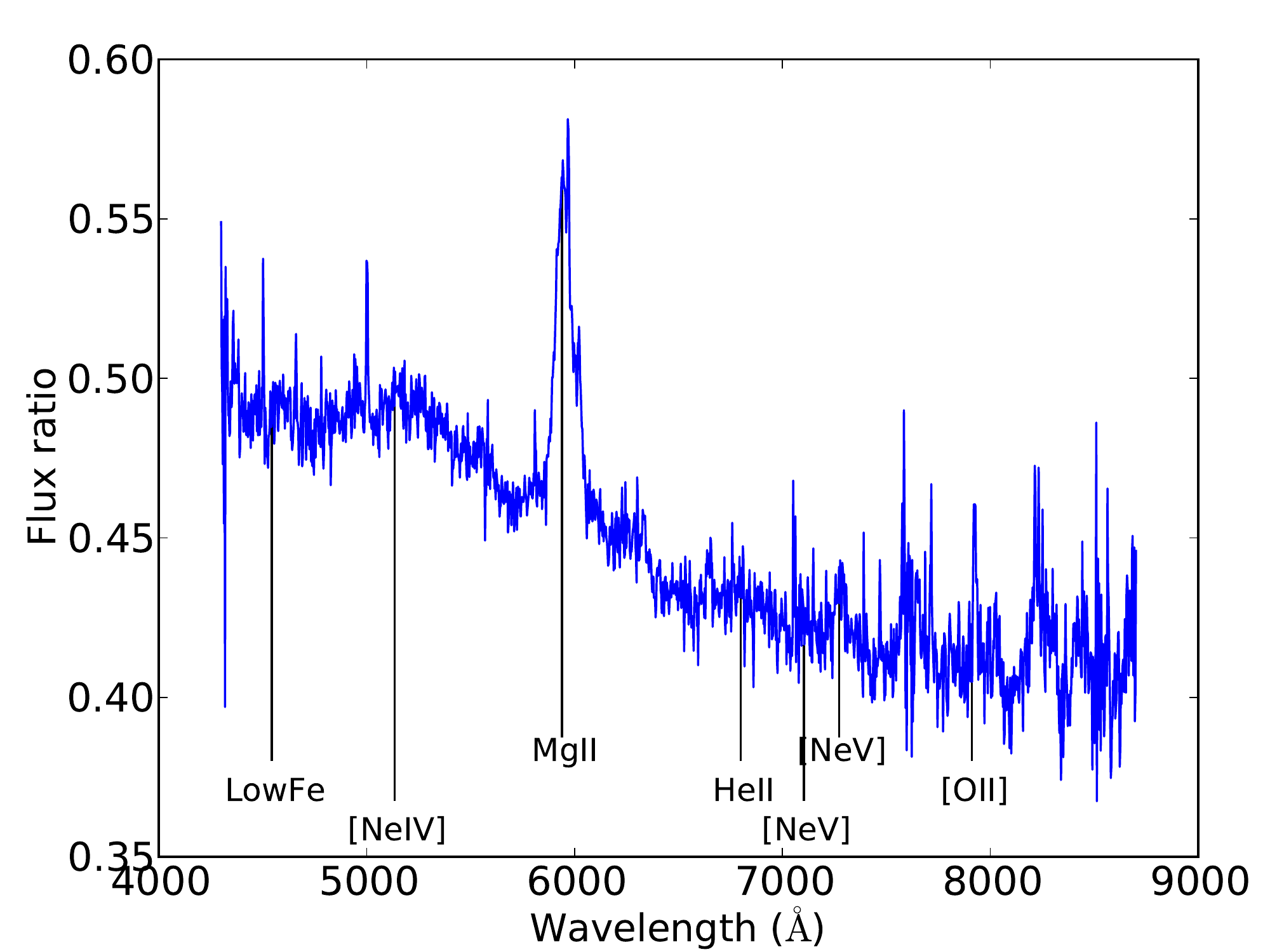}} 
\setcounter{subfigure}{9}
\renewcommand{\thesubfigure}{(\alph{subfigure})}
  \subfigure[SDSS~J1335+0118 (B/A)]{\includegraphics[scale=0.35]{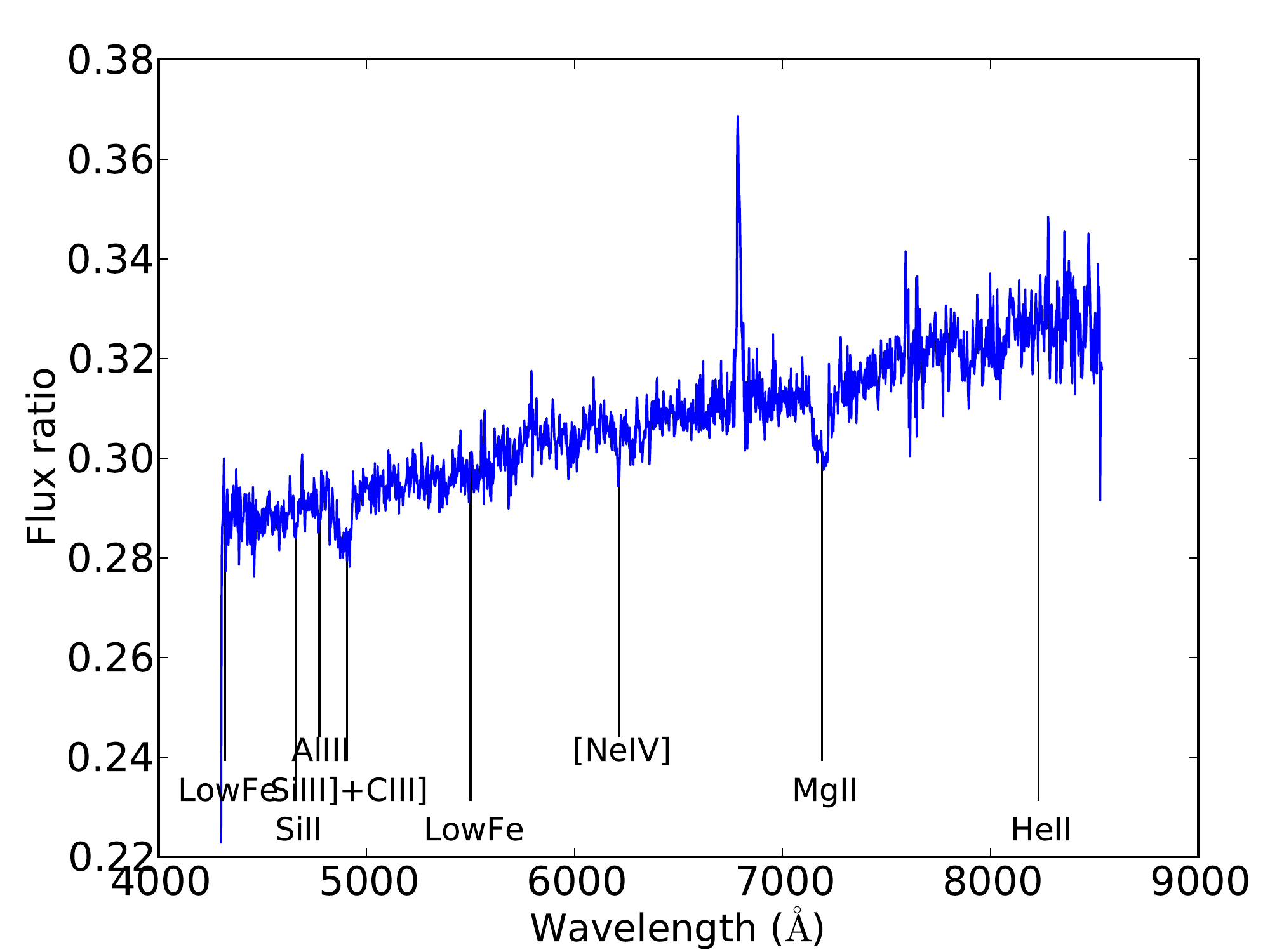}} 

\setcounter{subfigure}{10}
\renewcommand{\thesubfigure}{(\alph{subfigure})}
  \subfigure[Q1355-2257 (B/A)]{\includegraphics[scale=0.35]{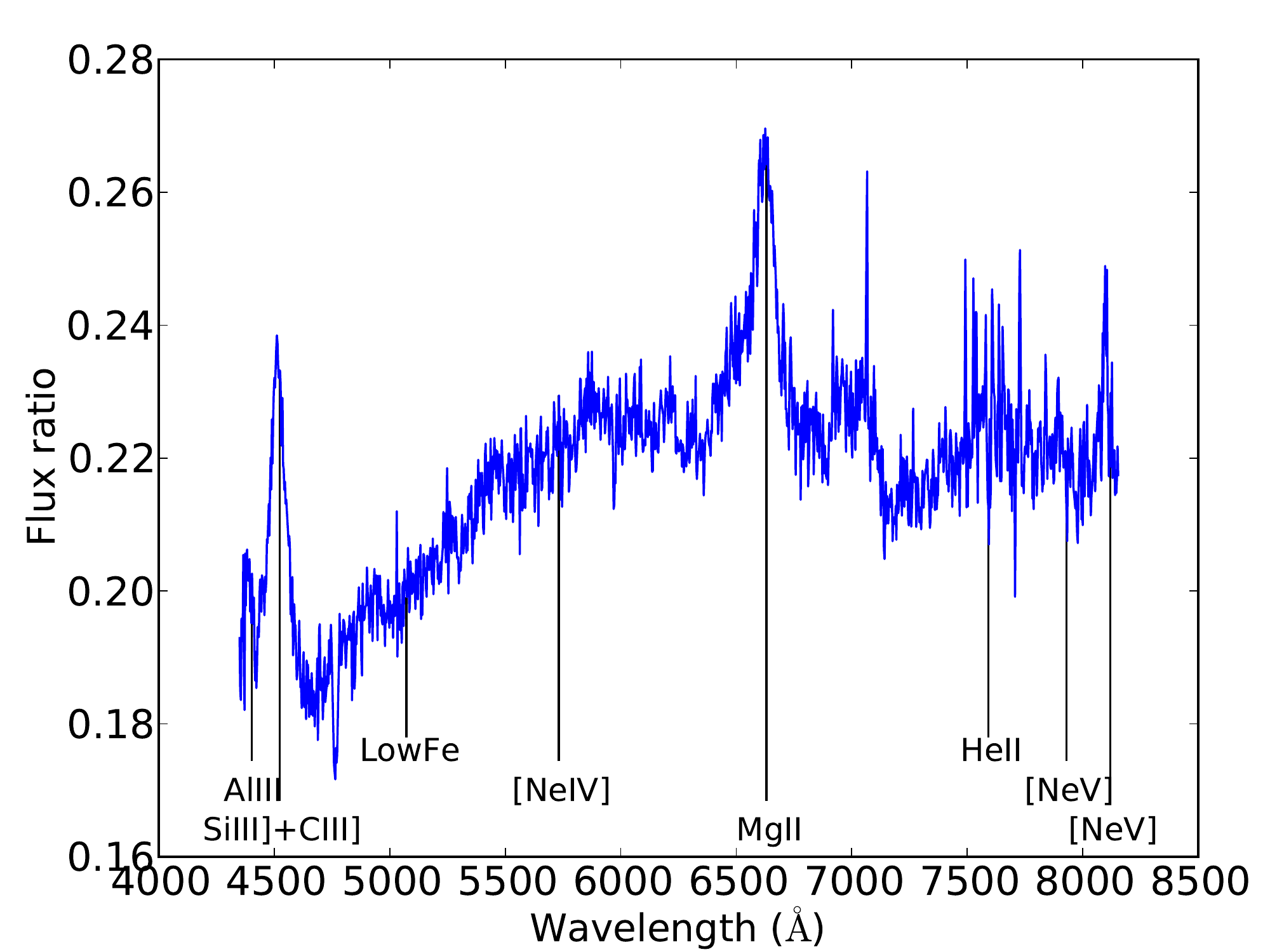}} 
\setcounter{subfigure}{11}
\renewcommand{\thesubfigure}{(\alph{subfigure})}
  \subfigure[WFI~2033-4723 (C/B)]{\includegraphics[scale=0.35]{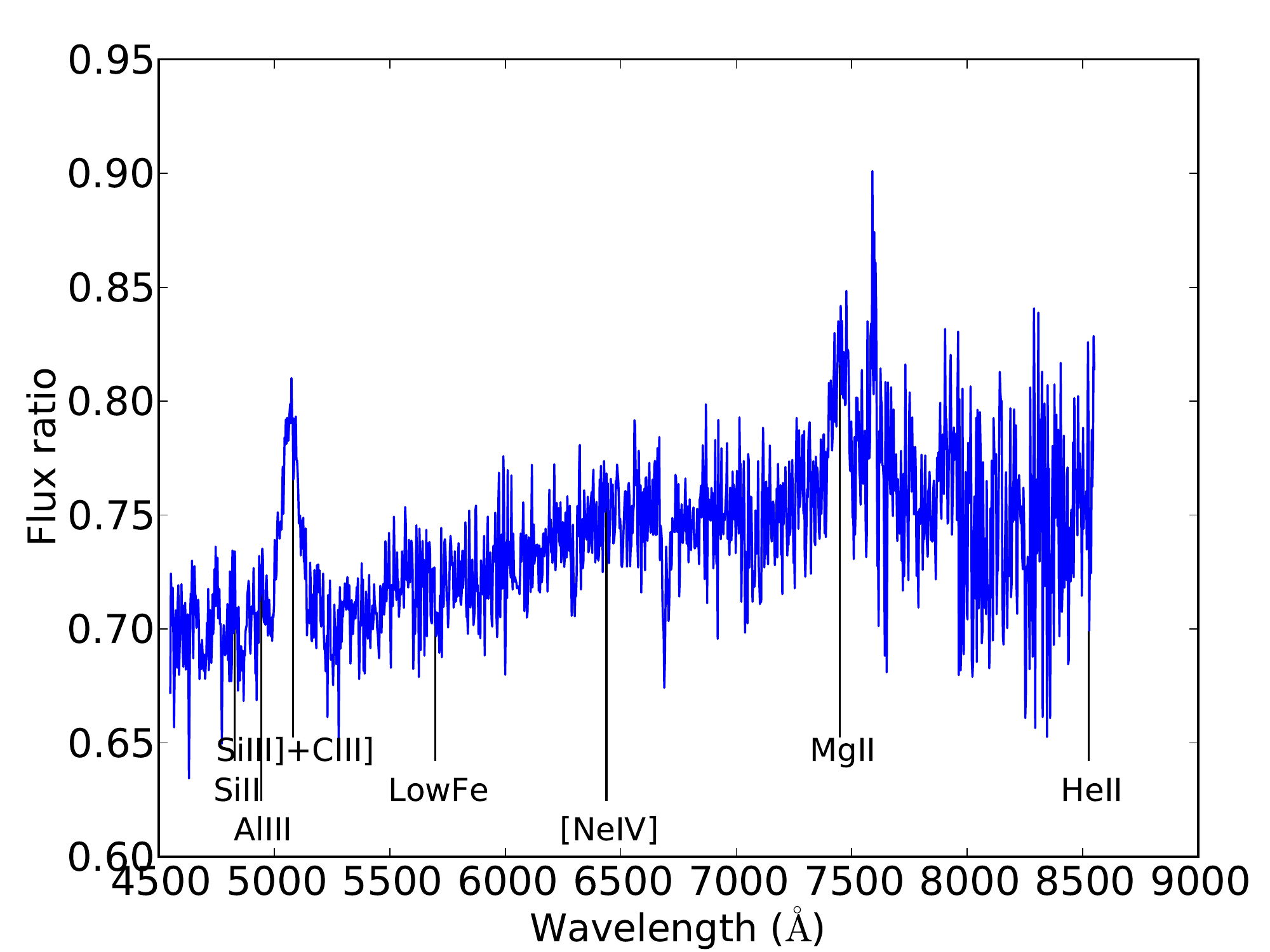}}              

\setcounter{subfigure}{12}
\renewcommand{\thesubfigure}{(\alph{subfigure})}
  \subfigure[HE2149-2745 (B/A)]{\includegraphics[scale=0.35]{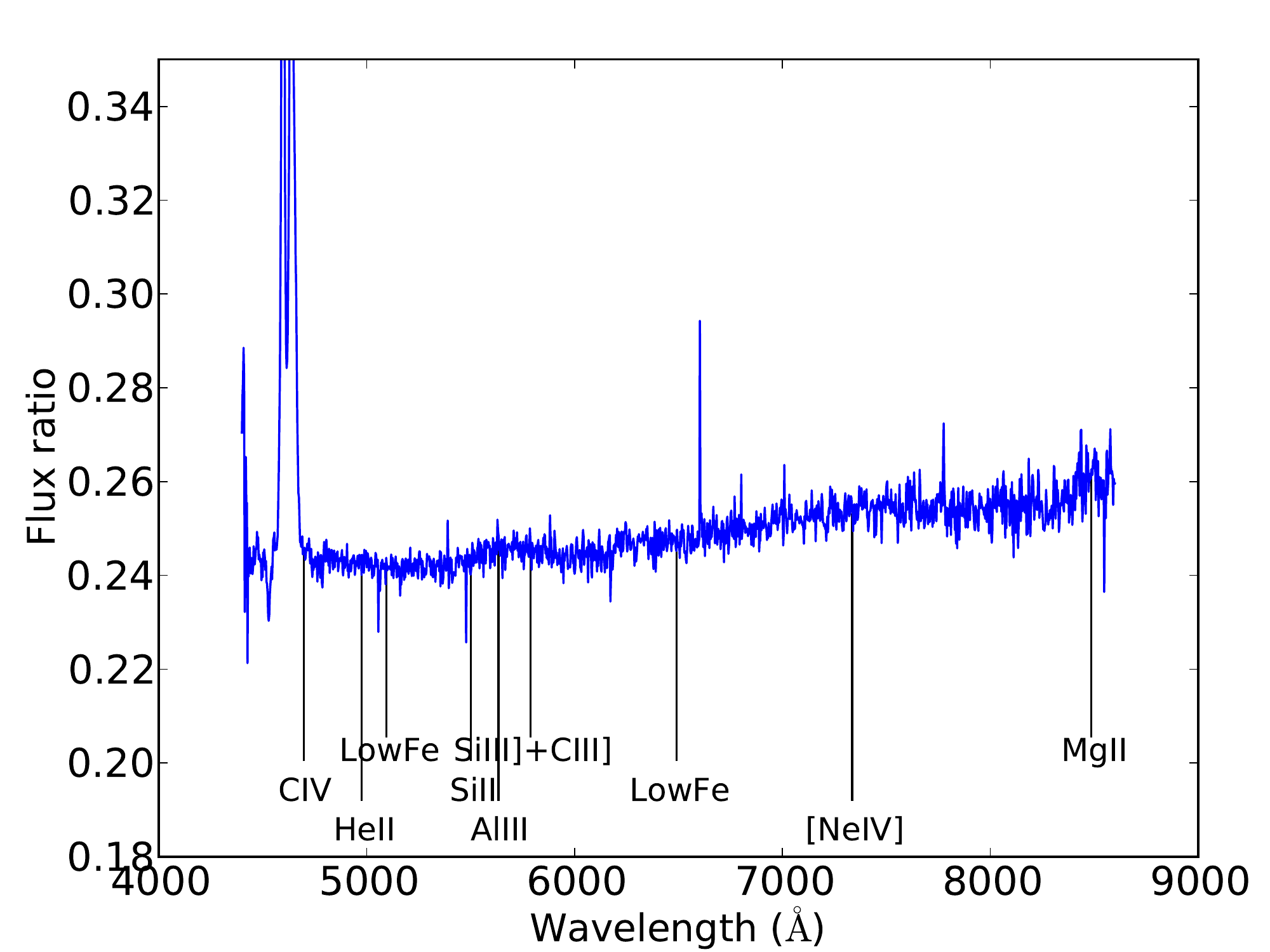}} 

\setcounter{figure}{0}
\caption[]{continued}
\end{figure*}

We provide hereafter detailed notes about the characteristics of the spectra of each object of the main sample. For each target we discuss i) the quality of the deconvolution, which might introduce spurious signal, ii) the chromatic changes observed in the spectra, iii) the microlensing-induced deformation of the emission lines, when a deformation is observed, and iv) important results from literature which shed light on the source of chromaticity and confirm/infirm our flux ratio measurements. We provide in Fig.~\ref{fig:Ratiomain}, the spectral flux ratios between the pairs of images and we discuss in the text four origins for the chromatic changes in the spectral ratios: differential extinction (DE), chromatic microlensing (CML), contamination by the lens (LC), or intrinsic variability (IV). In order to quantify the possible systematic error on our estimate of $M$ caused by intrinsic variability, we also provide in Table~\ref{tab:variability} the amplitude of variation $\Delta m$ of an object over the time-spent of the time-delay. This quantity has been derived using the $g-$band structure functions (divided in 6 bins of $M_{BH}$ and $L$, cf. their Eq. 2 and Table 3) of \cite{Wilhite2008} and the black hole masses and luminosities of Table~\ref{tab:MBH}. In the following, the references to the NIR flux ratios from literature are not systematically given. They can be found in Table~\ref{tab:macro}.

{\bf {(a) HE0047-1756:}} {\it {Deconvolution:}} The deconvolution is slightly less good than for other systems. The total flux left in the residual under the QSO images never exceeds 0.1\% of the flux of the QSO image. Although this is a very small amount of flux, this appears as a systematic feature suggesting that the PSF is less representative of the QSO images than in other systems. \\
{\it {Chromaticity:}}  The ratio B/A shows a monotonic decrease with increasing wavelength. This chromatic change is not due to contamination by the flux of the lensing galaxy. Indeed, in order to reconcile the shape of the spectrum in A \& B, one has to invoke a contamination of A (the brightest image) by $\sim$ 4 times the measured flux of G and a contamination 10 times smaller for image B. Therefore, the two most likely explanations are DE with ($M($blue$)$, $M($red$)$, $\mu$) = (0.231, 0.220, 1.173) and CML with ($M$, $\mu($blue$)$, $\mu($red$)$) = (0.220, 1.260, 1.170). \\
{\it {Broad Lines:}} Whatever the origin of the chromatic changes, the blue wing of \MgII\, is microlensed. Microlensing of the blue wing of \CIII\, is also tentatively observed but this is more uncertain owing to the proximity of this line from the red edge of the spectrum.\\ 
{\it {Notes:}} The observed monotonic decrease of B/A was also reported in the discovery spectra of \cite{Wisotzki2004}. For these spectra, obtained in Dec. 2001 and Sept. 2002, we measure $A=0.347 \pm 0.002$ and $M = 0.24 \pm 0.02$ at the wavelength of \CIII. The different value of $A$ found in our spectra is caused by time-variable microlensing. On the other hand, $M$ is compatible with our measurement, confirming that intrinsic variability may not be too large. The $H-$band flux ratio from \cite{Chantry2010}, does not follow the chromatic variation of $M$ observed in the spectra and is found 0.1 mag smaller than the value derived at the level of \CIII. This may be explained by intrinsic variability ($\Delta m \sim 0.07$ mag) and/or microlensing of the H-band continuum.

{\bf {(b) Q0142-100:}} {\it {Deconvolution:}} Because of the inaccurate PSF, residual signal above the noise is visible in the deconvolved image. However, this flux amounts only to 0.02-0.2\% of the total flux in the lensed images. The flux of the lens galaxy G, located only $\sim$ 0.4$\arcsec$ from B, reaches 20\% of the flux in image B in the red part of the spectrum. This could lead to significant contamination of image B by the lensing galaxy G. \\ 
{\it {Chromaticity:}} There is no differential ML between the continuum and the emission lines however the two spectra do not superimpose once scaled with the same magnification factor.  This chromatic effect is probably caused by residual contamination of image B by flux from the lens since only $\sim$ 13\% of the observed flux of G is needed to explain the observed chromatic trend. Intrinsic variability could also play a role. We discard DE because it involves a larger reddening of image A which is farther away from the galaxy than image B. \\
{\it {Notes:}} Our flux ratio is in good agreement with those obtained by \cite{Fadely2011a} in the $K-$ and $L'-$ bands. \cite{Koptelova2010} measured flux changes by 0.1 mag over a period of 100 days,which probably explain the flux ratio differences with $H-$band \citep{Lehar2000}. Color differences associated to IV might also play a role and have been reported for this system by \cite{Koptelova2010}.  The study of the extinction in the lensed images by \cite{Orstman2008} disfavours significant extinction in this system, in disagreement with \cite{Falco1999, Elliasdotir2006}. 

{\bf {(c) SDSS~J0246-0825:}} {\it {Deconvolution:}} The deconvolution is very good. The small systematic residual detectable in the vicinity of the brightest lensed image A contributes to less than 0.05\% of the flux of A. \\
{\it {Chromaticity:}} The observed chromatic trend in B/A is compatible with DE ($M$(blue)$=$0.32, $M$(red)$=$0.34, $\mu=$0.76) or with CML (M$=$0.34, $\mu$(blue)$=$0.73, $\mu$(red)$=$0.76).  Contamination of image A by the host galaxy, unveiled as a ring feature close to A in \cite{Inada2005} is plausible but is probably very low due to the slit clipping and to the relative faintness of this feature.  \\
{\it {Broad Lines:}} Whatever the origin of the chromatic changes, the broad component of \CIII~is microlensed but ML of \MgII\,is more tentative.\\
{\it {Notes:}}  There is a good agreement between our spectral-based estimate of $M$ and the $H-$ and $L-$ band flux ratios \citep{Inada2005, Fadely2011a}. The agreement in the $K-$band is slightly less good but still marginally consistent with the other measurements \citep{Fadely2011a}.

{\bf {(d) HE~0435-1223:}} {\it {Deconvolution:}} Very good results are obtained with the deconvolution. The lensing galaxy is relatively bright compared to the lensed images (only 4 times fainter than the flux of the lensed images above 6500$\AA$) and we may not exclude contamination of the latter by the lens. The symmetric location of the lensed images aside the lensing galaxy and their similar brightnesses however argue against large differential effects (i.e. if contamination takes place, the spectra of A \& D should be corrupted the same way by the lens). \\
{\it {Chromaticity:}} We observe a chromatic increase of B/D with increasing wavelength, in agreement with the chromatic changes observed by \cite{Fadely2011a} based on HST images. This trend may not be produced by contamination from the lensing galaxy because one needs a large contamination of image $B$ (by at least 30\% of the observed flux of G) and nearly no contamination of image $D$ to mimic this effect. Because we find approximately the same microlensing factor in the blue and in the red, DE ($M$(blue)$\sim$1.34, $M$(red)$\sim$1.47, $\mu\sim$0.81) is the most natural explanation, B being more reddened than D by the dust in the lensing galaxy. Alternatively, CML may be at work ($M\sim$1.47, $\mu($blue$) \sim$ 0.74, $\mu($red$)\sim$ 0.82). \\
{\it {Broad Lines:}} Whatever the origin of the chromatic changes, ML of the red wing of the \CIII\, and \MgII\, lines is observed. \\
{\it {Notes:}} Two fairly different flux ratios in $K-$band have been reported in literature. \cite{Fadely2011a} reported $B/D=$1.49$\pm$0.12 while \cite{Blackburne2011a} reported $B/D=$1.27$\pm$0.04. The latter estimate agrees well with the $H-$band and $L-$band flux ratios while the former one agrees with our spectroscopic estimate $M=$1.47. \cite{Fadely2011a, Fadely2011b} interpreted the $K-$band flux ratio as caused by microlensing. However, the simplest microlensing scenario is hard to reconcile with the non monotonic variations of B/D, if real. Additional data are needed to solve this puzzle. 

{\bf {(e) SDSS~J0806-2006:}} {\it {Deconvolution:}} The deconvolution is good but the spectra have a significantly lower signal to noise than for the other systems. The residual flux under the point-like images reaches up to 0.2\% of the flux of the QSO image. The lensing galaxy is well deblended from the QSO images. \\
{\it {Chromaticity:}} The estimate of $A$ at wavelengths shorter than \CIII~ is uncertain due to the proximity of the noisy edge of the spectrum. Despite of this, the measurements are consistent with a flat ratio $B/A$ in the continuum from 4000 to 8000\,$\AA$~and ML affecting only the continuum emission. \\
{\it {Notes:}} Our average macro-magnification ratio $M_{BA}\sim$ 0.435 agrees with the $H-$ and $K-$band ratios \citep{Sluse2008, Fadely2011a}, although the latter two ratios differ by 0.16 mag, suggesting that the $H-$band flux is still slightly microlensed. At larger wavelengths, \cite{Fadely2011a} find a $L'$-band flux ratio $B/A<$0.164, suggesting that the $H-$ and $K-$ band continua are microlensed as well, but probably by a massive substructure. \\
{\it {Broad Lines:}} They are apparently unaffected by microlensing unless $M$ is significantly underestimated as suggested by the literature data. If this scenario is correct, the broad \MgII\, and \CIII\, emission lines should then be significantly microlensed. 

{\bf {(f) FBQ~0951+2635:}} {\it {Deconvolution:}} The deconvolution is good. The residual flux in the vicinity of the lensed images amounts less than 0.03\% of the brightest lensed image A and less than 0.2\% of the flux of $B$. The lens galaxy is only $\sim$0.15$\arcsec \sim 4$\, pixels away from image $B$ and is only $\sim$ 5 times fainter in the reddest part of the spectrum (i.e. $> 7000$\AA). This might lead to residual contamination of $B$ by the lensing galaxy. The spectral regions 7200-7400\,\AA~and 7700-7950\,\AA\, are unreliable due to an enhanced level of noise.\\
{\it {Chromaticity:}} Despite the flat continuum flux ratio, we may not exclude possible residual contamination of image $B$ by the lens because of the small separation between $G$ and $B$.  We identify ML of the continuum based only on the \MgII~emission, which is not microlensed. The decomposition of the spectra around that line is however not entirely satisfactory because $F_{M\mu}$ under \MgII\, does not show the monotonic variation expected if only the power law continuum was microlensed. This might be associated to microlensing of \FeII\, or to spurious effect of contamination by the lens. \\
{\it {Notes:}} Microlensing of the continuum in this system, at different epochs, is supported by several other studies \cite[e.g.][and reference therein]{Schechter1998, Jakobsson2005, Munoz2011}. A low amplitude chromatic change of $B/A$ from 4000 to 9000\,\AA\, has been detected based on HST images obtained 2.5 years before our data \citep{Munoz2011}. There is a good agreement between our spectral-based estimate of $M$ and the radio flux ratio (which unfortunately lacks error estimates). The $H-$band ratio \citep{Falco1999} is larger by about 0.2 mag compared to our estimate. This offset is hardly explained by intrinsic variability (Tab.~\ref{tab:variability}) and therefore suggests that the continuum is still microlensed in $H-$band. 

{\bf {(g) BRI~0952-0115:}} {\it {Deconvolution:}} The deconvolution is very good. Spatially resolved narrow \Lyalpha~is visible in the background image. This emission is not produced in the QSO but in the host galaxy of this remote quasar. \\ 
{\it {Chromaticity:}} Because of the Lyman break, we do not estimate $A$ in both sides of the emission but only in the continuum redward of \Lyalpha. The ratio $B/A$ seems however flat from the blue to the red with only an imprint of the broad emission lines. Therefore, the continuum emission is microlensed. \\
{\it {Broad Lines:}} Our decomposition unveils ML of a significant fraction of the \Lyalpha~line. Because the flux leading to the absorption is microlensed as the continuum, the latter does not appear in $F_M$, which unveils a (nearly) symmetric emission roughly centered on the narrow emission. \\
{\it{Notes:}} The $H-$band flux ratio is similar to our continuum flux ratio, and therefore supports a significant microlensing ($\sim$0.45 mag) at that wavelength. This is not a surprise as $H-$band corresponds to rest-frame UV emission ($\lambda\sim$ 3300 \AA,) which is small enough to be significantly microlensed. 

{\bf {(h) SDSS~J1138+0314:}} {\it {Deconvolution:}} The deconvolution is good with some residual flux and background excess in the vicinity of image B. Its origin is possibly associated to the QSO host galaxy. \\  
{\it {Chromaticity:}} There is a small chromatic change of C/B from the blue to the red part of the spectrum which is hardly explained by contamination from the lensing galaxy. The amplitude of this effect is however very small and our measurement are compatible with no CML and no DE. \\
{\it {Broad Lines:}} The ML of the BLR is large in this system, isolating the narrow component of the \CIV~flux in the non microlensed fraction of the spectrum. The signal is less pronounced in \CIII. We do not detect ML of the \HeII\,and \OIIIb\, emission. If we use the $K$-band flux ratio as the correct value of $M$, we find significant microlensing of the \CIII\, line. \\
{\it{Notes:}} Our spectroscopic estimate of C/B differs by 0.25 mag from the $K-$band measurement \citep{Blackburne2011a}. This is hardly explained by intrinsic variability and suggests significant reddening of image C. This also explains the $H-$band ratio. Alternatively, we might have underestimated the amount of microlensing in our spectra. This has to be confirmed with additional data.

{\bf {(i) J1226-0006:}} {\it {Deconvolution:}} During the deconvolution process, we forced the separation between the 2 lensed images and the lensing galaxy to match the HST separation in order to reduce cross-contamination. Although a good deconvolution is obtained, we keep in mind that residual contamination shouldn't be excluded due to the small separation of the system ($\Delta$ AG$=$0.437\arcsec, $\Delta$ AB$=$1.376\arcsec) and of a pixel size twice larger than for the other lenses (i.e. 0.2\arcsec/pix.).  \\
{\it {Chromaticity:}} There is a strong change of the flux ratio from the blue to the red. We find evidence for unproper deblending of the QSO and of the lens flux, especially below $\lambda \leq 6500 \AA$ -corresponding to the $4000\AA$ break of the lens- where the continuum spectrum of the lens and of the QSO have similar shapes. The decomposition of the \MgII~line also shows imprint of the $H$ and $K$ absorption bands from the lens. This effect could modify the intrinsic shape of the microlensing signal. Based on the \MgII~and \OII\,$\lambda 3727$\AA~lines, we find that only DE{\footnote{Assuming the same $M$ over the whole wavelength range implies that ML is the lower in the blue, in disagreement with the smaller source size at that wavelength.}} ($M($blue$)=$0.850, $M($red$)=0.765$, $\mu=$0.536) may cause the observed chromatic change, with image $B$ being more reddened by the lensing galaxy. This is unexpected as the lens is closer from image A than B. Another possibility would be a color change associated to the intrinsic variability, as the time delay in system should be of the order of 25 days.\\
{\it {Broad Lines:}}  We observe ML of the blue component of \MgII\,but we are unsure of the role of the contamination by the lensing galaxy in our spectral decomposition. \\
{\it {Notes:}} The  $H-$band flux ratio is similar to the optical flux ratio  $A=$ 0.456 but deviates significantly from our line-based estimates of $M=$ 0.80. Although intrinsic variability and differential extinction may play a role in the explaining the discrepancy, it seems plausible that the $H-$band continuum is in fact microlensed nearly at the same level as the optical one.  

{\bf {(j) SDSS~J1335+0118:}} {\it {Deconvolution: }} During the deconvolution process, we forced the separation between the bright lens image $A$ and the lens galaxy component $G$ to be identical to the HST separation. Small residual flux left after deconvolution close to $A$ and $B$ amounts less than 0.03\% of the QSO flux. \\
{\it {Chromaticity:}} The change of $B/A$ from the blue to the red is incompatible with CML, as the latter has to be stronger at bluer wavelengths. Instead, we hypothesise DE ($M($blue$)=$0.21, $M($red$)=$0.23, $\mu=$1.396), image $A$ being more reddened than image $B$. We discard the possibility that the observed chromatic trend is caused by residual contamination from the lensing galaxy as we estimate that more than 70\% of the observed flux of the galaxy should contaminate image $B$ to flatten the spectral ratio.  \\
{\it {Broad Lines:}} We observe ML of the red wing of \CIII~ and of \MgII. For \MgII\, and \CIII~, there is a second (very)-broad component which is microlensed. \\
{\it {Notes:}} The photometry published by \cite{Oguri2004a}, associated to data obtained about 2 years before our spectra, shows a slow decrease by 0.3 mag of $\Delta m_{BA}$ from $g-$ to $K-$band, compatible with our results, but they argue this is not conclusive due to their photometric error bars. On the other hand, they find a flat spectral ratio $B/A$ without clear imprint of the emission lines. This contrasts with our higher signal to noise spectra were we observe differential microlensing between the continuum and the emission lines. Owing to the expected IV (Table~\ref{tab:variability}), our estimate of $M=0.23$ is compatible with the $M =$ 0.29 measured in $H-$band. It however deviates significantly from the $K-$band measurement $M=0.41$ \citep{Oguri2004a}. To explain these ratios we have to postulate significant differential extinction. \cite{Oguri2004a} observed a chromatic decrease of $\Delta m_{BA}$ by 0.15 mag from $r-$band to $K-$band. This amount would lead to a value of $M$ corrected from reddening $M=0.264$. This still disagree with the $K-$band value but the two values get marginally compatible provided the effect of intrinsic variability is 50\% larger than predictions from the structure function. 

{\bf {(k) Q1355-2257{\footnote{We found an error in the wavelength calibration of image B and therefore, we had to shift the original wavelength solution by 2 \AA~for that image.}}:}} {\it {Deconvolution:}} During the deconvolution process, we forced the separation between the bright lens image $A$ and the lens galaxy component $G$ to be identical to the HST separation. The deconvolution is good but low level residual flux, up to 0.2\% of the faintest image, is visible aside the faintest lensed image and the lensing galaxy. A small excess of flux appears in the background and in the PSF component of the galaxy at the wavelength of the peak of the \MgII~emission. This flux is likely associated to image B but amounts less than 0.5\% of the \MgII~flux in that image. \\
{\it {Chromaticity:}}  Contrary to what is observed for the other systems, the spectral ratio B/A in the continuum does not vary in a monotonic way (Fig.~\ref{fig:Ratiomain}k). The factor $A$ is roughly the same in the continuum for $\lambda >$ 5450\AA~ but decreases significantly for bluer wavelengths. This trend cannot be explained by contamination from the lensing galaxy. There is significant differences between the spectra of images $A$ \& $B$ at basically every wavelength suggesting a complex ML of the continuum and of the broad line region, including the region emitting \FeII. Because we derive a similar value of $M$ around \MgII~ and \CIII~, it seems plausible that the chromatic change of the flux ratio in the continuum is caused by ML rather than DE although we cannot rule out the influence of the latter. The measurement of $M$ from the \NeV~ narrow emission lines, although more uncertain, is compatible with the one obtained for \MgII.\\
{\it {Broad Lines:}} We observe ML of the red wing of \MgII. The \CIII~line being close to the red-edge of the spectrum, we estimate $A$ for this line in the range 4650-4680 \AA. We do not find ML of that line.   \\
{\it {Notes:}} Our estimate of $M$ is in rough agreement with the $K-$band flux ratio once we account of the possible effect of intrinsic variability (Table~\ref{tab:variability}). The difference of $\sim$ 0.15 mag between the $H-$ and $K-$band ratios seems to be too large to be caused by intrinsic variability and is compatible with the $H-$band continuum affected by a small amount of microlensing. We scanned the spectra published by \cite{Morgan2003} and applied our decomposition method to these ones. We find for \MgII\, $(M, \mu) = (0.33, 0.63)$ and $(M, \mu) = (0.38, 0.56)$ around \CIII. Like in our spectra, only the red wing of \MgII\, is microlensed but not \CIII. The small differences on the derived values of $M$ are easily explained by intrinsic variability.

{\bf {(l) WFI~2033-4723:}}  {\it {Deconvolution:}} The deconvolution is good. Residual flux under the point-like images is $<$ 0.1\% of the QSO flux. \\
{\it {Chromaticity:}} The amplitude of the chromatic differences between the continuum in image $B$ and $C$ is small and probably caused by small uncertainties in the deblending of the QSO images and of the lens galaxy.  \\
{\it {Broad Lines:}} There is ML of the blue wing of \CIII~ or/and of \SiIII~but no ML of the \MgII~line.\\
 {\it {Notes:}} Our estimate of $M$ is in excellent agreement with the $H-$band and $K-$band measurements \citep{Vuissoz2008, Blackburne2011a}.

{\bf {(m) HE~2149-2745:}} {\it {Deconvolution:}}  The deconvolution is good. Residual flux under the point-like images is $<$ 0.09\% of the QSO flux. The background flux retrieved by the deconvolution process is probably associated to unproperly subtracted sky.  \\
{\it {Chromaticity:}} The measurement of $A$ at the level of the \CIV~ line is difficult because of its vicinity from the blue edge of the spectrum. Our estimate of $A$ for this line is performed in the range 5070-5120\,\AA~where \FeII~emission is minimal. There is no differential ML between the continuum and the emission lines. However the two spectra do not superimpose once scaled with the same magnification factor. First, there is a chromatic effect which may be caused by a small amount of DE, image $B$ being more extinguished than image $A$, in agreement with its location closer to the lens galaxy. Intrinsic variability combined with the time delay of $\sim$ 103 days \citep{Burud2002a} might also explain the observed color difference. Second, the absorbed fraction of the \CIV\,emission do not superimpose once scaled by $B/A \sim$ 0.242. This is likely caused by time-variable broad absorption which is seen in images $A$ and $B$ at two different epochs separated by the time delay. \\
{\it {Notes:}} Comparison of our spectra with those of \cite{Burud2002a} confirm our estimate of $M=0.245$ at the level of the \CIII~line. This value is also in agreement with the $H-$band flux ratio. 

\begin{table}[t!]
\begin{center}
\begin{tabular}{l|cccc}
\hline
Object & z& $\Delta t$ (days)  & $\Delta m$ (SF)  \\
\hline
(a) HE~0047-1756     & 1.678 & (12)   & 0.073  \\
(b) Q0142-100        & 2.719 & 131    & 0.354  \\
(c) SDSS~J0246-0825  & 1.689 & (5.5)  & 0.053  \\
(d) HE~0435-1223     & 1.693 & 6.5    & 0.058  \\
(e) SDSS~J0806+2006  & 1.540 & (50)   & 0.188  \\
(n) SDSS~J0924+0219  & 1.524 & (5.7)  & 0.061  \\
(f) FBQ~0951+2635    & 1.247 & (14)   & 0.072  \\
(g) BRI~0952-0115    & 4.426 & (13)   & 0.116  \\
(o) J1131-1231    & 0.657 & 12$^\dagger$ & 0.07 \\
(h) SDSS~J1138+0314  & 2.438 & (5)    & 0.067  \\
(i) J1226-0006       & 1.123 & (26)   & 0.123  \\
(j) SDSS~J1335+0118  & 1.570 & (49)   & 0.162  \\
(k) Q1355-2257       & 1.370 & (73)   & 0.194  \\
(p) H1413+117        & 2.55  & (20)   & 0.116  \\
(l) WFI~2033-4723    & 1.662 & 62     & 0.215  \\
(m) HE~2149-2745     & 2.033 & 103    & 0.273  \\
\hline
\end{tabular}
\end{center}
\vspace{0.2cm}
{\tiny{ {\bf {Notes:}} $\dagger$: Preliminary time delay estimate between A \& B by \cite{Morgan2006}.}}
\caption{Time delays for the observed systems and typical flux variation expected during this time-spent. When time delays have not been observed, we give in parentheses the predicted value from our lens model. We do not quote the values for Q2237+0305 and for the image pair B-C in J1131-1231 because the time delay is likely $<$1 day. }
\label{tab:variability}
\end{table}

\section{Characteristics of the extended sample}
\label{appendixB}

\begin{figure*} 
\centering
\setcounter{subfigure}{13}
\renewcommand{\thesubfigure}{(\alph{subfigure})}
  \subfigure[SDSS~J0924+0219 (A/B)]{\includegraphics[scale=0.35]{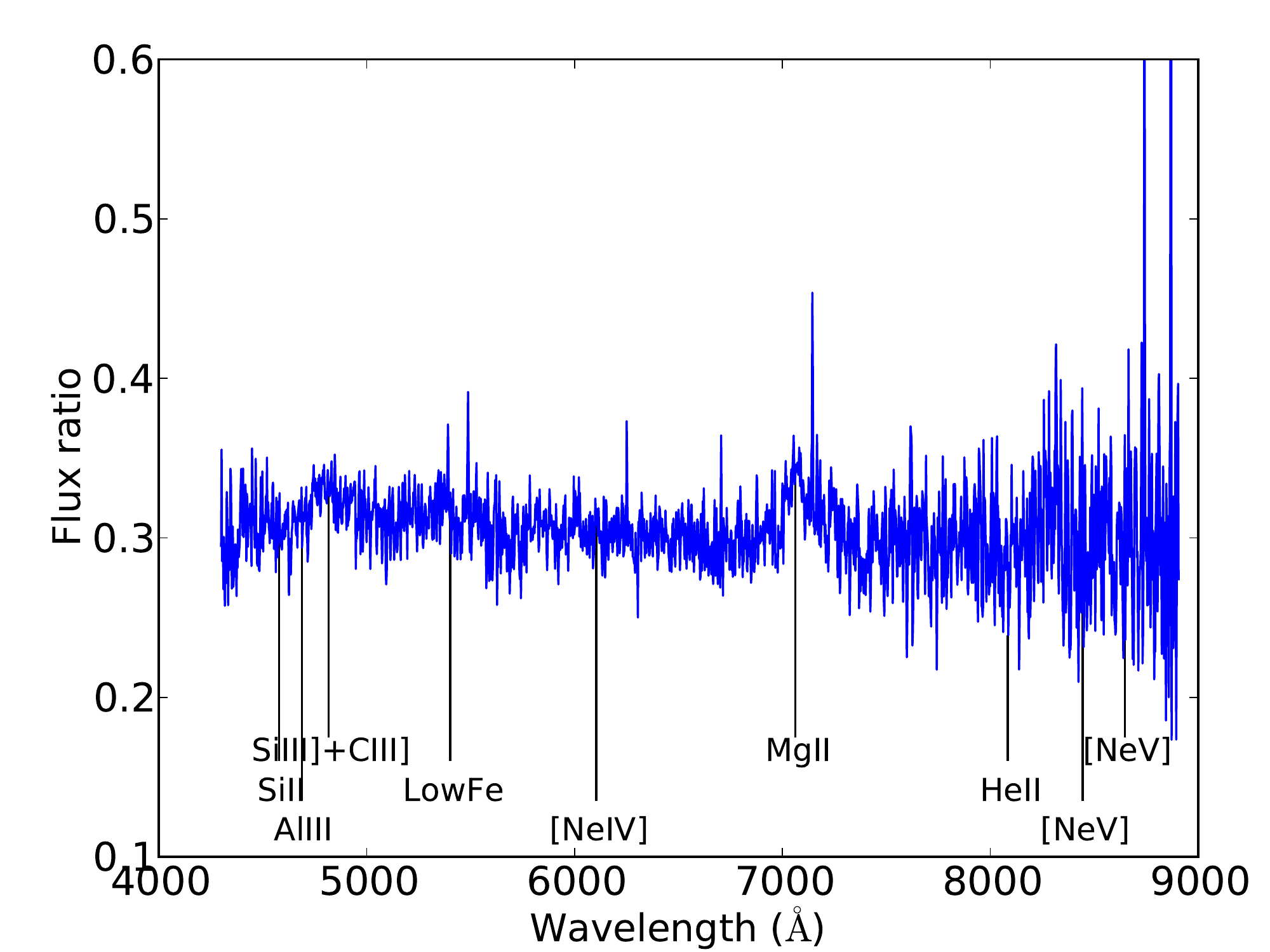}} 
\setcounter{subfigure}{14}
\renewcommand{\thesubfigure}{(\alph{subfigure}1)}
  \subfigure[J1131-1231 (B/C)]{\includegraphics[scale=0.35]{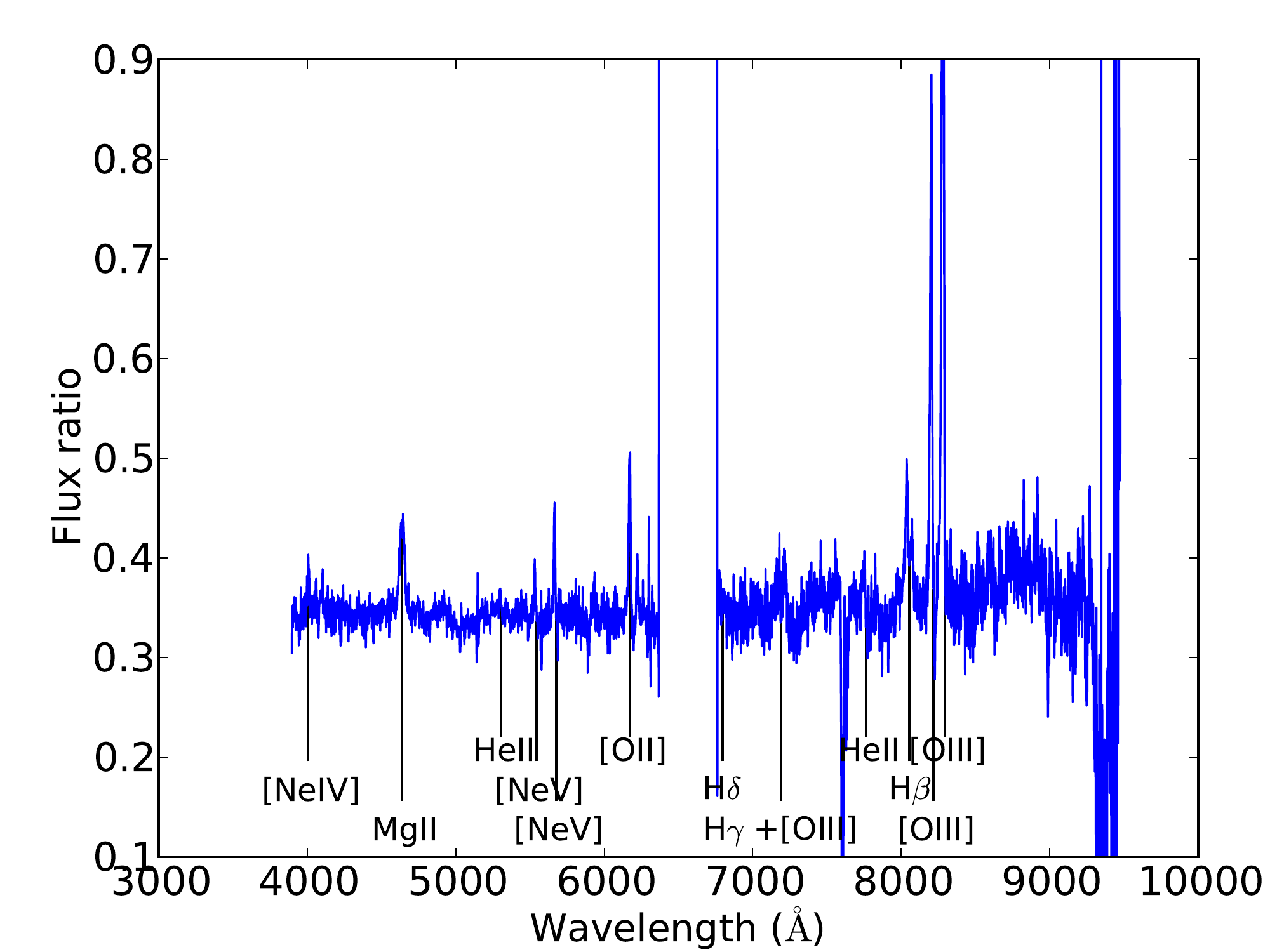}} 

\setcounter{subfigure}{14}
\renewcommand{\thesubfigure}{(\alph{subfigure}2)}
  \subfigure[J1131-1231 (A/B)]{\includegraphics[scale=0.35]{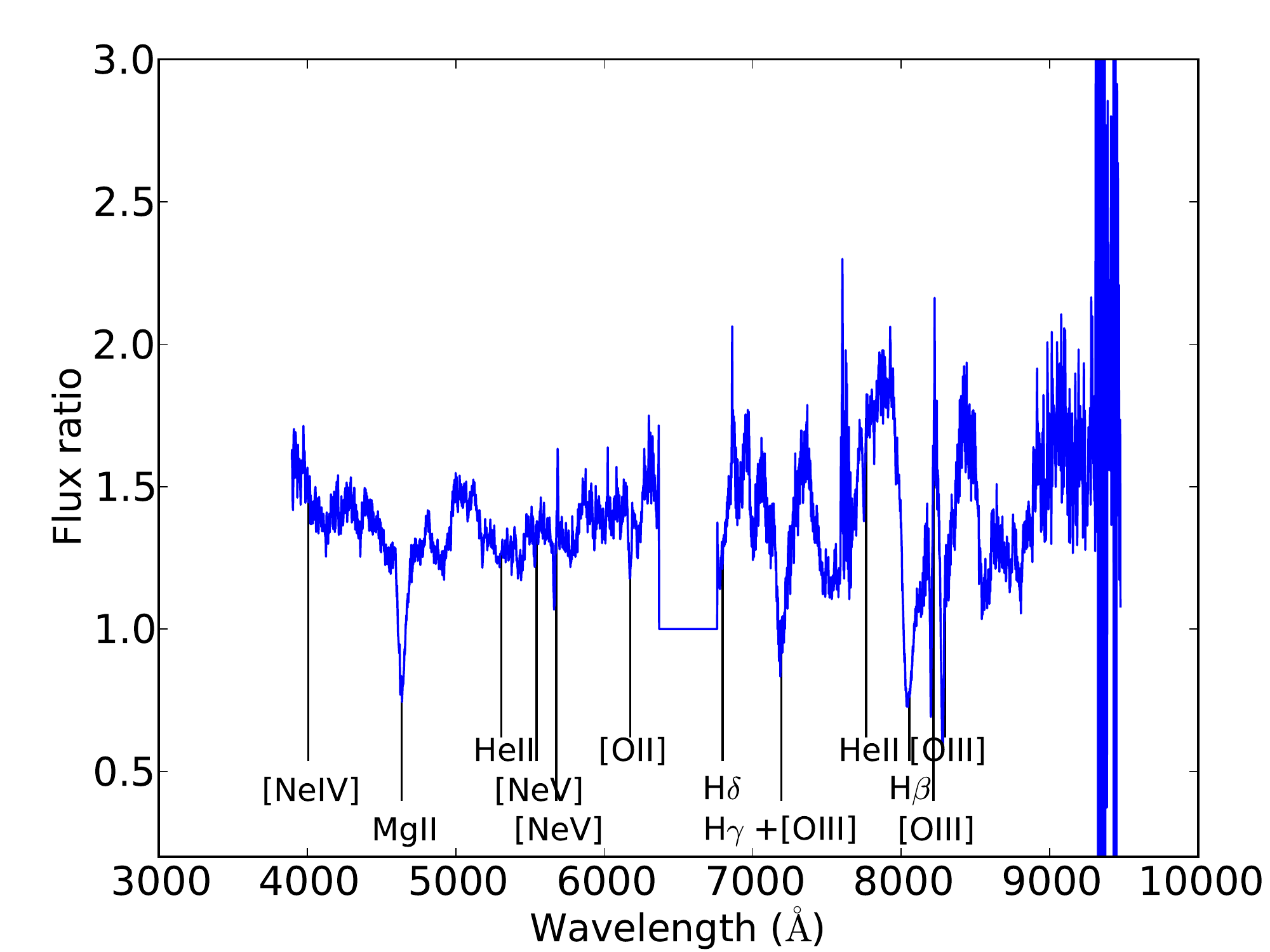}} 
\setcounter{subfigure}{15}
\renewcommand{\thesubfigure}{(\alph{subfigure}1)}
  \subfigure[H1413+117 (AB/D)]{\includegraphics[scale=0.35]{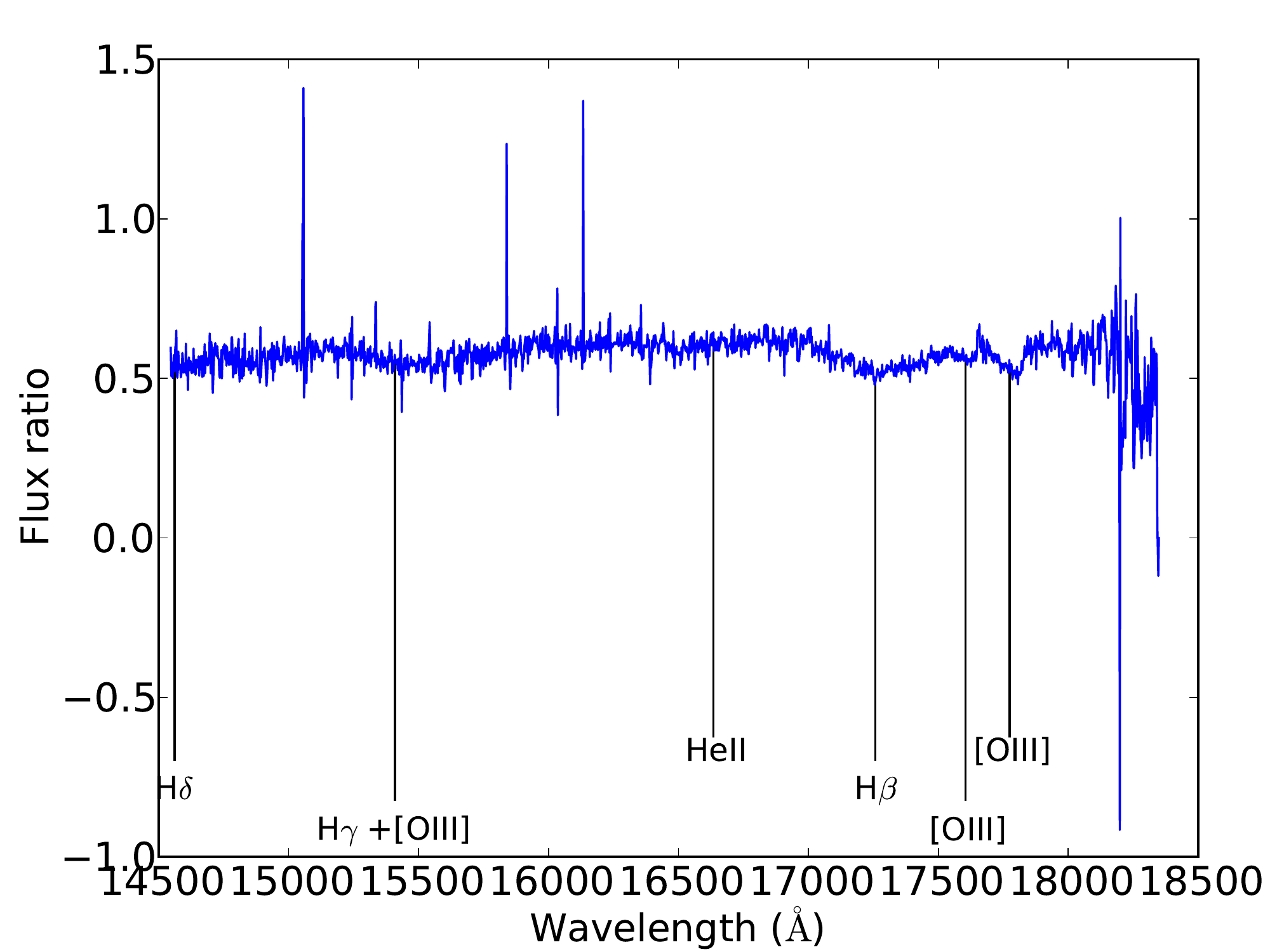}}              

\setcounter{subfigure}{15}
\renewcommand{\thesubfigure}{(\alph{subfigure}2)}
  \subfigure[H1413+117 (AB/D)]{\includegraphics[scale=0.35]{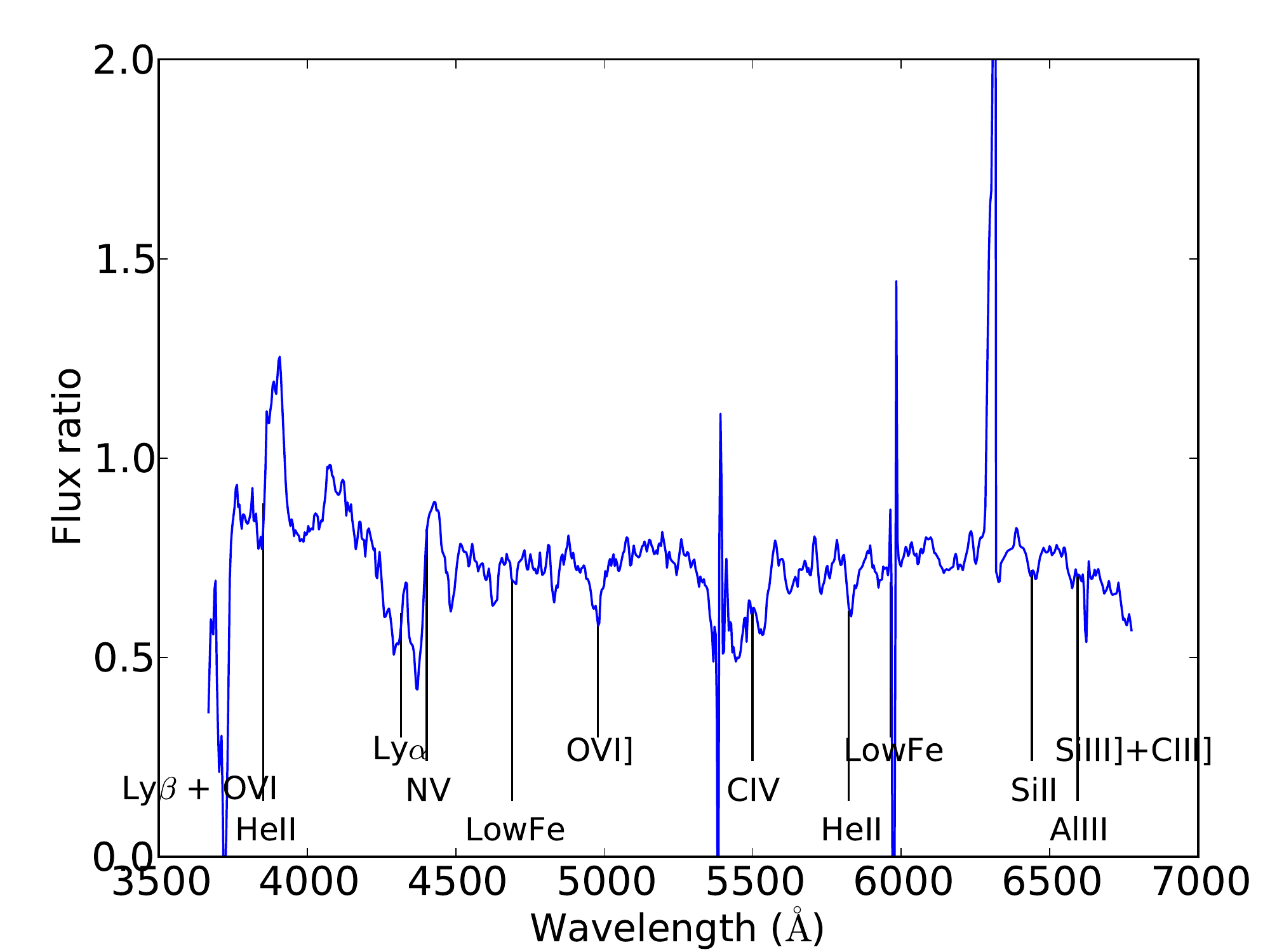}} 
\setcounter{subfigure}{16}
\renewcommand{\thesubfigure}{(\alph{subfigure})}
  \subfigure[Q2237+0305 (A/D)]{\includegraphics[scale=0.35]{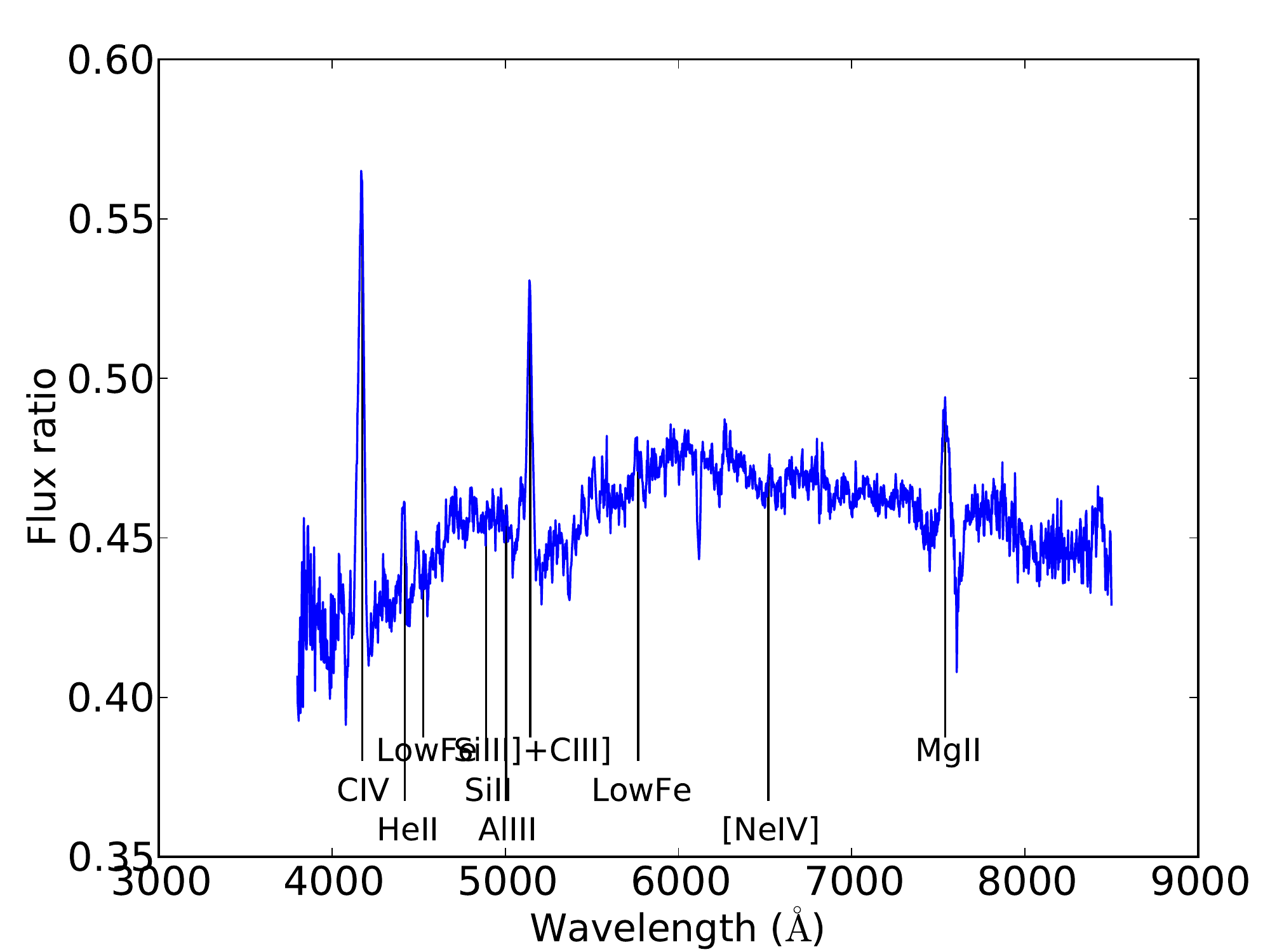}}              

\setcounter{subfigure}{17}
\renewcommand{\thesubfigure}{(\alph{subfigure})}
  \subfigure[HE 2149+0305 (A/B); Burud et al. (2000)]{\includegraphics[scale=0.35]{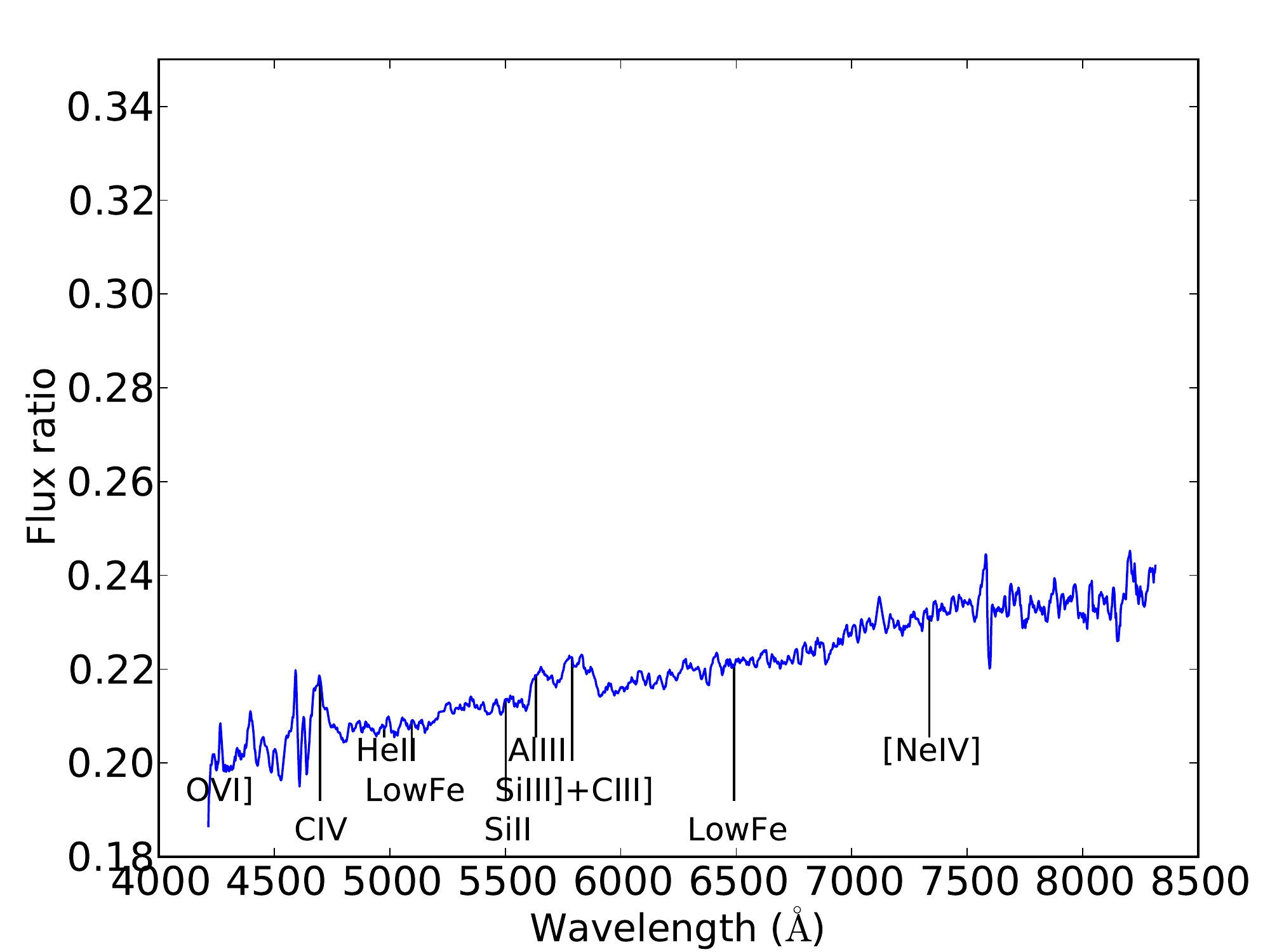}} 

\caption{Flux ratio for all the image pairs of our extended sample.}
\label{fig:Ratioext}
\end{figure*}

We summarize here the main characteristics of the extended sample of objects introduced in Sect.~\ref{subsec:addsys}. We provide the spectral ratio between the image pairs in Fig.~\ref{fig:Ratioext}. For each object, we also discuss our estimate of $M$ from the spectra at the light of literature data (except for HE 2149-2745 which was discussed in Appendix~\ref{appendixA}). 

{\bf {(n) SDSS~J0924+0219}} ($z_l = 0.394$, $z_s = 1.524$): The VLT-FORS spectra of images $A$ and $B$, obtained in Jan. and Feb. 2005 by \cite{Eigenbrod2006a}, show only a weak differential ML between the continuum and the emission lines (\CIII\, and \MgII) but no clear ML deformation of the lines (Fig.~\ref{fig:MmD}). The spectral ratio is flat over the spectral range of our spectra. On the other hand, \cite{Keeton2006a} identified flux ratios in disagreement with macro model predictions in both the continuum and emission lines (\Lyalpha, \CIV\, and \CIII) based on low resolution spectra obtained with the Advanced Camera for Survey (ACS) on 29.05.2005. They demonstrated, based on microlensing simulations, that the observed anomalies in the image pair $A-D$ could easily be explained by ML. Unfortunately, their spectra have insufficient spectral resolution to allow a proper investigation of the line profile differences. \\
{\it {Notes:}} The flux ratio $A/B=0.44\pm0.04$ in $H-$band \citep{Eigenbrod2006a} is in good agreement with our estimate of $M$. Although $K-$band observation of this system exist \citep{Faure2011}, they were not able to derive accurate photometry of the lensed images due to the ring. No other NIR/MIR photometric data of this system are available in the literature. 

{\bf {(o) J1131-1231}} ($z_l=0.295$, $z_s=0.657$): \cite{Sluse2007} presented single epoch spectra of this system obtained in April 2003. They identified ML of the broad emission lines for images $A$ and $C$ and presented MmD of the \Hbeta\,and \MgII\, emission lines using the image pairs $A-B$ and $C-B$. They also identified contamination of the spectra by flux from the host galaxy and empirically corrected for this effect. In image $C$, only the core of the emission line is not microlensed while microlensing affects the broad component of the lines in image $A$. Because of the uncertainty on the host contamination, exact values of $M$ and $\mu$ may be more prone to systematic uncertainties{\footnote{Because the narrow \OIII\, lines are resolved and possibly macro-magnified by a different amount than the broad lines, their use to set-up $M$ might be another source of systematic errors.}} than in other systems. Nevertheless, the microlensing of the emission lines is a robust result which qualitatively remains even if $M$ or $\mu$ are under/over-estimated by up to 40\%.\\  
{\it {Notes:}} The flux ratios in $K-$band disagree with those derived from the narrow \OIII~emission lines \citep{Sluse2007, Sugai2007}. Differential extinction is not a plausible explanation because of the lack of monotonic change of the flux ratios with wavelength and because of the lack of strong hydrogen absorption in X-ray \citep{Chartas2009}. Therefore, it is likely that the $K-$band flux ratio is significantly microlensed. 

{\bf {(p) H1413+117}} ($z_l$ unknown, $z_s=2.55$): This system is the first broad absorption line (BAL) quasars where ML has been unambiguously observed \citep{Angonin1990, Hutsemekers1993}. \cite{Hutsemekers2010} presented the MmD for this system at four different epochs spanning a 16-years time-range. They identified that ML was affecting mostly image $D$, roughly in the same way, along the time-spent of the observations. The \Lyalpha, \CIV, \Hbeta, and \Halpha~emission have been analysed. In addition to the ML in the broad absorption, they found evidence for ML of the central core of the \CIV\, and of the \Lyalpha\, (i.e. the wings are not microlensed), but no ML in the Balmer lines (see their figures 6, 7, 8). We show in Fig.~\ref{fig:MmD} the decomposition for the \CIV\, and for the \Hbeta\, lines for the spectra obtained in 2005. Chromatic changes of the flux ratios are observed in this system \citep{Hutsemekers2010, Munoz2011}. On one hand, there is differential extinction between images $A$ \& $B$, and on the other hand, there is chromatic microlensing of image $D$ consistent with microlensing of a standard accretion disk \citep{Hutsemekers2010}. Note that microlensing also affects component C \citep{Popovic2005}, but the line profile differences are more subtle \citep{Hutsemekers2010}. \\

{\bf {(q) HE~2149-2745}} ($z_l = 0.603$, $z_s = 2.033$): Our spectra of this system do not show evidence of ML, however, \cite{Burud2002a} presented spectra of the two lensed images of this BAL quasar, obtained on 19.11.2000, where they observed chromatic changes of the continuum. They mentioned subtle differences in the line profile of \CIII\, but we fail to detect clear ML signature of the lines using the MmD on these spectra. The MmD is however difficult to perform because of the significant \FeII\, emission blueward of \CIII\, and of the significant chromaticity of the spectral ratio which is sensible even on the small wavelength range covered by the line. We display in Fig.~\ref{fig:MmD} the decomposition of \CIII\, using $\mu(\lambda)$ instead of the average $\mu$ between the blue and red part of the line. The chromatic changes observed in these spectra have been discussed by \cite{Burud2002a} as possibly caused by differential extinction or chromatic microlensing. Although we may not rule out that differential extinction is present, we are now able to say based on the new spectra that the slope in the spectra of \cite{Burud2002a} was mostly caused by CML. \\

{\bf {(r) Q2237+0305}} ($z_l=0.0394$, $z_s=1.695$): There is a clear ML of the broadest component of the \CIV\, and \CIII\, lines which has been observed in image $A$ over the 3 years time-spent of the spectrophotometric monitoring presented in \cite{Eigenbrod2008a}. This signal has been used by \cite{Sluse2011a} to derive a size of the BLR in agreement with the size-luminosity relation obtained by reverberation mapping studies for other systems. Image $D$ has been found to be affected by differential extinction. The MmD applied to the extinction corrected spectra, averaged over the first year of the monitoring (Oct. 2004-Sept. 2005), is shown in Fig.~\ref{fig:MmD}. \\ 
{\it {Notes:}} Although we originally used the lens model flux ratio $A/D=1.0$ to make the MmD, we are also able to derive $M$ empirically using the MmD. Following that procedure, we derive a very similar value of $M$ (Table~\ref{tab:analysis}), but we disfavour $M<0.87$ because they lead to the appearance of a clear dip, that we consider as unphysical, in the center of the $F_{M\mu}$ fraction of the \CIII~line. \cite{Falco1996} published radio flux ratios $A/D=$0.77$\pm$0.23 and \cite{Minezaki2009} published $D/A=0.87\pm0.05$ at 11.67 $\mu $m. These values are in good agreement with our spectroscopic estimates. We should however notice that small differences between the MIR flux ratios of \cite{Minezaki2009}, \cite{Agol2009} and \cite{Agol2001}, as well as chromatic changes in the MIR, suggest that a small amount of microlensing may still affect these wavelengths.

\end{document}